\documentclass[fleqn,usenatbib]{mnras}

\usepackage{newtxtext,newtxmath}

\usepackage[T1]{fontenc}
\usepackage{ae,aecompl}

\usepackage[T1]{fontenc}
\usepackage{ae,aecompl}
\usepackage{natbib}
\usepackage{graphicx}	
\usepackage{amsmath}	
\usepackage{booktabs}

\newcommand{\cinnzero}{88131}
\newcommand{\coutzero}{32677}

\newcommand{\nfog}{967}
\newcommand{\nfogrestr}{677}
\newcommand{\totrel}{94604}
\newcommand{\totpoor}{55619}
\newcommand{\totsou}{150223}
\newcommand{\totsouwdist}{120808}
\newcommand{\perctotsouwdist}{80}

\newcommand{\percprotoinn}{40}

\newcommand{\percprotoout}{47}

\newcommand{\nprotoperc}{23}

\newcommand{\sffmin}{0.26}
\newcommand{\sffmax}{0.42}
\newcommand{\sffslope}{-0.040}
\newcommand{\sffsloperr}{0.003}
\newcommand{\sjumpe}{2.9}
\newcommand{\sjumpo}{3.7}
\newcommand{\gjumpe}{4.1}
\newcommand{\egjumpe}{0.5}
\newcommand{\gjumpo}{5.0}
\newcommand{\egjumpo}{0.7}
\newcommand{\percores}{19.7}
\newcommand{\perclumps}{80.1}
\newcommand{\perclouds}{0.3}
\newcommand{\maxmassratio}{4.3}
\newcommand{\minmassratio}{1.1}
\newcommand{\andersonmatches}{2560}
\newcommand{\andersonmatchesnoq}{2192}
\newcommand{\ntesttbol}{419}
\newcommand{\testtbolfiftyttwo}{6.7}
\newcommand{\testtbolninetyttwo}{16.3}
\newcommand{\testtbolfiftytfour}{1.8}

\newcommand{\mediantempinnproto}{15.2}
\newcommand{\mediantempinnpre}{11.4}
\newcommand{\mediantempoutproto}{15.3}
\newcommand{\mediantempoutpre}{10.5}
\newcommand{\mediantempinnhii}{24.6}
\newcommand{\mediantempouthii}{23.9}
\newcommand{\mediantempinnnomir}{14.6}
\newcommand{\mediantempoutnomir}{15.3}
\newcommand{\tenpercenttempinnhii}{20.4}
\newcommand{\tenpercenttempouthii}{20.3}
\newcommand{\percinterstinnproto}{57}
\newcommand{\percinterstinnpre}{39}
\newcommand{\percinterstoutproto}{44}
\newcommand{\percinterstoutpre}{39}
\newcommand{\mediansdinnproto}{0.21}
\newcommand{\mediansdinnpre}{0.14}
\newcommand{\mediansdinnnomir}{0.24}
\newcommand{\mediansdinnhii}{0.07}
\newcommand{\mediansdoutproto}{0.12}
\newcommand{\mediansdoutpre}{0.10}
\newcommand{\mediansdoutnomir}{0.15}
\newcommand{\mediansdouthii}{0.05}
\newcommand{\nhiiinner}{2806}
\newcommand{\nhiiouter}{869}
\newcommand{\medianlminnproto}{2.6}
\newcommand{\medianlminnpre}{0.2}
\newcommand{\medianlmoutproto}{3.1}
\newcommand{\medianlmoutpre}{0.1}
\newcommand{\nlmrzamsinn}{21}
\newcommand{\nlmrzamsout}{24}
\newcommand{\percinterslminnproto}{39}
\newcommand{\percinterslminnpre}{27}
\newcommand{\percinterslmoutproto}{29}
\newcommand{\percinterslmoutpre}{25}
\newcommand{\medianlsubinnpre}{5.7}
\newcommand{\medianlsubinnproto}{30.4}
\newcommand{\medianlsuboutpre}{4.4}
\newcommand{\medianlsuboutproto}{36.9}
\newcommand{\percinterslsubinnproto}{32}
\newcommand{\percinterslsubinnpre}{22}
\newcommand{\percinterslsuboutproto}{24}
\newcommand{\percinterslsuboutpre}{21}
\newcommand{\mediantbinnproto}{39.5}
\newcommand{\mediantbinnpre}{17.6}
\newcommand{\mediantboutproto}{43.4}
\newcommand{\mediantboutpre}{16.2}
\newcommand{\mediantbinnhii}{50.5}

\newcommand{\mediantbouthii}{51.5}

\newcommand{\percinterstbinnproto}{14}
\newcommand{\percinterstbinnpre}{9}
\newcommand{\percinterstboutproto}{5}
\newcommand{\percinterstboutpre}{5}
\newcommand{\perckauffproto}{62}
\newcommand{\perckauffpre}{56}
\newcommand{\perckrumhtot}{4}
\newcommand{\perckauffprototest}{55}
\newcommand{\perckauffpretest}{47}
\newcommand{\nbressert}{57}
\newcommand{\nbressertouter}{18}

\newcommand{\iaps}{$^1$}
\newcommand{\chil}{$^2$}
\newcommand{\lamm}{$^3$}
\newcommand{\iuf}{$^4$}
\newcommand{\toron}{$^5$}
\newcommand{\oafi}{$^6$}
\newcommand{\calg}{$^7$}
\newcommand{\cardiff}{$^8$}
\newcommand{\liver}{$^9$}
\newcommand{\stsi}{$^{10}$}
\newcommand{\caltech}{$^{11}$}
\newcommand{\porto}{$^{12}$}
\newcommand{\esa}{$^{13}$}
\newcommand{\irabo}{$^{14}$}
\newcommand{\koln}{$^{15}$}
\newcommand{\unile}{$^{16}$}
\newcommand{\inafle}{$^{17}$}
\newcommand{\chalm}{$^{18}$}
\newcommand{\charlot}{$^{19}$}
\newcommand{\nwest}{$^{20}$}
\newcommand{\colo}{$^{21}$}
\newcommand{\buda}{$^{22}$}
\newcommand{\oar}{$^{23}$}
\newcommand{\torv}{$^{24}$}
\newcommand{\diet}{$^{25}$}

\def\pdeg{\ifmmode $\setbox0=\hbox{$^{\circ}$}\rlap{\hskip.11\wd0 .}$^{\circ}
          \else \setbox0=\hbox{$^{\circ}$}\rlap{\hskip.11\wd0 .}$^{\circ}$\fi}

\title[Full catalogue of properties of Hi-GAL clumps]{The Hi-GAL compact source catalogue - II. The 
360$^{\circ}$ catalogue of clump physical properties}

 \author[D. Elia et al.]{Davide Elia,\iaps\thanks{E-mail: davide.elia@iaps.inaf.it}
M. Merello,\chil~
S. Molinari,\iaps~
E. Schisano,\iaps~
A. Zavagno,\lamm~,\iuf~
D. Russeil,\lamm~
P. M\`{e}ge,\lamm~
\newauthor
P.~G. Martin,\toron~
L. Olmi,\oafi~
M. Pestalozzi,\iaps~
R. Plume,\calg~
S.~E. Ragan,\cardiff~
M. Benedettini,\iaps~
D.~J. Eden,\liver~
\newauthor
T.~J.~T. Moore,\liver~
A. Noriega-Crespo,\stsi~
R. Paladini,\caltech~
P. Palmeirim,\porto~
S. Pezzuto,\iaps~
G.~L. Pilbratt,\esa
\newauthor
K.~L.~J. Rygl,\irabo~
P. Schilke,\koln~
F. Strafella,\unile~,\inafle~
J.~C. Tan,\chalm,\charlot~
A. Traficante,\iaps~
A. Baldeschi,\nwest~
J. Bally,\colo~
\newauthor
A. M. di Giorgio,\iaps~
E. Fiorellino,\buda,\oar,\torv~
S.~J. Liu,\iaps~
L. Piazzo,\diet~
D. Polychroni\iaps~
\\ \\
Author affiliations are listed at the end of the paper}

\date{Accepted XXX. Received YYY; in original form ZZZ}

\pubyear{2020}

\begin{document}
\label{firstpage}
\pagerange{\pageref{firstpage}--\pageref{lastpage}}
\maketitle

\begin{abstract}
We present the $360^\circ$ catalogue of physical properties of Hi-GAL
compact sources, detected between 70 and 500~$\umu$m. This 
release not only completes the analogous catalogue previously produced by the Hi-GAL 
collaboration for $-71^\circ \lesssim \ell \lesssim 67^\circ$, 
but also meaningfully improves it thanks to a new set of heliocentric distances, \totsouwdist\ in total.
About a third of the \totsou\ entries are located in the newly added portion of the Galactic plane.
A first classification based on detection at 70~$\umu$m as a signature of ongoing star-forming activity distinguishes between protostellar sources (\nprotoperc~per cent of the total) and starless sources, 
with the latter further classified as gravitationally bound (pre-stellar) or unbound.
The integral of the spectral energy distribution, including ancillary photometry
from $\lambda=21$ to 1100~$\umu$m, gives the source luminosity and other
bolometric quantities, while a modified black body fitted to data for $\lambda \geq 160~\umu$m 
yields mass and temperature.
All tabulated clump properties are then derived using photometry and heliocentric
distance, where possible. Statistics of these
quantities are discussed with respect to both source Galactic
location and evolutionary stage. No strong differences in the distributions of evolutionary
indicators are found between the inner and outer Galaxy. However, masses and densities in the inner Galaxy
are on average significantly larger, resulting
in a higher number of clumps that are candidates to host massive star formation. Median behaviour of
distance-independent parameters tracing source evolutionary status is examined as 
a function of the Galactocentric radius, showing no clear evidence 
of correlation with spiral arm positions.

\end{abstract}

\begin{keywords}
{Stars: formation -- ISM: clouds -- ISM: dust -- Galaxy: local interstellar matter -- Infrared: ISM -- Submillimeter: ISM}
\end{keywords}



\section{Introduction}\label{intro}
\defcitealias{eli17}{Paper~I}
The observational study of star formation makes use of both analysis of single objects and small regions
and analysis of large surveys. These two approaches are complementary, because surveys 
provide observers with numerous targets to be inspected in more detail, e.g., by means of
interferometric techniques. At the same time, large surveys produce a Galactic-scale 
view of star formation, which in turn represents a fundamental bridge between our knowledge 
of this phenomenon in the Milky Way and in external galaxies.
Moreover, almost every recent article presenting studies of early phases of star formation 
based on infrared/sub-mm surveys begins with highlighting the importance of a statistical approach
to address the formation of massive stars, which is still quite elusive because of
intrinsically low incidence and relatively fast time scales. Notwithstanding, massive stars have significant
feedback on the surrounding environment and hence the great interest.
Among these surveys, Hi-GAL
\citep[\textit{Herschel}\footnote{\textit{Herschel} is an ESA space observatory with science 
instruments provided by European-led Principal Investigator consortia and with important 
participation from NASA.} InfraRed Galactic Plane Survey,][]{mol10a} has a unique combination of characteristics favourable for systematically 
observing the early stages of star formation throughout the Milky Way. Hi-GAL was an Open Time
Key Project that was granted about 1000~hours of observing time with the \textit{Herschel} 
Space Observatory \citep{pil10}, drawn from all three \textit{Herschel} Announcements of Opportunity (KPAO, AO1,
AO2) supplemented by Director's Discretionary Time (DDT). Hi-GAL data were taken in parallel mode, using the two cameras on 
board \emph{Herschel}: PACS \citep[70 and 160~$\umu$m bands,][]{pog10} and SPIRE 
\citep[250, 350 and 500~$\umu$m bands,][]{gri10}. The uniqueness of Hi-GAL 
with respect to other Galaxy plane surveys in the continuum 
is threefold:
($i$) the wavelength range covered was crucial for studying the spectral energy distribution (SED) of cold dust, whose peak is expected to fall at $\lambda > 100~\umu$m; 
($ii$) the unprecedented sensitivity and dynamical range of the satellite-borne \textit{Herschel} observations enabled the detection of both diffuse emission and faint compact sources; 
and ($iii$) the unbiased observation of the entire Galactic plane probed a statistically significant variety of environmental conditions across the Milky Way. 

The first Hi-GAL instalment of observing 
time (AO) spanned the Galactic coordinate range $-71.0^{\circ} \lesssim \ell \lesssim 67.0^{\circ}$,
$|b|< 1.0^{\circ}$. For this area \textit{towards} the inner Galaxy (somewhat inappropriately dubbed ``inner Galaxy'', 
see Section~\ref{innerouter}), 
single-band photometric catalogues of compact sources (namely objects unresolved or poorly 
resolved in the maps) were delivered by \citet{mol16a}. 
Using this photometry, \citet[][hereafter Paper~I]{eli17} compiled a catalogue of 
physical properties of sources with a reliable spectral energy distribution (SED).

In the present context of completion of the $360^{\circ}$ coverage of Hi-GAL,
the work of Molinari et al. (in preparation) will represent the completion of the photometric catalogues of
\citet{mol16a}, while this paper represents the completion of the physical source catalogue
of \citetalias{eli17}, giving a global view of early phases of star formation across the whole Milky Way.
Quantitatively, with this work we extend the longitude coverage of \citetalias{eli17} by
a factor $\sim 2.6$ and increase the number of catalogued reliable compact sources from 100922 to
\totsou.

The main improvements and advances are as follows:
\begin{itemize}
\item For longitudes outside the range already published in \citetalias{eli17}, we 
present for the first time the catalogue of compact source
physical properties and discuss their statistics, similarly to \citetalias{eli17}.
\item For the longitude range already published in \citetalias{eli17}, we use
a new set of source kinematic distances delivered by \citet{meg21}, 
accordingly rescaling all of the distance-dependent parameters (such as sizes, masses, luminosities).
\item It is now possible to discuss the distribution of such physical properties as a function of position
in the Galaxy. In particular, we focus on comparison of overall statistics
for the inner Galaxy and outer Galaxy, defined here by the radial zones in the latitude range covered that are inside and outside the Solar circle, respectively.
\end{itemize}

In Section~\ref{building} the procedures followed to build the catalogue are 
described briefly, referring the reader to \citetalias{eli17} for further details. 
Sections~\ref{distparam} and~\ref{nodistparam} focus on the statistics of distance-dependent 
and distance-independent parameters, respectively. In Section~\ref{galactotrends},
global trends for such quantities as a function of the Galactocentric distance are 
discussed. Section~\ref{summary} summarises our conclusions. Finally,
Appendix~\ref{catdescription} gives a detailed description of each catalogue column and Appendix ~\ref{mirappendix} 
contains a brief analysis
of flux distributions for mid-infrared (MIR) counterparts of our Hi-GAL sources.

\section{Building the catalogue}\label{building}

\subsection{SED selection, classification, and fitting}\label{sedsel}

The Hi-GAL SEDs were assembled, starting from the single-band photometry lists of 
Molinari et al. (in preparation) and adopting the same procedure used for \citetalias{eli17}.
Here we briefly summarise the main steps 
and subsequent filtering, referring the reader
to \citetalias{eli17} for further details.

\begin{itemize}
\item 
Only sources detected in the common area surveyed by both PACS and SPIRE cameras
are considered for subsequent steps. For the observations already considered in \citetalias{eli17}, the boundaries 
of this area have been refined, being slightly but systematically enlarged. At the end of the selection 
process described below, this results in the inclusion of more than 1000 new sources at longitudes
already covered in \citetalias{eli17}. This, and especially the adoption of a new set of heliocentric distances,
lead us to discuss a new the statistics of source properties in this longitude range, an important part of the full catalogue.

\item
Sources detected at different \textit{Herschel} bands are associated as counterparts
of the same object simply based on position matching \citep[see also][]{eli13}. Possible cases of 
multiplicity are resolved simply by keeping the closest counterpart. Only for the 70~$\umu$m and 
MIR ancillary bands (see below) is the total flux of all possible counterparts also
computed, for use in calculating bolometric parameters.
\item To filter SEDs as being suitable for fitting with a modified black body (hereafter MBB),
SEDs are accepted as reliable by having at least three adjacent fluxes in the spectral range 
$160~\umu\mathrm{m} \leq \lambda \leq 500~\umu\mathrm{m}$, a 
concave-down shape, and $F_{350}-F_{500}>0$. This selects SEDs of
\totsou~sources.

\begin{figure}
\centering
\includegraphics[width=8.4cm]{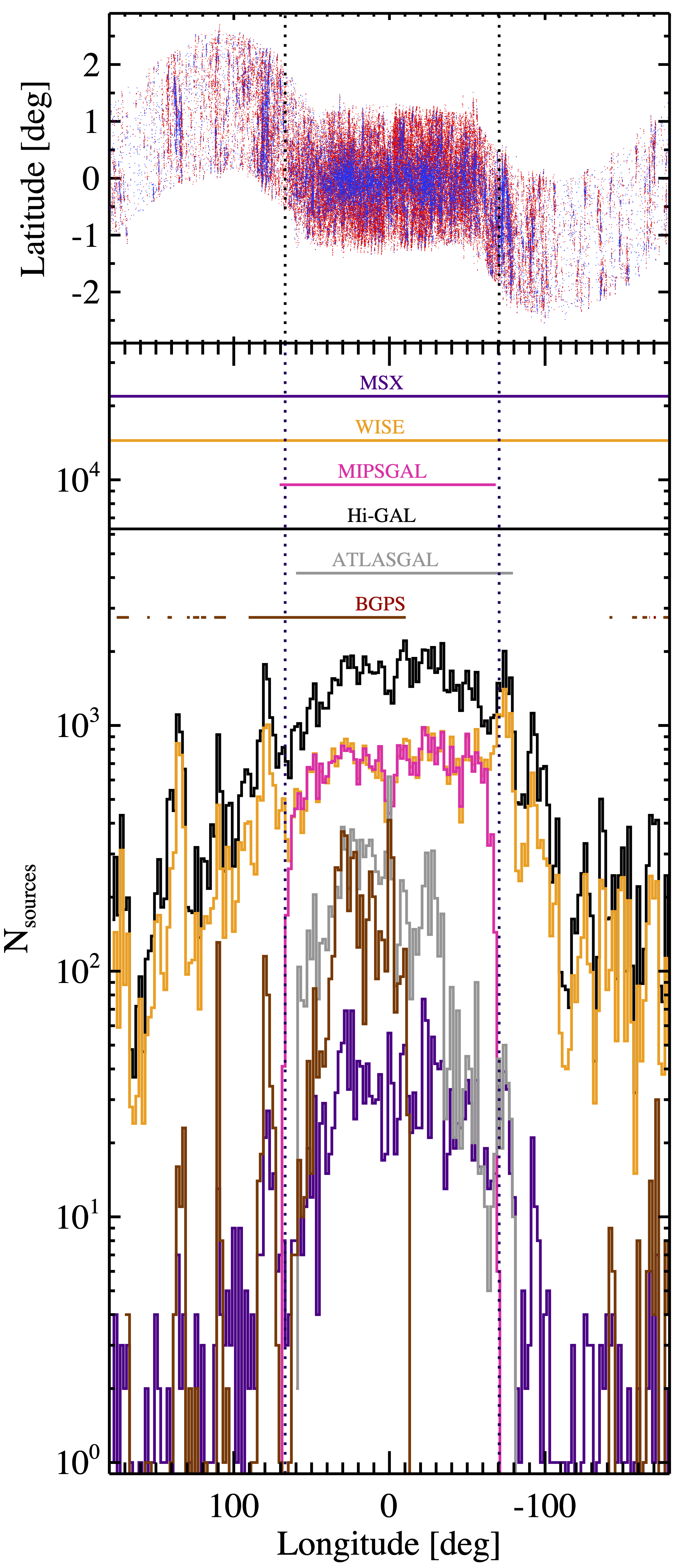}
\caption{Top panel: Positions ($\ell,b$) of Hi-GAL sources. To reduce crowding in the plot, only
protostellar (blue) and pre-stellar (red) sources in the ``high-reliability'' catalogue
are displayed. Bottom panel: histogram of Hi-GAL sources (black) in $2^{\circ}$-bins of
Galactic longitude, together with histograms of counterparts found in the
MSX (purple), WISE (orange), MIPSGAL (magenta), ATLASGAL (grey), and BGPS (brown) surveys,
respectively. Local peaks (in logarithmic scale) in the outer Galaxy at about $-168^\circ$,
$-154^\circ$, $-136^\circ$, $-92^\circ$, $-74^\circ$, $+80^\circ$, and $+136^\circ$, can be 
attributed to Gem~OB1, Rosette, CMA~OB1, Vela~C, Cygnus-X, and W3/W4/W5 regions, 
respectively. In the upper part of the panel, the longitude coverage for each survey is also shown. 
Dotted vertical lines crossing both panels delimit the longitude range already presented
in \citetalias{eli17}.}
\label{surveycoverage}
\end{figure}

\item
With the same procedure used in \citetalias{eli17}, counterparts to the Hi-GAL sources at 21, 
22, 24, 870 and 1100~$\umu$m have been found in the MSX \citep{ega03}, WISE \citep{wri10},
MIPSGAL \citep{gut15}, ATLASGAL \citep{sch09,cse14}, and BGPS \citep{ros10,gin13} catalogues, respectively.
The coverage of the outer Galaxy is full for MSX and WISE, and poor or even missing for the
other surveys (Fig.~\ref{surveycoverage}). While we used photometry from sub-millimetre 
surveys to better constrain the MBB fit (see below), MIR fluxes, where available, were used
instead to quantify the excess of emission with respect to such a fit, which can significantly increase the 
estimates of bolometric quantities (see below). In this respect, MSX and WISE compensate for the unavailability
of MIPSGAL in the outer Galaxy, ensuring a uniformity between the inner and outer Galaxy, at least for 
sources brighter than 0.1~Jy (see Appendix~\ref{mirappendix}, Fig.~\ref{mirfluxes}). The same applies 
for areas severely saturated in MIPSGAL (see, e.g., the dip around $\ell=0^{\circ}$ in the distribution of MIPSGAL
sources in Fig.~\ref{surveycoverage}, bottom).

\begin{figure*}
\centering
\includegraphics[width=16cm]{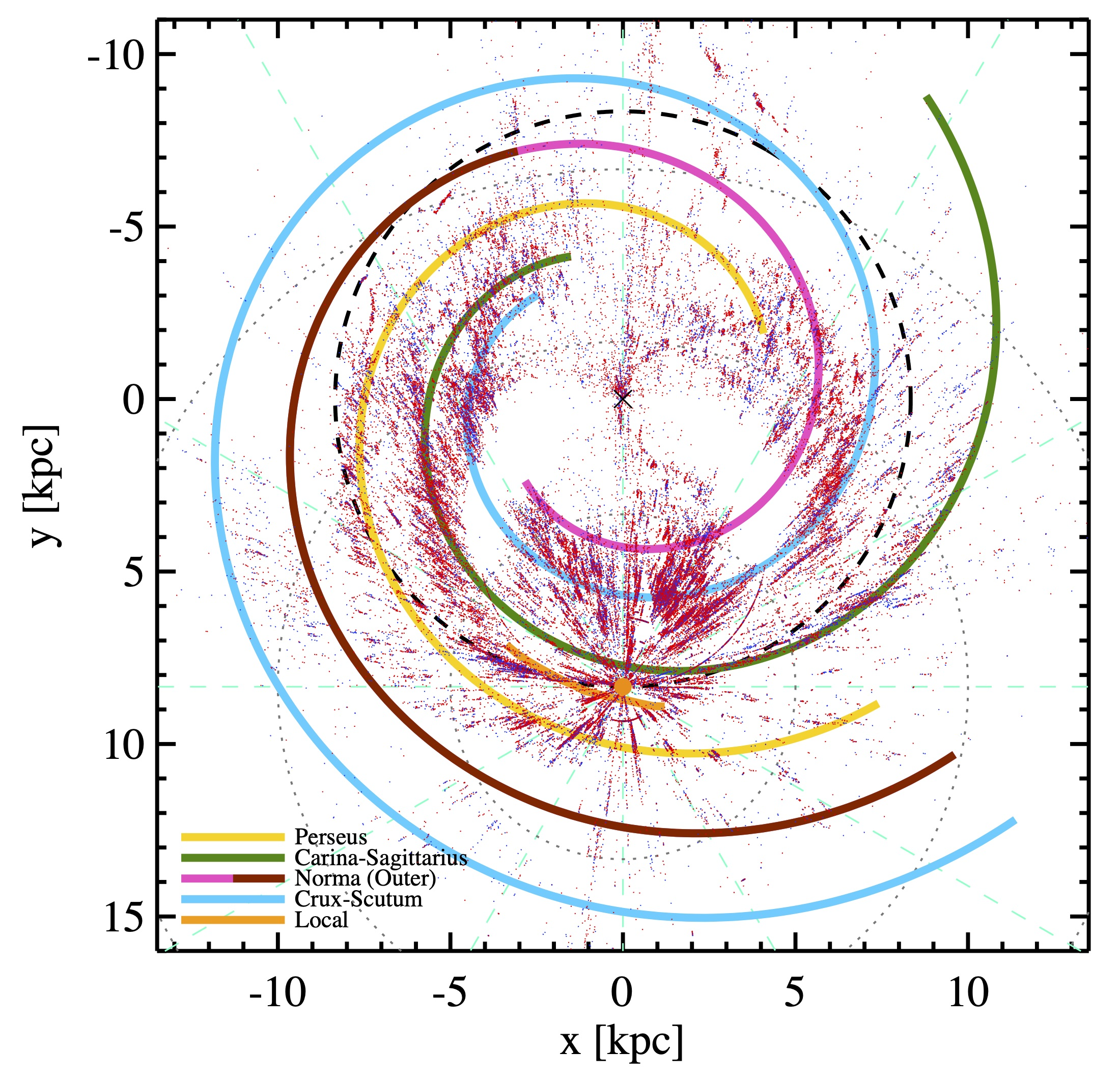}
\caption{Positions projected in the Galactic plane for the pre-stellar (red dots) and protostellar (blue dots) Hi-GAL objects with a known distance
(unbound objects not shown to reduce crowding). The Galactic centre at coordinates $[x,y]=[0,0]$ is indicated with a $\times$, and the Sun at [0,8.34]~kpc with an orange dot. 
Some unnatural delineation of an imaginary circle passing through the Sun and Galactic centre arises from sources placed arbitrarily at the tangent point in the heliocentric distance estimates by \citet{meg21}, see the text. Cyan dashed lines indicate Galactic longitude in steps of $30^{\circ}$.
The Solar circle, separating the ``inner'' from ``outer'' Galaxy, is represented with a black dashed circle.
Grey dotted circles represent heliocentric distances of~5, 10, and 15~kpc. 
Spiral arms from the four-arm Milky Way prescription of \citet{hou09} are 
plotted, except for the Local arm taken from \citet{xu16}, with the arm-colour 
correspondence at the bottom left.
The Norma arm is represented using two colours: magenta for the inner part; and brown for the outer part, which is 
generally designated as the Outer arm, whose starting point is established in agreement
with \citet{mom06}. The uniform line thickness is not representative of the actual arm widths.}
\label{galpos}
\end{figure*}

\item
Source heliocentric distances were determined by \citet{meg21},
who developed a new code for assigning a $V_{\mathrm{LSR}}$ to each Hi-GAL source, using all the available spectroscopic data complemented by a morphological analysis to choose the best velocity in presence of multiple spectral components along the line of sight. This analysis is based on considerations on the spatial distribution of molecular emission, rather than simply on its brightness. Once the velocity is determined, if no stellar or maser parallax distance is known, the kinematic distance is calculated and the near/far distance ambiguity inside the Solar circle is solved with the H\,\textsc{i} self-absorption method or from distance–extinction data. This procedure is similar to that of \citet{rus11} which provided us with distances for \citetalias{eli17}, but with substantial improvements in the $V_{\mathrm{LSR}}$ assignment criteria.  Furthermore, the spectroscopic data base was considerably updated, and a new rotation curve adopted \citep{rus17}. In particular, we adopted their distance list based on line detection in molecular spectra at a $3-\sigma$ level (where $\sigma$ represents the noise level of each spectrum).
From this set of distances, we rejected ($i$) those corresponding to a Galactocentric distance
$R_\mathrm{GC} > 20$~kpc, and ($ii$) those placed 
at the distance of the tangent
point because of a kinematically forbidden $V_{\mathrm{LSR}}$, but having a velocity differing by more than 
10~km~s$^{-1}$ from that of the tangent point. After this selection, valid distances were assigned to 
\totsouwdist~sources, whose positions are shown in Fig.~\ref{galpos}.
Statistics are reported in Tables~\ref{catstats} and~\ref{rej_catstats}.
A more detailed discussion about distances is postponed to Section~\ref{heliocentric}.

\item A MBB with constant
emissivity index $\beta=2$ was fitted to SED data at $\lambda 
\geq 160~\umu$m to estimate the total mass and average temperature of the clump. 
Two different expressions for the MBB were used: one explicitly containing the optical depth $\tau_\lambda$
\citep[e.g.,][their Equation~3]{eli16} and one assuming optically thin emission at all wavelengths
(their Equation~8). The former is preferred if the free parameter, wavelength $\lambda_0$ for which
$\tau_{\lambda_{0}}=1$ is less than than 50.6~$\umu$m, based on
considerations described in \citetalias{eli17}. Otherwise, the physical parameters are derived
with the latter. Subsequently, for calculating bolometric temperature and luminosity,
integrals based on the analytical best-fitting MBB are considered for pre-stellar sources.
However, for a protostellar source the integral of the MBB for only 
$\lambda \geq 160~\umu$m is combined with fluxes observed at shorter wavelengths (PACS 
at 70~$\umu$m and, if available, MSX, WISE, and MIPSGAL). For sources without
a distance estimate, the fit was still performed in order to compute the distance-independent 
parameters. Distance-dependent parameters were calculated for a hypothetical distance of 1~kpc and
appropriately flagged in the catalogue (see Appendix~\ref{catdescription}).

\item 
The properties calculated through the fit of SEDs with fluxes in at least four 
\textit{Herschel} bands are generally considered ``highly reliable''. The corresponding
sources are included in the main catalogue, with the exception of cases in which the results of the best-$\chi^2$
fit corresponded to the extreme values explored for the temperature, namely 5 and 40~K, which might be the result of a failed fit.
These exceptions are relegated to the ``low-reliability'' source catalogue.
Sources with only three fluxes (which are necessarily starless, by 
construction) have properties derived from a poorly constrained fit and are also reported in the 
``low-reliability'' source catalogue. These catalogues contain
\totrel~and \totpoor~sources, respectively. Subsequent discussion in this paper is based entirely 
on the ``high-reliability'' catalogue, except in cases where it is stated explicitly that \textit{both} catalogues are used.
\item An overall classification distinguishing clumps containing star formation activity (protostellar)
versus quiescent clumps (starless) is based on the presence or not, respectively, of a detection at 
$70~\umu$m. This classification makes use of only \textit{Herschel} photometry and, 
as widely discussed in \citetalias{eli17} and \citet{bal17a}, can be affected by confusion and/or lack of sensitivity at
$70~\umu$m (for $F_{70} \lesssim 0.2$~Jy). The first effect is due to lack of spatial resolution at increasing distance, so that the 
$70~\umu$m-emission produced by a limited fraction within a given clump is assigned to the entire 
object, to which a protostellar classification is then given. The second effect goes opposite to the previous one, because it leads to misclassify sources with a true star formation content as starless. In particular, low-mass Class~0 objects are known to have low luminosities \citep{dun13}, so their possible contribution to the clump emission at 70~$\umu$m may remain undetected. In fact, 
spectroscopic signatures of ongoing star formation have been found in small samples of 
$70~\umu$m-quiet \textit{Herschel} clumps, both infall \citep[e.g.,][]{tra17} and outflow \citep[e.g.,][]{dua13}. In this respect, starless sources should be more rigorously named candidate starless clumps. We refer the reader to Appendix~C of \citetalias{eli17} and to Section~\ref{heliocentric} of this paper for a further discussion of the combined effect of these two biases.
\item
Finally, starless clumps are further classified as gravitationally bound (hereafter pre-stellar) 
or unbound, using the so-called ``Larson's third law'',
$M(r) = 460~M_{\odot}~(r/\mathrm{pc})^{1.9}$ \citep{lar81}, as a threshold to divide the $M$-$r$ plane, where $M$ and $r$ are the source mass and physical radius, respectively. 
It is necessary to point out that this classification is based only on considerations about gravitational stability. However, gravitationally unbound sources can be confined by external pressure \citep[e.g.,][]{kai11}. A complete virial analysis for clumps \citep[see, e.g.,][]{pat16} would require information about external pressure, together with magnetic field and total internal kinetic energy (including turbulence), and is beyond the scope of this paper, which is essentially based on photometric observations only.
\end{itemize}

\begin{figure}
\centering
\includegraphics[width=8.5cm]{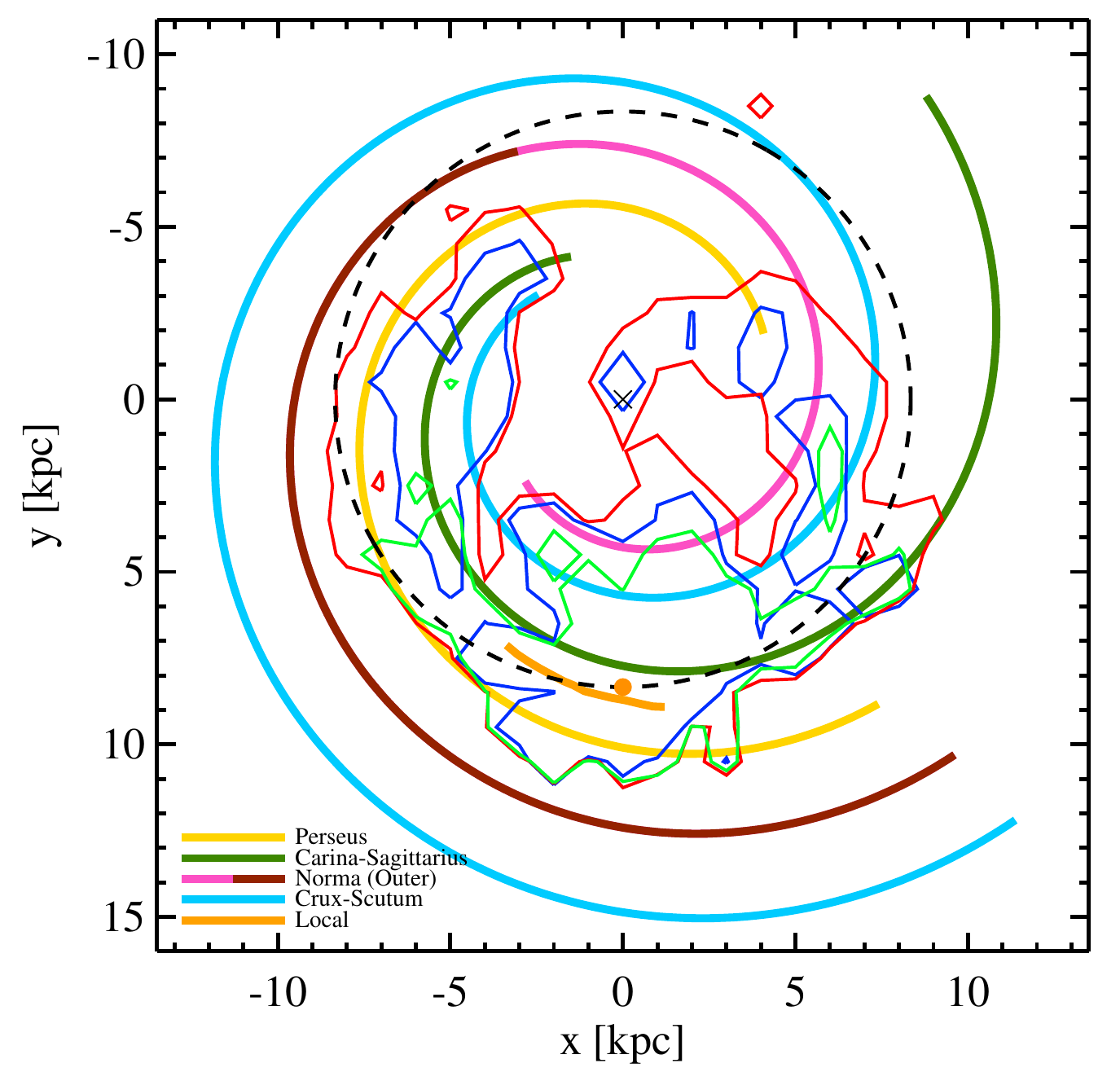}
\caption{The same as Figure~\ref{galpos}, but with source density contours displayed instead of source positions. Sources are counted in boxes of 1 kpc$^2$. To avoid figure crowding, only one contour, corresponding to a level of 100~sources~kpc$^{-2}$ is plotted for each source class: green for starless unbound sources (not shown in Figure~\ref{galpos}), red for pre-stellar, and blue for protostellar, respectively.}
\label{galposcontours}
\end{figure}

The coordinates of protostellar and pre-stellar clumps contained in the ``high-reliability catalogue'' are shown
in the top panel of Fig.~\ref{surveycoverage}, from which it is evident how the Hi-GAL
coverage followed the Galactic warp in the outer Galaxy.
Furthermore, in Fig.~\ref{galposcontours} the distribution of the Hi-GAL sources in the Galactic plane (already shown in Fig.~\ref{galpos}) is rendered through source density contours for the different classes. No particular behaviour is seen for different source populations with respect to spiral arm locations.

\subsection{``Inner'' vs ``outer'' Galaxy}\label{innerouter}

In this paper, a systematic comparison between properties of Hi-GAL sources in the ``inner Galaxy'' and 
``outer Galaxy'' is carried out. In previous Hi-GAL literature ``inner Galaxy'' has generally been used to 
indicate the first tranche of the survey corresponding to the \textit{Herschel} KPAO cycle, and published in
\citetalias{eli17}. These observations, initially intended to span the 
$|\ell|< 60^{\circ}$ longitude range, were actually extended to the $-71.0^{\circ} \lesssim \ell 
\lesssim 67.0^{\circ}$ range, corresponding to Hi-GAL tiles from $\ell$290 to $\ell$066 (according 
to the Hi-GAL nomenclature). Keeping this definition of ``inner Galaxy'' would allow a direct and easy comparison with \citetalias{eli17}.

However, this longitude range also contains sources located outside the 
Solar circle. Therefore, to avoid a quite arbitrary and counter-intuitive inner/outer division
here we prefer to
adopt a different more natural definition, ``inner Galaxy'' and ``outer Galax'' being the regions respectively inside or outside the 
Solar circle, 8.34~kpc as adopted by \citet{meg21}.
This definition, which applies to sources with a distance (and then Galactocentric radius) determination,
gives \cinnzero~and \coutzero~sources in the inner and in the outer Galaxy, respectively.

A classification for sources without a distance estimate can be attempted as follows. First, sources in the second and 
third Galactic quadrants are definitely located outside the Solar circle. However, sources in the first and fourth quadrants can belong to either the inner or 
outer Galaxy.

\begin{figure}
\centering
\includegraphics[width=8.5cm]{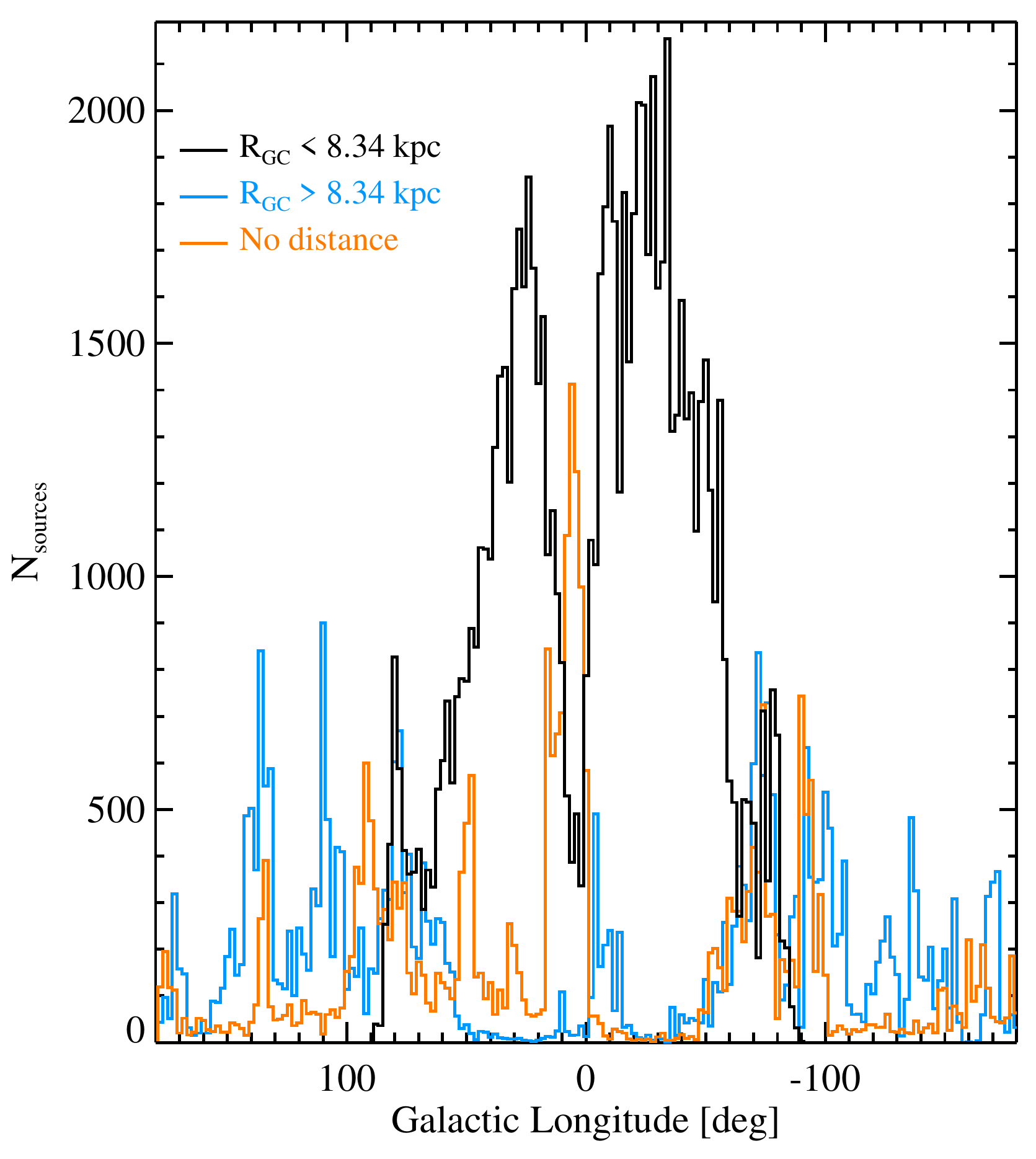}
\caption{Histograms in $2^{\circ}$-bins of
Galactic longitude of Hi-GAL sources located inside the Solar circle (black), outside (light blue), and
lacking a distance/Galactocentric radius estimate (orange), respectively.}
\label{longinnerouter}
\end{figure}

Histograms of the number of sources in $1^{\circ}$-bins of
longitude (see Fig.~\ref{longinnerouter}, in which for the sake of clarity $2^{\circ}$-bins
are shown), for sources with distances inside or outside the Solar circle, and for sources with no distance estimate, can be used
to establish a rough criterion for assigning the latter sources in the first and fourth quadrants to either the inner or outer Galaxy for further analyses. In longitude bins in which the inner Galaxy sources 
outnumber the outer Galaxy sources ``inner'' is assigned, otherwise ``outer'' (which as might be expected occurs only near $\ell \pm 90^{\circ}$).

\begin{table*}
\centering
\caption{Number of sources in the 360$^{\circ}$ catalogue with high-reliability parameters, subdivided by evolutionary class and inner/outer Galaxy location.}
\label{catstats}
\begin{tabular}{lccccc}
\hline
 & \multicolumn{2}{c}{Inner Galaxy} & \multicolumn{2}{c}{Outer Galaxy} & Total\\
 & w/ distance & w/o distance & w/ distance & w/o distance & \\
\hline
Protostellar &        22132 &         5476 &         7572 &            6 &        35186\\
Pre-stellar &        32013 &         8589 &         8683 &            4 &        49289\\
Unbound &         3476 &         3033 &         3613 &            7 &        10129\\
Total &        57621 &        17098 &        19868 &           17 &        94604\\
\hline
\end{tabular}
\end{table*}

\begin{table*}
\centering
\caption{Number of sources in the 360$^{\circ}$ catalogue with low-reliability parameters, subdivided by evolutionary class and inner/outer Galaxy location.}
\label{rej_catstats}
\begin{tabular}{lccccc}
\hline
 & \multicolumn{2}{c}{Inner Galaxy} & \multicolumn{2}{c}{Outer Galaxy} & Total\\
 & w/ distance & w/o distance & w/ distance & w/o distance & \\
\hline
Protostellar &          166 &           27 &           33 &            0 &          226\\
Pre-stellar &        20705 &         5250 &         5308 &            2 &        31265\\
Unbound &         9639 &         7007 &         7468 &           14 &        24128\\
Total &        30510 &        12284 &        12809 &           16 &        55619\\
\hline
\end{tabular}
\end{table*}

The total numbers of sources, separated by different
evolutionary classes and by inner/outer Galaxy location, are reported in Table~\ref{catstats} for the 
``high-reliability'' catalogue, and in Table~\ref{rej_catstats} for the ``low-reliability'' catalogue, respectively.

\section{Distance-dependent parameters}\label{distparam}
\subsection{Heliocentric distance and Galactocentric radius}\label{heliocentric}
The heliocentric distance is a crucial parameter for characterizing the detected compact 
sources, not only to compute quantities depending directly on distance, such as physical
size, mass and luminosity \citep[see a dedicated discussion in][]{bal17a,bal17b}, but also to 
understand what meaning we can ascribe to other quantities that are formally 
distance-independent, especially when distant sources are actually the combination/blending of unresolved 
structures.

\begin{figure}
\centering
\includegraphics[width=8.5cm]{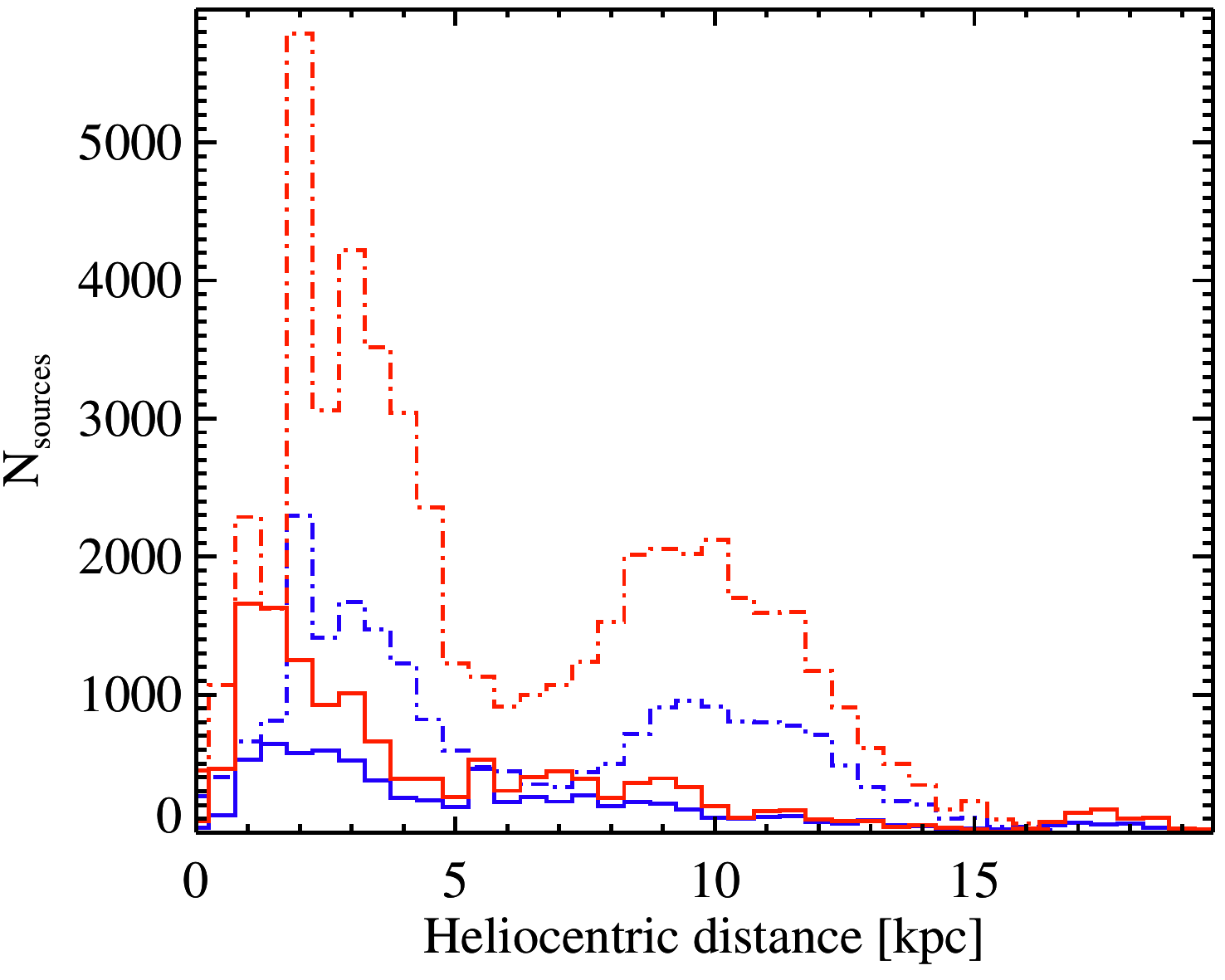}
\includegraphics[width=8.5cm]{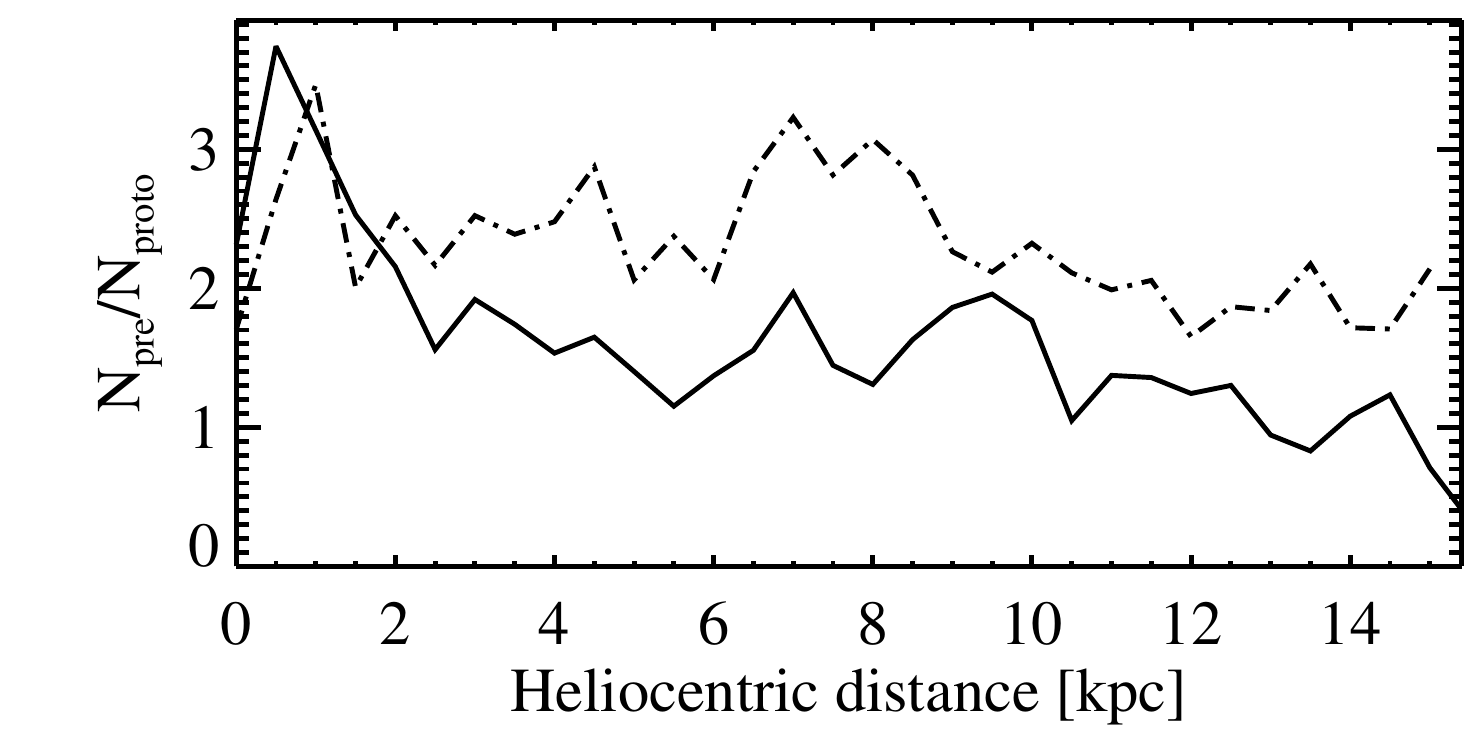}
\caption{Top panel: number counts vs heliocentric distance for Hi-GAL pre-stellar
(red) and protostellar (blue) clumps taken from both catalogues.
Dot-dashed lines are used for distributions corresponding
to the inner Galaxy, and solid lines for the outer one.
Bottom panel: number ratio of pre-stellar to protostellar sources in the distance bins defined in the top
panel, for the inner (dot-dashed) and the outer Galaxy (solid). To ensure statistical reliability,
only bins in which both numerator and denominator are larger than 50 are reported, resulting in a foreshortened $x$-axis.}
\label{hist_dist}
\end{figure}

Adoption of the new distance set of \citet{meg21} significantly increases 
the number of catalogue sources with a known distance, in both absolute and relative terms: now \totsouwdist~out of \totsou\ (\perctotsouwdist~per cent) compared with \citetalias{eli17}, 57065 out of 100922 ($57$~per cent). 

Figure~\ref{hist_dist}, top, shows number counts vs heliocentric distance for our sample of 
objects (top panel), divided according to pre-stellar vs protostellar and inner vs outer Galaxy. Sources are included from both catalogues, i.e., regardless of the reliability of their SEDs, because distance 
determination is independent of the SED fit. The 
histograms for the inner Galaxy look bimodal \citep[cf.][]{urq13a}, mostly due to near/far distance 
ambiguity present at those longitudes; this is not seen in the outer Galaxy, where most 
sources are found to be located within 10~kpc.

Figure~\ref{hist_dist}, bottom, shows that pre-stellar sources are generally more abundant than protostellar sources
at most distances in both the inner and outer Galaxy.
However, the pre-stellar/protostellar number ratio seems to decrease at increasing distance, albeit with considerable
scatter. 
Two competing effects can affect this ratio at large distances: on the one hand,
insufficient PACS sensitivity at 70~$\umu$m may lead to misclassifying
protostellar sources as pre-stellar; on the other hand, source confusion with possible blending of 
pre-stellar and protostellar sources may result in a single object classified 
as protostellar \citepalias[see][Appendix~C1]{eli17}. The trend seen is consistent with the latter effect being
predominant. 
Considering all sources in the outer Galaxy, the percentage classified as protostellar is \percprotoout~per cent; this is slightly higher than the corresponding number in the inner Galaxy, \percprotoinn~per cent.

\begin{figure}
\centering
\includegraphics[width=8.5cm]{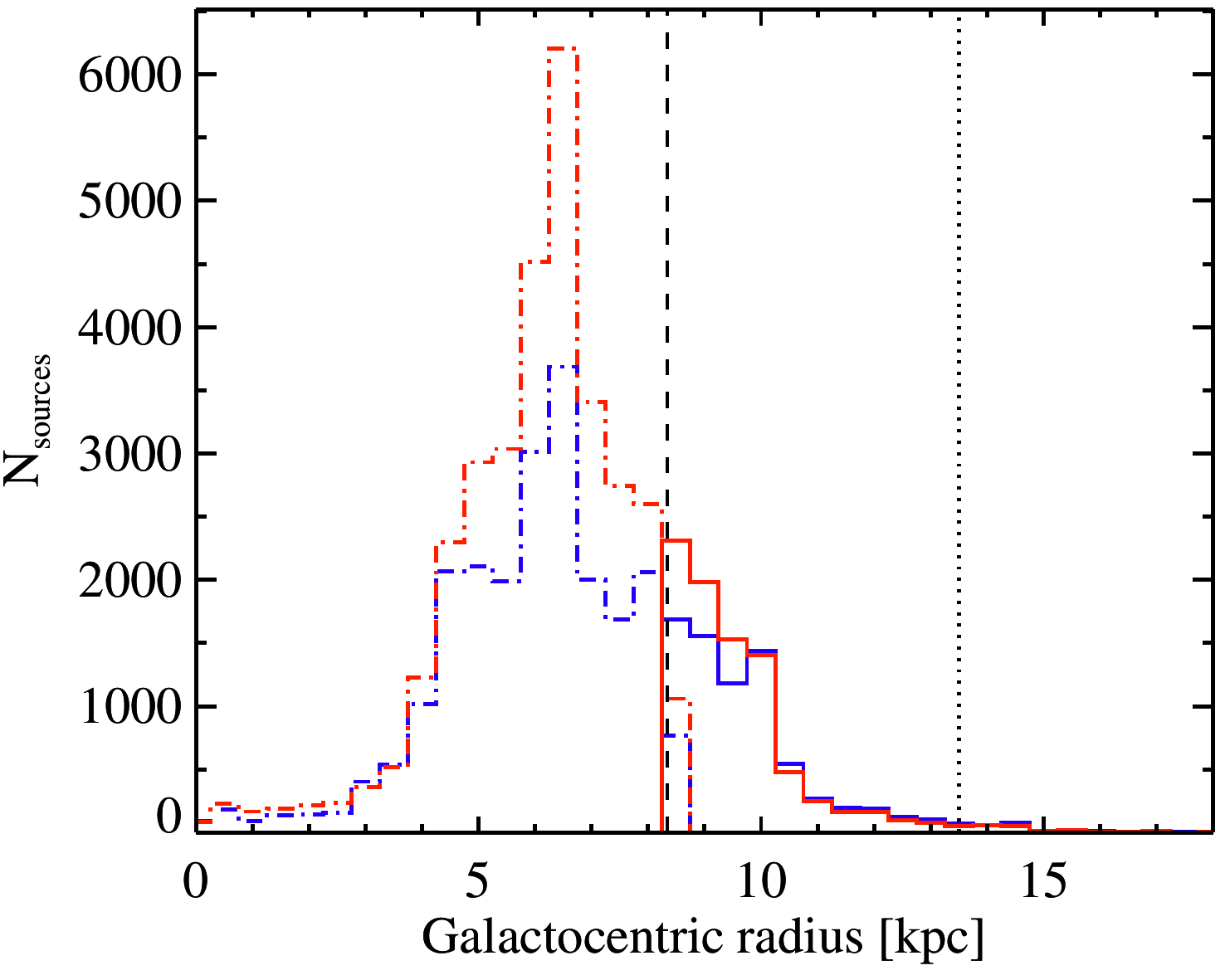}
\caption{Same as Fig.~\ref{hist_dist}, top panel, but for Galactocentric radius. The vertical 
dashed and dotted lines delimit the outer Galaxy and the so-called far outer Galaxy ($R_\mathrm{GC} > 13.5$~kpc),
respectively.}
\label{hist_rgal}
\end{figure}

Figure~\ref{hist_rgal} presents number counts vs Galactocentric radius $R_\mathrm{GC}$, which
is not affected by the near/far distance ambiguity.
The peak at about 6~kpc in the inner Galaxy, seen also by \citet{rag16} 
based on the previous set of Hi-GAL distances and by \citet{wie15} for ATLASGAL sources, 
is compatible with the position of the so-called ``molecular ring'' \citep[e.g.,][]{dob12,miv17}. 
Other local peaks are present at about 8.5~kpc and 10-11~kpc in the outer Galaxy. In
\citet{sch11} these three features, observed over a sample of a few hundred BGPS sources, are associated with
the Sagittarius, Local, and Perseus arms, respectively. However, a feature around 4.5~kpc
attributed by \citet{sch11} and \citet{wie15} to the closer tip of the Galactic bar is not prominent 
in our data, which in general appears smoother due to the large number of inter-arm sources in 
our catalogue (Fig.~\ref{galpos}).

\begin{figure}
\centering
\includegraphics[width=8.5cm]{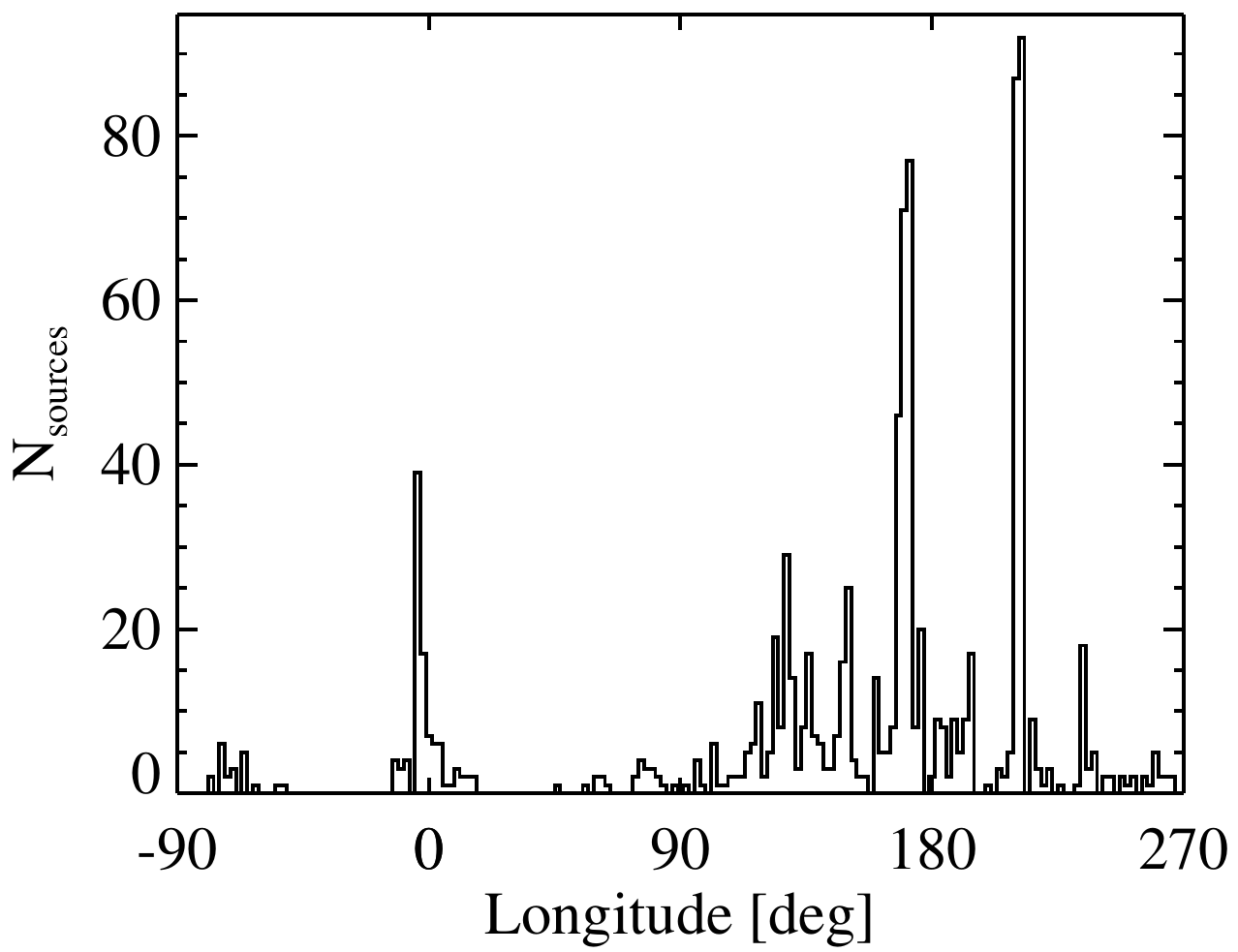}
\caption{Number counts in $2^{\circ}$-bins of Galactic longitude of Hi-GAL sources 
located in the FOG ($R_\mathrm{GC} > 13.5$~kpc). Both pre-stellar and protostellar 
sources are counted, from both 
catalogues. 
The $x$-axis range is set from $-90^{\circ}$ to $270^{\circ}$ in order to place the 
local peaks around $0^{\circ}$ and $180^{\circ}$ well inside the plot area.}
\label{fogfig}
\end{figure}

Finally, we discuss the ability of the PACS and SPIRE cameras (but also of line 
surveys used to determine distances) to detect sources in the far outer Galaxy (hereafter FOG). 
Various boundaries for the FOG, in terms on $R_\textrm{GC}$, are found
in the literature, such as 13.5 \citep{hey98}, 15 \citep{hon11}, and 16~kpc \citep{urq13a},
all of which are well outside the ranges of $R_\textrm{GC}$ probed by the aforementioned ATLASGAL and BGPS surveys.
But for Hi-GAL, Fig.~\ref{hist_rgal} suggests that a small but meaningful number
of clumps, essentially contained in the range 13.5~kpc~$< R_\textrm{GC} < $~15~kpc, 
deserves attention. Considering sources at $R_\textrm{GC}\geq 13.5$~kpc,
and including both catalogues, 
we find that \nfog~sources lie in the FOG. However, it is evident from
Fig.~\ref{fogfig} that many of these large $R_\textrm{GC}$ values come from
lines of sight close to the Galactic centre and anti-centre, areas suffering from large
uncertainties in kinematic distances. Neglecting sources with $|\ell|< 10^{\circ}$ or 
$|\ell-180^{\circ}|< 10^{\circ}$, \nfogrestr~sources remain. 
The prominent peak around 
$\ell \sim 212^{\circ}$ is from sources found by \citet{meg21} to be associated with the Sh2-284 H\,\textsc{ii} region; the adopted heliocentric distance 6.6~kpc yields $R_\textrm{GC}= 14.4$~kpc.
Most of the remaining sources are concentrated in the second quadrant.
A brief analysis of the physical properties for sources in the FOG is given in 
Section~\ref{galactotrends}.

\subsection{Physical size}\label{physsize}
\begin{figure*}
\centering
\includegraphics[width=18cm]{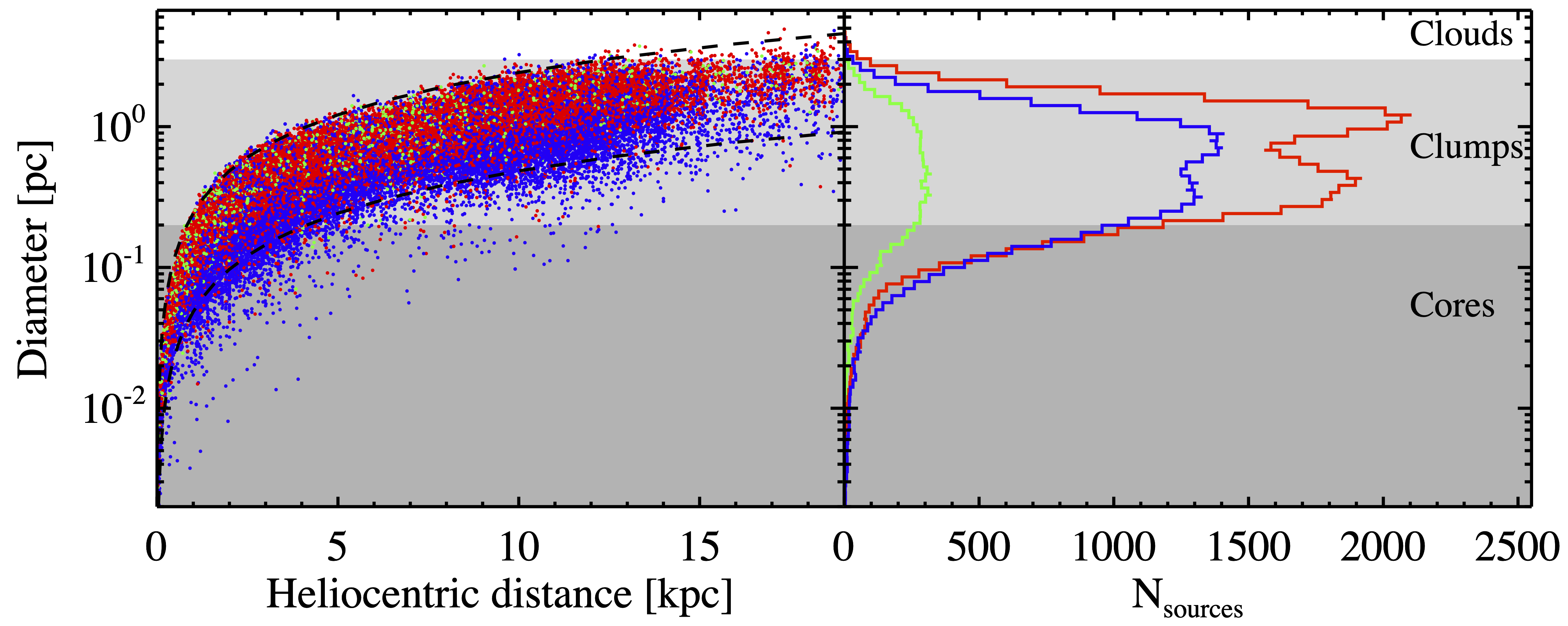}
\caption{Left-hand panel: Hi-GAL clump linear diameters, obtained by combining distance and angular size estimated at 250~$\umu$m as explained in Appendix~\ref{catdescription}, versus 
distances (blue: protostellar; red: pre-stellar; green: starless unbound).
Different background tones of grey indicate ranges of diameter corresponding to different 
object classifications labelled at the right (see Section~\ref{physsize}). Upper and lower dashed lines correspond to
an angular size of 50 and 10~arcsec, respectively. Right-hand panel: distribution 
of source diameters for protostellar, pre-stellar, and starless unbound sources, rotated to share the $y$-axis and background colours with the left-hand panel.
Line colours also the same encoding of source classification.}
\label{sizefig}
\end{figure*}
Estimating the physical size of compact sources is of great importance to understanding the nature 
of objects investigated. \citetalias{eli17} showed that most Hi-GAL sources fulfil 
the definition of clumps 
\citep[$0.2~\mathrm{pc} < D < 3~\mathrm{pc}$, based on][where $D$ is the diameter of the structure]{ber07}, 
while a smaller fraction
of nearby sources can be classified as cores, i.e., condensations supposed
to host (or be progenitors of) formation of a single star or small stellar system.

Figure~\ref{sizefig} reports the same information as in \citetalias{eli17}, updated with 
new distances and extended to the entire $360^{\circ}$ survey coverage. 
Trends seen in \citetalias{eli17} 
are basically confirmed. 
On average, protostellar sources are more 
compact than pre-stellar sources (left-hand panel). Most sources (\perclumps~per 
cent) can be classified as clumps, so that hereafter we often refer to all sources as 
``clumps''. Only \percores~per cent and \perclouds~per cent 
of sources fulfill the \citet{ber07} definition of cores and clouds, respectively 
(right-hand panel). The bi-modality that appears in the distributions of both pre-stellar and protostellar 
sources is a direct consequence of the bi-modality in distances seen in the upper panel of
Fig.~\ref{hist_dist}.

\subsection{Mass vs radius}\label{masses}

\begin{figure}
\centering
\includegraphics[width=8.5cm]{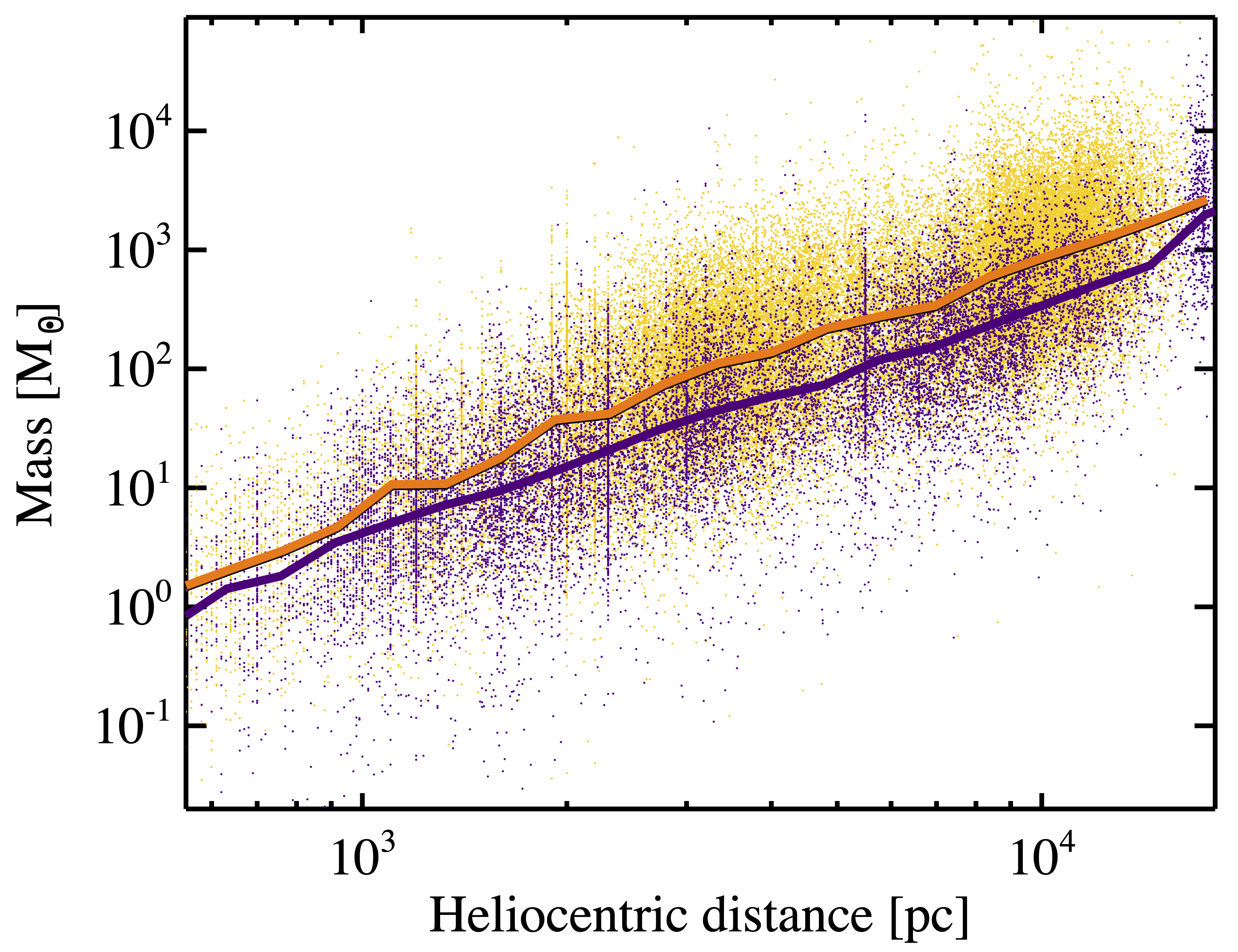}
\caption{Clump mass vs heliocentric distance for sources in the inner (orange dots)
and in the outer Galaxy (purple dots). Orange and purple thick lines represent
the median mass in logarithmic bins of distance 
for these two populations, respectively.}\label{massdist}
\end{figure}

The clump mass depends on the square of the estimated heliocentric distance. Moreover, the selection effect, 
known as Malmquist bias \citep[see, e.g.,][]{zet18}, favours
detection of larger and larger masses and luminosities at increasing
distance. Such bias 
affects not only the completeness of the observed sample, but also the nature
of the objects included: for a very distant source whose internal 
structure cannot be resolved with \textit{Herschel}, the total derived mass (even 
$M>10^5 M_{\odot}$) does not describe an entity forming a single star, but rather a large 
and complex structure hosting several compact sources \citep[see, e.g.,][]{bal17a}.
For this reason, here we avoid considering an overall mass function for 
Galactic clumps regardless to their distance \citepalias[it is more reasonable to consider it within
bins of distance as in][]{eli17} or drawing up any ranking 
of the most massive clumps in the Galaxy. 

Distance bias has to be taken into account in a global comparison of the 
masses encountered in the inner and outer Galaxy, given the different ranges of 
distances found in Section~\ref{heliocentric} for inner versus outer.
However, the deficiency of large masses in the outer
Galaxy compared to the inner Galaxy appears to be intrinsic: considering common 
bins of heliocentric distance, Fig.~\ref{massdist} shows that the median mass of sources in the outer Galaxy is always smaller (by a factor 
ranging from \minmassratio~to \maxmassratio) than the corresponding median for the inner
Galaxy. The largest distance bin, $d \gtrsim 16$~kpc, could be misleading: sources
far behind the Galactic centre, thus entirely in the outer Galaxy, also have
high mass estimates due to their relatively large distances, and so any highly uncertain distance assignments could lead to the upturn in the purple curve.

This general trend of larger clump masses in the inner Galaxy compared with
the outer Galaxy was already suggested by \citet{zah16} (but based on a much smaller statistical
sample), and can be understood as being the result of the essentially different regime of surface density found
in Section~\ref{surfdenspar}. This point will be addressed in 
Section~\ref{galactotrends}.

\begin{figure*}
\centering
  \begin{tabular}{@{}cc@{}}
    \includegraphics[width=8.0cm]{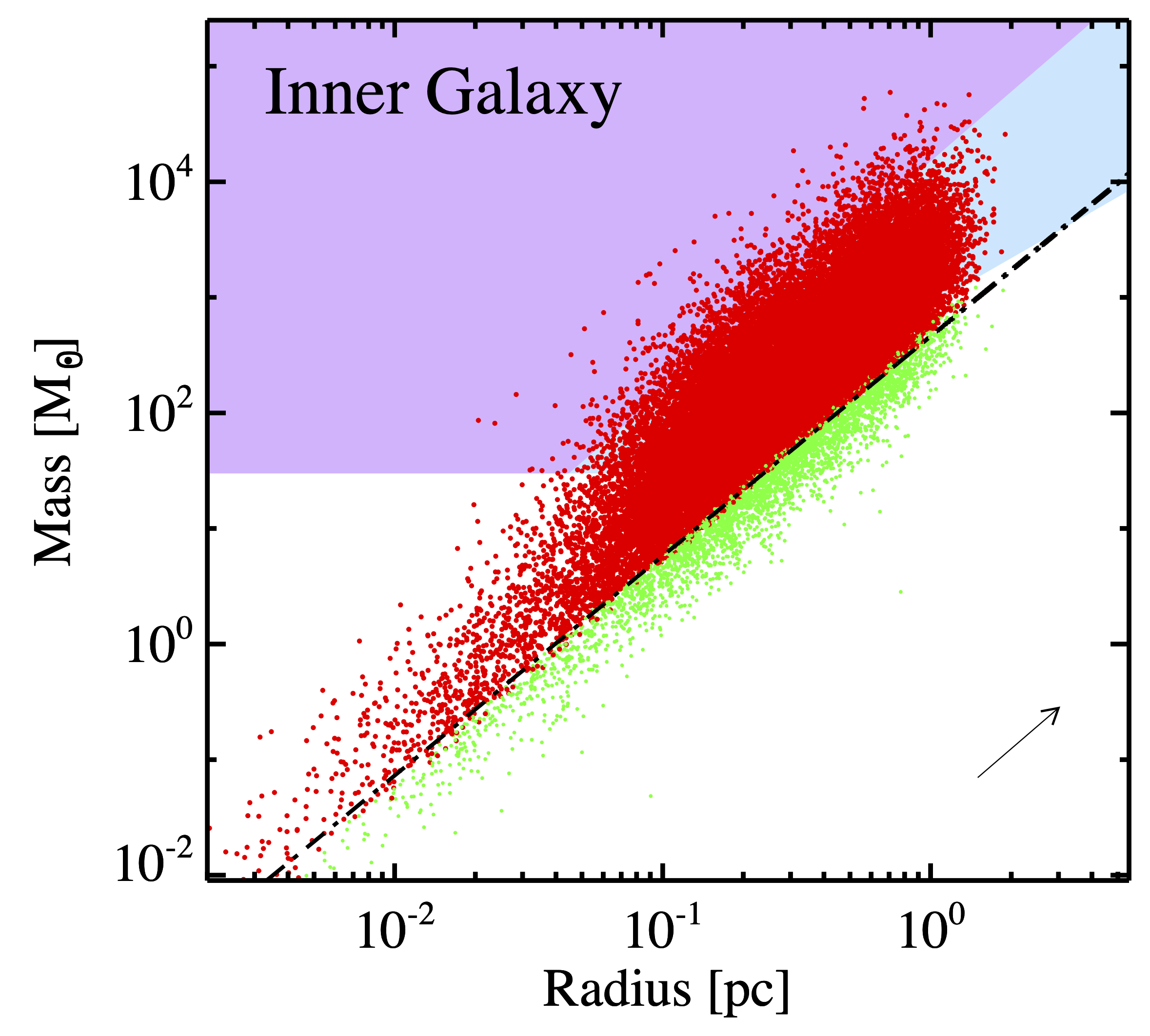} &
    \includegraphics[width=8.0cm]{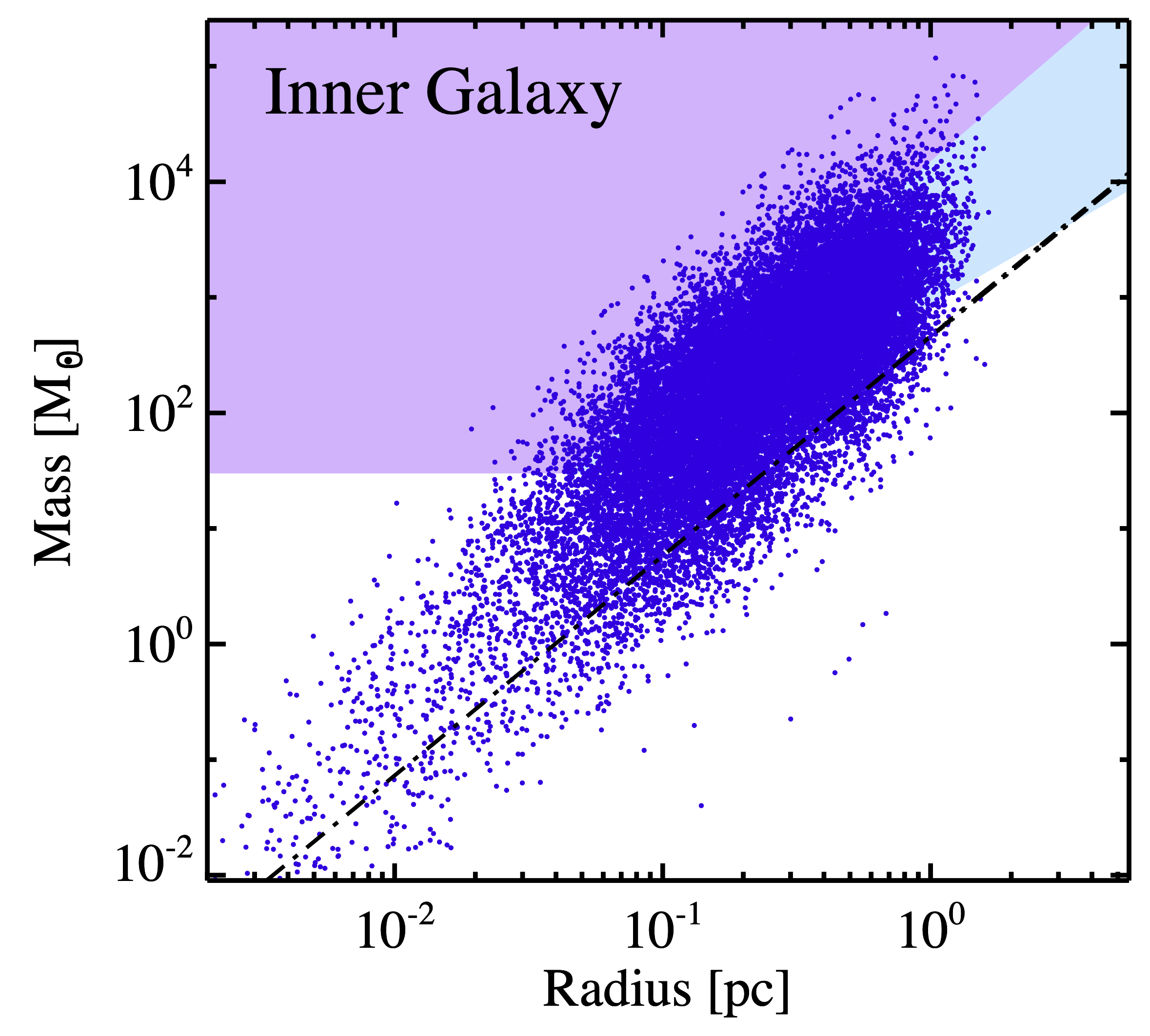} \\
    \includegraphics[width=8.0cm]{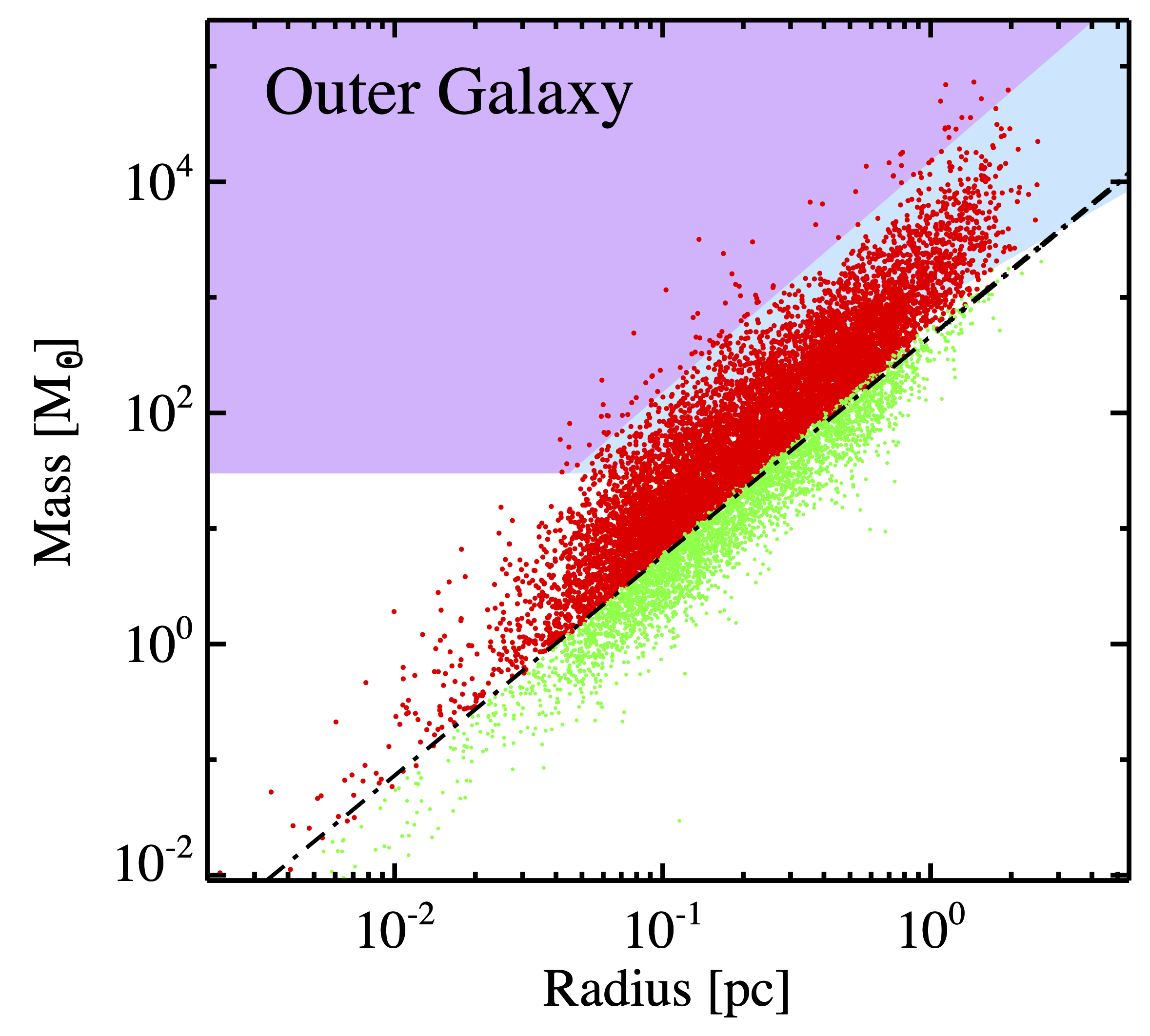} &
    \includegraphics[width=8.0cm]{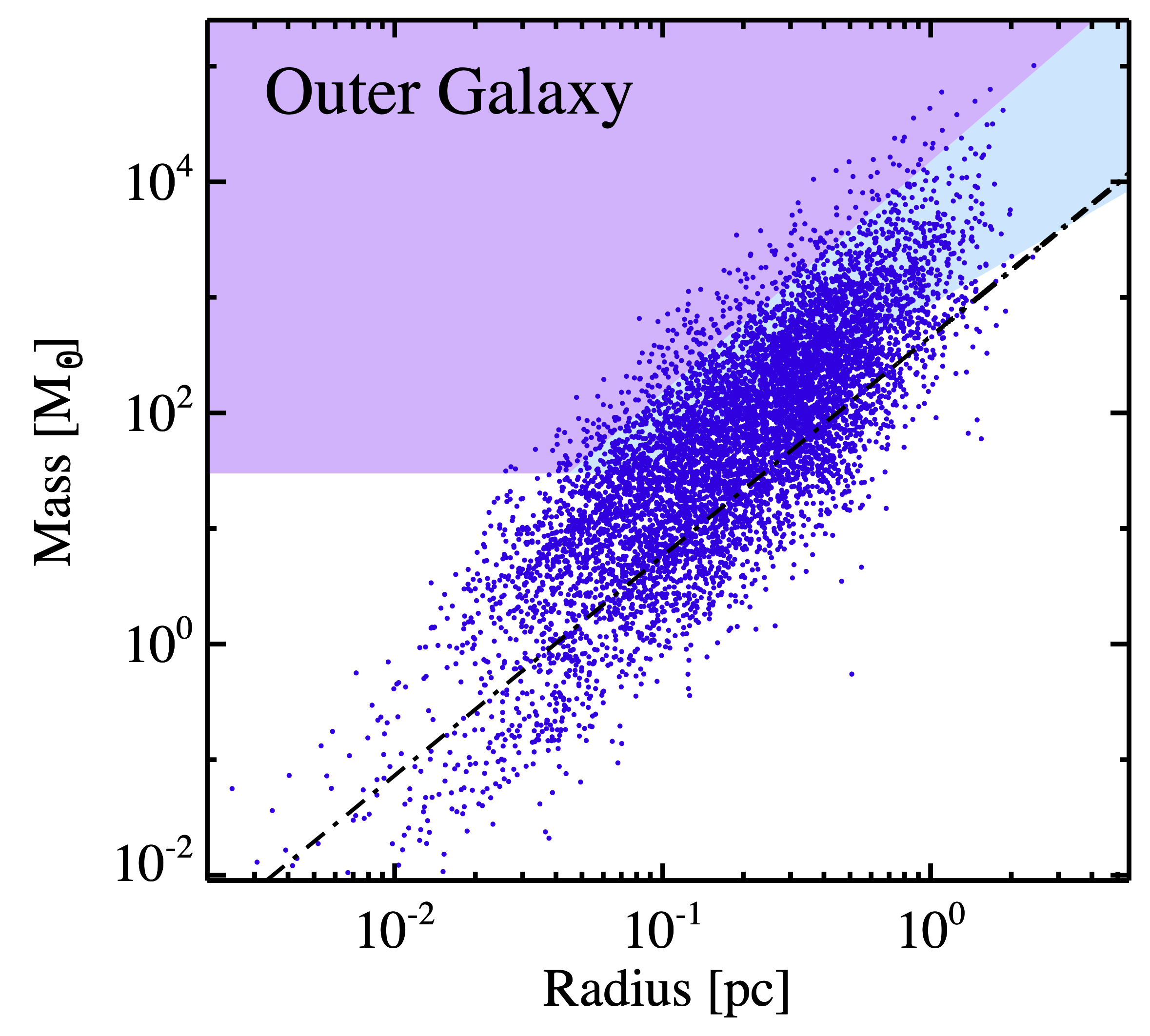} \\
  \end{tabular}
  \caption{Mass vs radius plot for starless sources in the inner (top left-hand panel)
  and outer Galaxy (bottom left), and for protostellar sources in the inner 
  (top right) and outer Galaxy (bottom right). 
  Uncertainty in distance is the main source of error on both displayed
  quantities and the shift corresponding to a hypothetical distance increase of a factor~2 is shown as an 
  arrow at the bottom right corner of the top left panel.
  In the left two panels containing starless sources, the dot-dashed black line $M(r) = 460~M_{\odot}~(r/\mathrm{pc})^{1.9}$ 
  \citep[][see Section~\ref{sedsel}]{lar81} separates pre-stellar (red) and unbound (green) sources. 
  The area in each diagram fulfilling the \citet{kru08} threshold for
  compatibility with high-mass star formation is shaded purple, and the less demanding \citet{kau10} threshold contains this and extends it as indicated in light blue. 
  Adopting $10~M_{\odot}$ as a lower limit
  for a massive star, and a star formation efficiency factor of $1/3$ for the 
  core-to-star mass transfer as in \citet{eli13}, these shaded areas cannot extend below $30~M_{\odot}$.}
  \label{masssize}
\end{figure*}

Studying the gravitational stability of sources or their ability to form high-mass stars requires
a combination of information about mass and size, as we will discuss with Fig.~\ref{masssize}. Note that the relation between these 
two estimated quantities is still distance-dependent, because they 
scale with distance quadratically and linearly, respectively. 
Moreover, in most cases the catalogued radius and mass represent overall summary observables
for clumps hosting an unresolved complex morphology, so that any inference about
gravitational stability should be taken as a large-scale description of a clump, while also
keeping in mind that large fluctuations in density are possible inside the object.

By plotting mass versus radius in Fig.~\ref{masssize} separately for inner and outer Galaxy sources
we can appreciate that larger values of mass are achieved in the inner Galaxy, 
as already seen in Fig.~\ref{massdist}.
In the two panels on the left for starless sources it is possible to visualise how the 
``Larson's third law'' is used to separate gravitationally bound (pre-stellar) from unbound
sources (see Section~\ref{sedsel}), determining the statistics reported in Tables~\ref{catstats}
and~\ref{rej_catstats}.
We point out that, strictly speaking, this kind 
of analysis should apply only to the pre-stellar sources, whose properties correspond to conditions
prior to the onset of star formation. Protostellar sources, on the other hand, have already experienced a mass transfer on to the forming star(s) and partial envelope dissipation, whose extent depends in principle on their individual evolutionary stage. Their masses therefore represent lower limits for the original ones.

\begin{table*}
\centering
\caption{Number of Hi-GAL sources in the inner or outer Galaxy having a mass-radius combination that is compatible with massive star formation according to five different increasingly-demanding thresholds (see the text).}
\label{himasstab}
\begin{tabular}{lcccccccccc}
\hline
 & \multicolumn{2}{c}{Ur14} & \multicolumn{2}{c}{KP10} & \multicolumn{2}{c}{Ba17} & \multicolumn{2}{c}{KM08} & \multicolumn{2}{c}{Br12}\\
 & inner & outer & inner & outer & inner & outer & inner & outer & inner & outer  \\
\hline
Protostellar &        16596 &         4013 &        13728 &         2827 &        12369 &         2330 &         3027 &          298 &           27 &            9\\
Pre-stellar   &        25503 &         5101 &        18207 &         2756 &        14758 &         1974 &         1156 &           88 &           12 &            9\\
Total &        42099 &         9114 &        31935 &         5583 &        27127 &         4304 &         4183 &          386 &           39 &           18\\
\hline
\end{tabular}
\end{table*}

As in \citetalias{eli17}, we discuss the regions of the mass versus radius plot corresponding
to conditions that from time to time have been considered necessary but not sufficient to have high-mass star 
formation inside the clumps. In particular, in Fig.~\ref{masssize} we show the area defined by the theoretical 
threshold of \citet{kru08}, corresponding to a clump surface density $\Sigma=1$~g~cm$^{-2}$, 
and how this area is extended using the empirical and less demanding threshold by \citet{kau10}.
In Table~\ref{himasstab}, we report statistics of pre-stellar and protostellar sources, in the inner
and in the outer Galaxy, fulfilling these two thresholds (indicated with ``KM08'' and ``KP10'',
respectively). Notice that the total number of sources above the KM08 threshold represents the \perckrumhtot~per cent of the entire catalogue, which is comparable with the 6~per cent level found by \citet{mer15} on a sample of 286 sources observed with SHARC-II \citep{dow03}.

Subsequently, KM08 has been demonstrated to be too conservative when compared with 
observations: \citet{lop10}, \citet{but12}, \citet{per13}, \citet{tan13}, \citet{urq14c}, and \cite{tra18}
report high-mass star formation even for surface densities in the range 
$0.05 \leq \Sigma \leq 0.5$~g~cm$^{-2}$. In particular, the numbers 
of sources fulfilling the less demanding value, 0.05~g~cm$^{-2}$ by \citet{urq14c}
are reported in Table~\ref{himasstab}, column ``Ur14''.

It is to notice that this proliferation of thresholds reflects a variety of observational conditions, and of adopted criteria. Instead of making a comparison with them, it may be possible, in turn, 
to extrapolate a threshold directly from our data, by comparison with a sample of well-known high-mass star forming objects. However, this is beyond the scope of this paper. Anyway, a 
specific analysis for the case of Hi-GAL observations was carried out by \citet{bal17a}, who evaluated the bias introduced by distance in classifying \textit{Herschel} sources as potentially able to form
high-mass stars. They suggested another power-law threshold with slope 
1.42, i.e., between 2 (KM08) and 1.33 (KP10) but much closer to the latter.
Source numbers corresponding to this threshold are also reported in Table~\ref{himasstab},
column ``Ba17''.

For all four thresholds there are
impressively high numbers of sources, both pre-stellar and protostellar, that can be considered as candidates for 
high-mass star formation. In the inner Galaxy
the numbers of candidates for the KP10, Ba17, and KM08 thresholds have all increased systematically compared with
\citetalias{eli17}, despite a slightly smaller total number of sources because of
the different definition of inner Galaxy adopted in this paper (Section~\ref{innerouter}). 
These increased numbers are due mostly to the increase of the number of sources having a distance estimate
in this work, rather than an increasing fraction of sources being candidates.
For example, for sources having a distance estimate in \citetalias{eli17} the fraction of sources fulfilling the KP10 threshold was 
 71~per cent and 65~per cent of the total protostellar 
and pre-stellar sources, respectively. The new
numbers reported in Table~\ref{himasstab} correspond to \perckauffproto~per cent~and 
\perckauffpre~per cent, respectively, of the larger totals having distances, i.e., lower percentages than in \citetalias{eli17}.

While this direct comparison with \citetalias{eli17} is perfectly feasible and consistent since the same dust opacity has been used, further comparisons with similar mass-radius plots
\citep[e.g., that of][]{svo16} are more complicated because they would imply to re-scale our masses taking into account different opacities adopted. \citet{svo16} clearly showed that numbers of
sources compatible with massive star formation according to a given threshold should be computed as a function of the adopted opacity. Furthermore, \citetalias{eli17} highlighted that the range of
typically used reference opacities would correspond to a scaling factor from 0.6 to 6 for masses. Therefore if, for example, we simply change the value of the reference opacity \citepalias[which is 0.1~cm$^2$~g$^{-1}$ at 300~$\mu$m, see][]{eli17} 
to that predicted at the same wavelength by the OH5 model of \citet{oss94} adopted by \citet{svo16}, a factor 0.77 should be applied to our masses. The consequent fraction of sources fulfilling the KP10 threshold would drop from \perckauffproto~to \perckauffprototest~per cent for the protostellar class, and from \perckauffpre~to \perckauffpretest~per cent for the pre-stellar class, respectively.

\begin{figure}
\centering
\includegraphics[width=8.5cm]{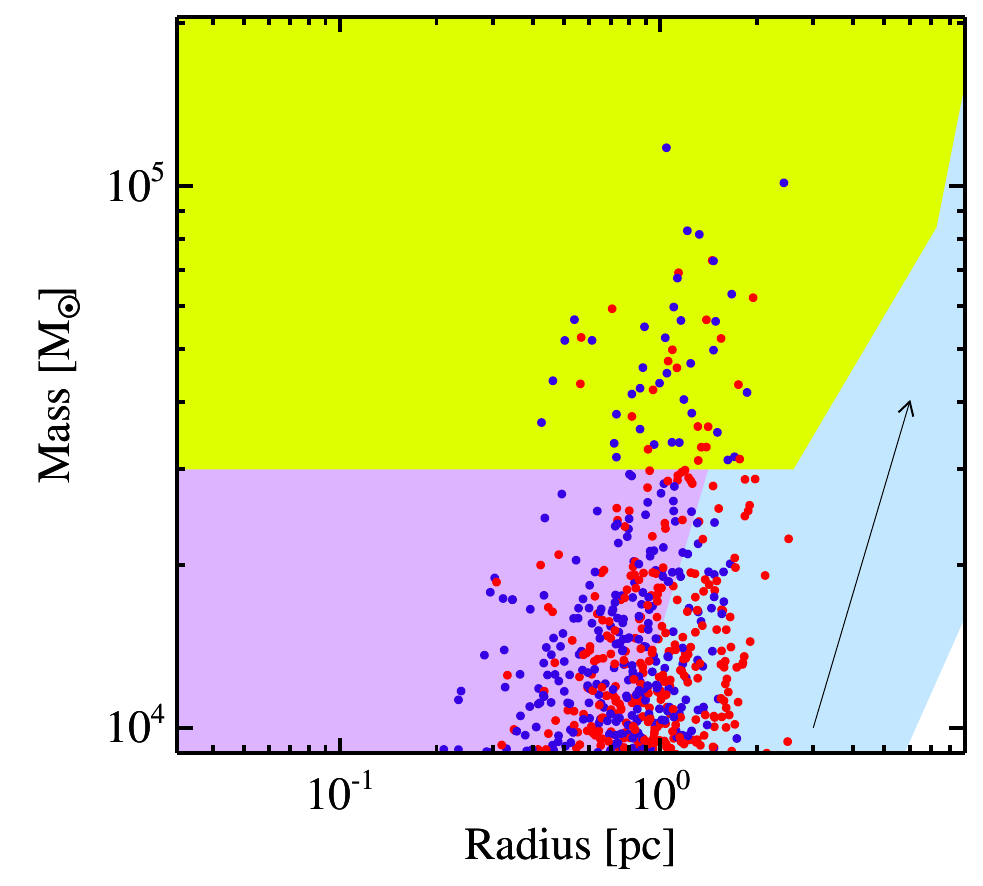}
\caption{Mass vs radius for pre-stellar (red) and protostellar (blue) sources in the high-mass regime. As in Fig.~\ref{masssize}, the shift corresponding to a hypothetical distance increase of a factor~2 is shown as an 
arrow. Purple and light blue areas are as
introduced in Fig.~\ref{masssize}. The overlying yellow sub-area towards high mass corresponds to the \citet{bre12} threshold
for identifying candidate massive proto-clusters.}\label{protoclusters}
\end{figure}

Finally, we investigated the threshold proposed by \citet{bre12} for identifying 
massive proto-cluster candidates, such that the content in stars would amount to $>10^4~M_\odot$
\citep{por10}. For $r< 2.6$~pc, they establish a minimum mass of $3 \times 10^4~M_{\odot}$
(corresponding to star formation efficiency 1/3), which is shown in Fig.~\ref{protoclusters}. Clumps in our sample have $r< 2.6$~pc, similar to the Galactic sources of \citet{bre12}, 
because surveys like Hi-GAL or ATLASGAL, resolve regions with a size 
of a several pc into smaller structures.\footnote{Though not 
relevant here, at larger $r$ the threshold increases, as $r^1$ up to 
$\sim 7.1$~pc, corresponding to the balance between the gravitational potential of the gas 
clump and the kinematics of the photo-ionized gas, and then as $r^3$, given by 
the condition of virial equilibrium observed in such structures.}
As recorded in Table~\ref{himasstab}, last column ``Br12'', we found 
\nbressert~sources fulfilling this criterion, \nbressertouter~of which are located in the outer Galaxy. 
Note that uncertainties
in the source distances can have a strong influence on this classification. For example, in 
Fig.~\ref{protoclusters} a decrease of a factor~2 in the distances of all sources would empty
the \citet{bre12} area almost completely (see magnitude of arrow), while the opposite would populate it with
hundreds more sources. A specific analysis of the
\nbressert~candidate proto-clusters is
reserved for future work, being beyond the scope of this paper

\subsection{Luminosity vs mass}\label{lumpar}

In this section we expand on the discussion of the bolometric luminosity 
versus envelope mass ($L_\mathrm{bol}$ vs $M$) diagram 
given in \citetalias{eli17}, to which the reader is referred 
for further details and previous literature.
This diagram is useful as an evolutionary diagnostic tool, when theoretical evolutionary 
tracks, taking into account an accretion phase and a clean-up phase \citep{mol08,smi14}, are over-plotted for
comparison to the data.

\begin{figure}
\centering
\includegraphics[width=8.5cm]{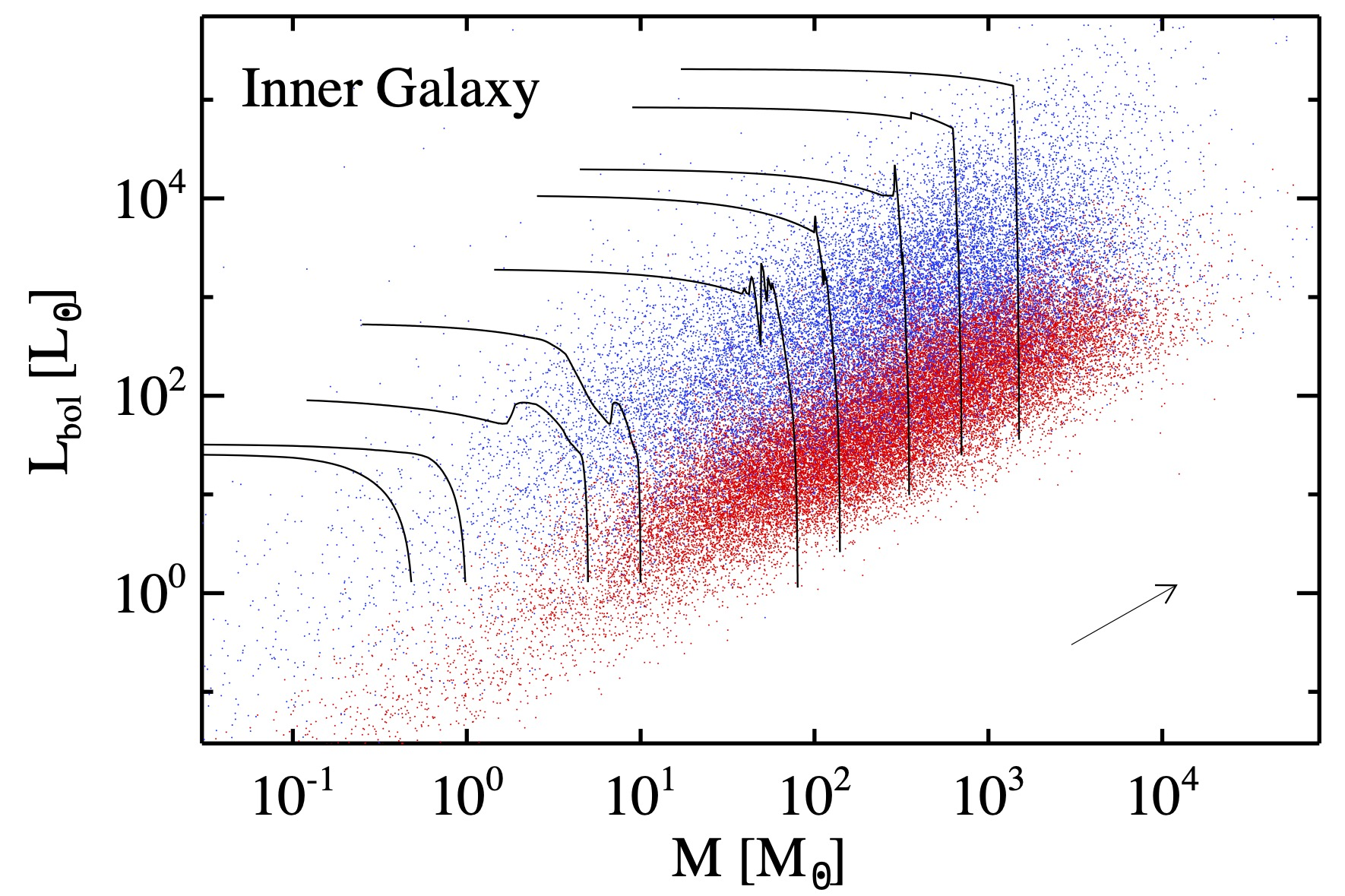}
\includegraphics[width=8.5cm]{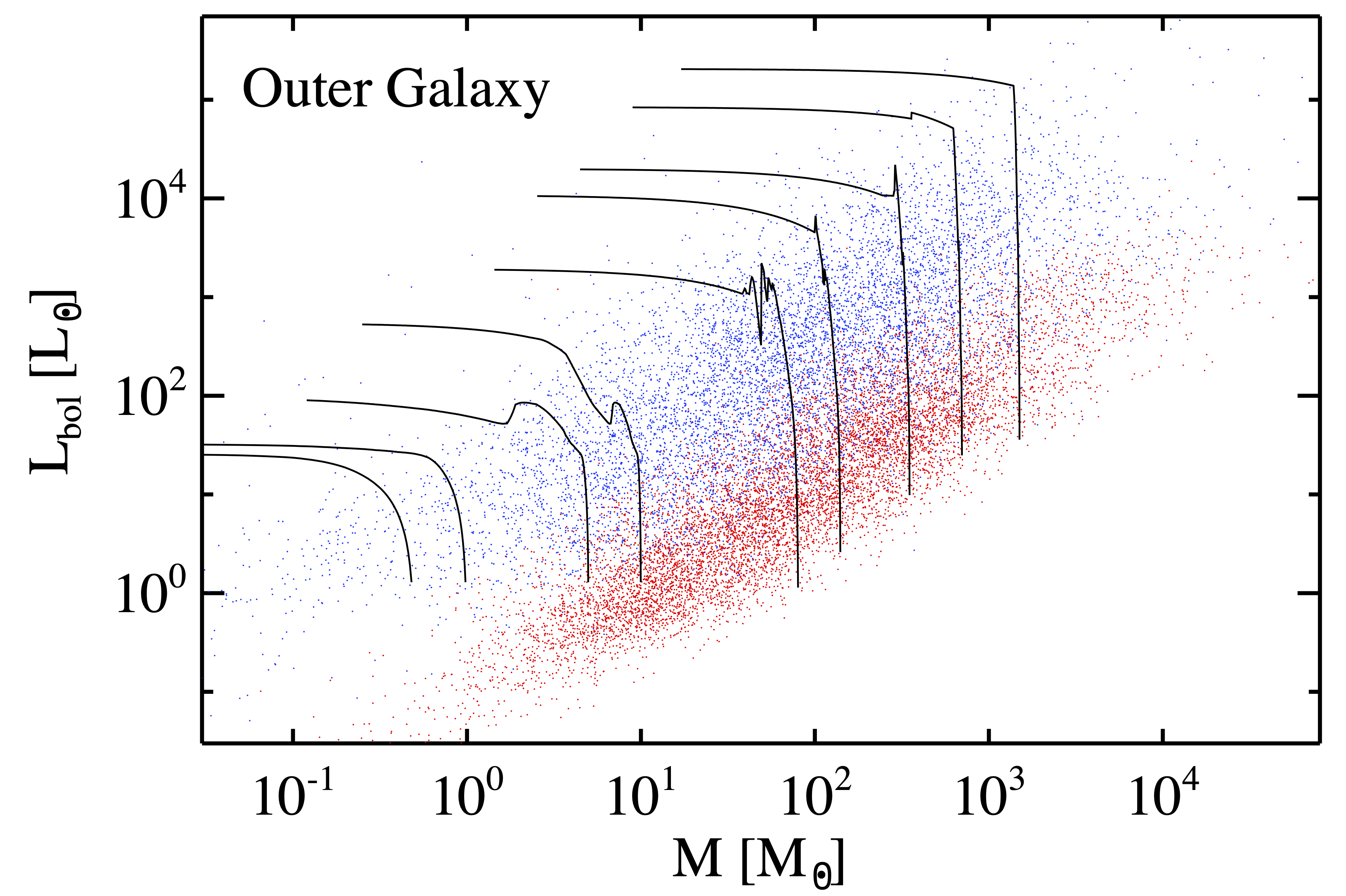}
\caption{Top panel: bolometric luminosity vs envelope mass for pre-stellar (red)
and protostellar (blue) sources in the inner Galaxy. The black lines represent
evolution, upwards and then to the left, on tracks from \citet{mol08}. 
As in Fig.~\ref{masssize}, since the distance estimate is the main source of uncertainty for both $M$ and $L_\mathrm{bol}$, a hypothetical distance increase of a factor 2 is represented as an arrow at the bottom right corner.
Bottom panel: same as top panel, but for the outer Galaxy.}\label{l_vs_m}
\end{figure}

The $L_\mathrm{bol}$ vs $M$ plot for sources in the inner Galaxy is presented again here
(Fig.~\ref{l_vs_m}, top), because of changes in distances and the different operational
definition of inner Galaxy adopted here. Postponing quantitative considerations to 
Section~\ref{lmr_par}, in which the ratio of $L_\mathrm{bol}$ to $M$ is used to summarise
the relation between these two quantities for different populations, here we simply note that Fig.~\ref{l_vs_m}
is qualitatively very similar to the corresponding plot in 
\citetalias{eli17}. Again, a high degree of segregation is found between pre-stellar sources, that populate the bottom
part of the diagram corresponding to the beginning of evolutionary tracks of \citet{mol08},
and protostellar sources, that are located in a higher area of the diagram corresponding to
more evolved stages and bordering the area occupied by H\,\textsc{ii} regions (see Section~\ref{lmr_par}).

Residual confusion between the two classes arises from the 
observed scatter in luminosity of pre-stellar clumps. This scatter corresponds to the relatively wide 
range of temperatures found (Section~\ref{temppar}), which 
depends, in turn, on different levels of external irradiation (Section~\ref{radarsec}) combined
with the absence of a central energy source. For
example, recently \citet{zha20}, focusing on massive starless clumps, showed that those associated with an
H\,\textsc{ii} region generally exhibit larger $L_\mathrm{bol}/M$ values, more typical of protostellar sources.

\begin{figure}
\centering
\includegraphics[width=8.5cm]{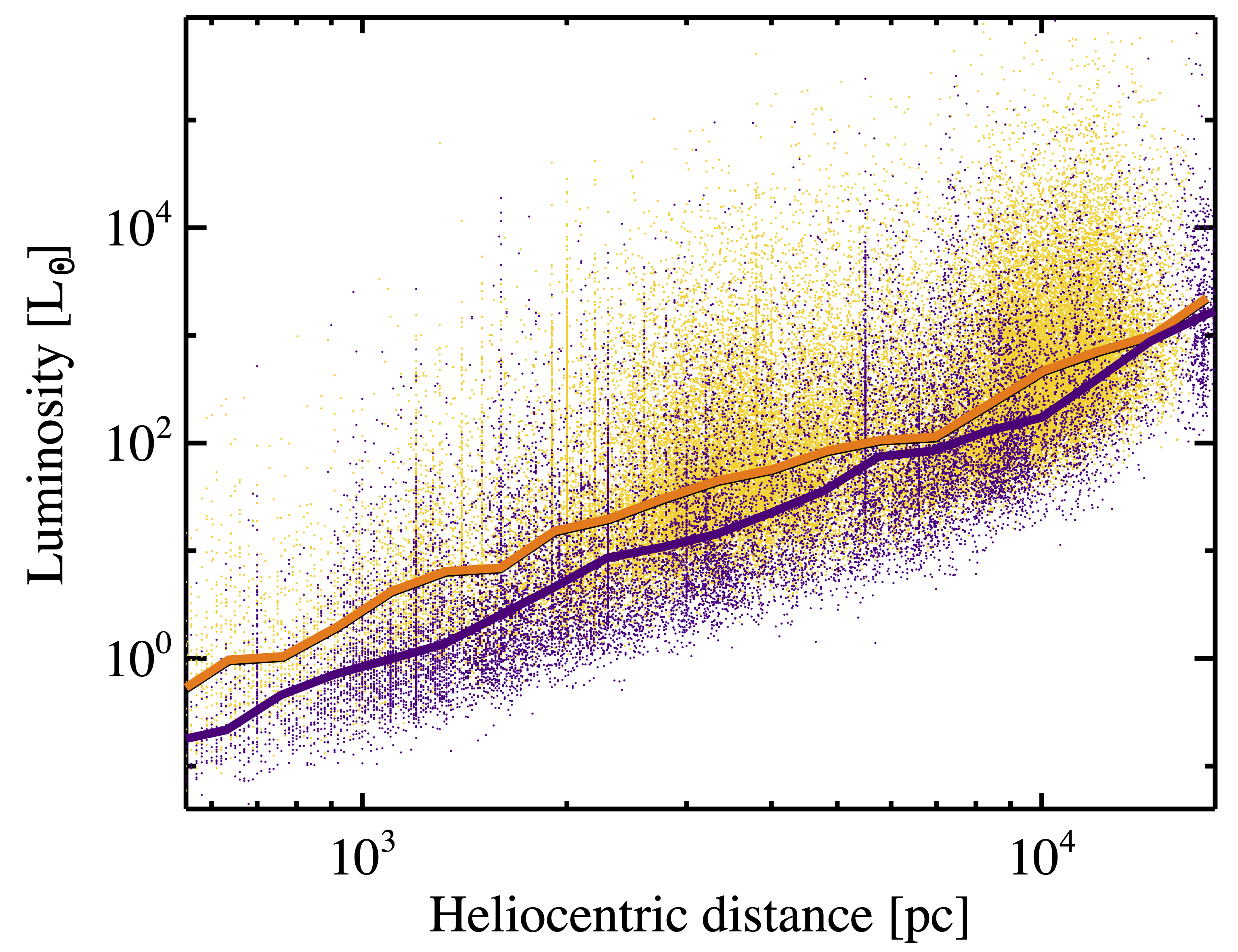}
\caption{Same as Fig.~\ref{massdist}, but for bolometric luminosity. 
}\label{lumdist}
\end{figure}

The $L_\mathrm{bol}$ vs $M$ diagram for the outer Galaxy sources (Fig.~\ref{l_vs_m}, bottom) exhibits qualitatively
similar behaviour, but spans different ranges in mass and luminosity, most of which
remain below $10^3~M_\odot$ and $10^3~L_\odot$, respectively. A different range for masses
in the outer Galaxy, which can be explained only partially by different distances involved, has been discussed in 
Section~\ref{masses}. Similarly, here we use median luminosities calculated in bins of 
distance (Fig.~\ref{lumdist}) to show that like for masses, on average luminosities are also intrinsically
lower in the outer Galaxy than in the inner Galaxy. 

\subsection{Clump lifetimes}\label{lifepar}
Information about the bolometric luminosity might be used, in principle, to infer clump lifetimes
similarly to \citet{urq18}. They establish a relation between the H\,\textsc{ii} region lifetime 
$t_{\textrm{HII}}$ and the luminosity of their sources through the function
$\log(t_{\textrm{HII}}/\textrm{yr})=(-0.13\pm 0.16) \times \log (L/L_\odot)+(6.1 \pm 0.8)$ of \citet{mot11}.
A further link with the source mass is established, based on a mild $L_\mathrm{bol}$-$M$ power-law relation they
 recognise in their data. Finally, lifetimes of different evolutionary classes 
\citep[quiescent, protostellar, young stellar objects, and massive star-forming regions, according to 
the classification of][]{kon17}
are derived as a function of mass bin by subdividing the total $t_{\textrm{HII}}(M)$ in proportion
to the relative populations of these classes in each bin. 
In other words, the H\,\textsc{ii} region lifetimes are required to convert relative lifetimes, typically obtained from population ratios \citep[cf., e.g.,][]{bat17}, to absolute ones.

Our approach contains some slight differences. First, we prefer not to identify a trend in the 
$L_\mathrm{bol}$-$M$ relation, given the high degree of degeneracy seen in Fig.~\ref{l_vs_m}.
In \citetalias{eli17} a conservative threshold of $22.4~L_{\odot}/M_{\odot}$ was established to 
identify, in the absence of radio observations, a robust sub-sample of protostellar sources that are candidates to host an H\,\textsc{ii}
region. This threshold was based on the $L_\mathrm{bol}/M$ 
distribution of Hi-GAL counterparts of CORNISH \citep{hoa12,pur13} radio sources obtained by
\citet{ces15}. We prefer to insert that single-valued threshold in the aforementioned function
of \citet{mot11} to establish the relation between mass and the corresponding $t_{\textrm{HII}}$.
The estimated lifetimes for the H\,\textsc{ii} regions are quite
uncertain due to the error bars in the function of \citet{mot11} and
our choice of a constant value, $22.4~L_{\odot}/M_{\odot}$, as representative of this evolutionary stage.
Second, we want to take into account differences among sources in terms of size (see discussion in 
Section~\ref{physsize}) and thus underlying unresolved structure, 
which depends in turn on the heliocentric distance. 
Note also that
in this analysis we do not include unbound clumps in general or all starless clumps in the low-reliability
catalogue.

\begin{figure}
\centering
\includegraphics[width=8.5cm]{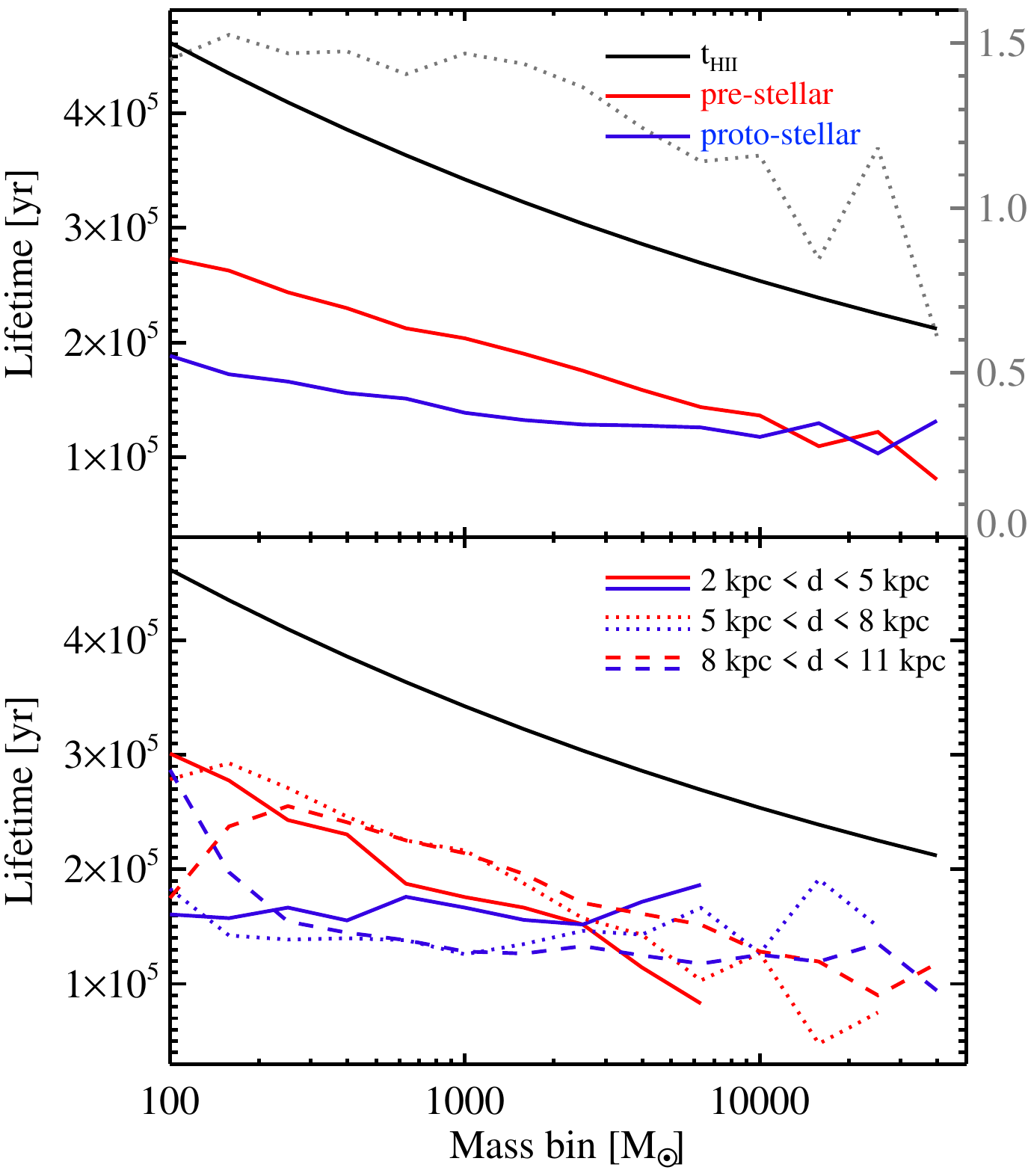}
\caption{Lifetimes vs mass. The black line in each panel represents the mass dependence of the lifetime $t_\textrm{HII}$ for a clump
hosting an H\,\textsc{ii} region according to 
$\log(t_{\textrm{HII}}/\textrm{yr})=-0.13 \times \log (L/L_\odot)+6.1$ \citep{mot11} and assuming $L_\mathrm{bol}/M=22.4~L_{\odot}/M_{\odot}$.
By definition $t_\textrm{HII}$ is the sum of the lifetimes of the pre-stellar and protostellar phases, and so these lifetimes can be
derived from the relative populations of these phases in each
bin of mass (see the text). The numeric ratio of pre-stellar to protostellar clumps is also plotted as a dotted grey line, referring to the grey $y$-axis on the right.
\textit{Top}: lifetimes of pre-stellar (red line) and protostellar (blue line) clumps for the entire sample.
\textit{Bottom}: same as the top, but dividing the
clumps into three different ranges of heliocentric distance: 2-5~kpc (solid lines), 5-8~kpc (dotted lines),
and 8-11~kpc (dashed lines).}\label{timescales}
\end{figure}

In Fig.~\ref{timescales} the results of this analysis are shown for the whole sample (upper panel) and for three
different ranges of distances, 2-5, 5-8, and 8-11~kpc (lower panel).
Unlike \citet{urq18}, in our case quiescent pre-stellar sources represent the majority of the sample and this translates into a longer lifetime compared to that of protostellar clumps, for masses up to
$\gtrsim 10^4~M_\odot$. The lower panel of the figure, however, shows how quantitatively different 
the relative behaviour of these two lifetimes becomes if a smaller range of distances, and hence masses, is considered. At closer distances, 
i.e., in the case less biased by distance, the behaviour is not unlike that seen overall, but the two 
curves cross at $M \sim 2 \times 10^3~M_\odot$. The next case, from 5 to 8~kpc, is also similar to the overall curves, but there is a relative deficit of pre-stellar sources at low masses ($M \sim 100~M{\odot}$) and exaggerated change at the highest $M$.
For the most distant case, 8-11~kpc, the deficit at low $M$ is much more pronounced so that the curves cross at $M \sim 140~M_\odot$.
These differences serve as a caution
that objects with the same mass but having a large range of distances might correspond to different kinds of structures, requiring separate analysis and conclusions.

Two further comments to this analysis are required. First, the constant $L_\mathrm{bol}/M$ ratio assumed for calculating H\,\textsc{ii} region lifetimes was determined in origin as a very
conservative threshold. Adopting a reasonably lower value for it \citep[for example, by a factor $\sim 2$, cf.][]{ces15} would imply, for a fixed mass bin, to linearly
decrease also the luminosity appearing in the reported relation by \citet{mot11} and, consequently, to estimate a systematically longer lifetime.

Second, it is to notice that the analysis above would remain qualitatively identical, in terms of relative proportions of pre-stellar and protostellar sources in single mass bins, if
another set of total lifetimes was adopted to absolutely scale the clump lifetimes. For example, while here we used H\,\textsc{ii} region lifetimes similarly to \citet{urq18}, \citet{svo16} used lifetimes
of CH$_3$OH masers, and \citet{bat17} used both. Considering only the behaviour of the class mutual proportions, we see that in our case the pre-stellar to protostellar ratio decreases at increasing mass bin (Fig.~\ref{timescales}, top) as in \citet{svo16}, but
with a shallower slope, essentially in the range between $10^2$ and $10^3~M_\odot$. This is due to a relevant amount of pre-stellar clumps also at relatively large
masses for which, in turn, two explanations can be given: $i$) the larger fraction of pre-stellar sources in the Hi-GAL catalogue compared to
the BGPS case, and $ii$) the lower temperatures obtained for many Hi-GAL pre-stellar clumps from MBB fit, compared with the kinetic ones adopted by \citet{svo16}, which typically imply higher
masses. Interestingly, the pre-stellar to protostellar number ratio of $\sim 1.5$ observed in Fig.~\ref{timescales} for masses up to $\sim 10^3~M_\odot$ corresponds to a relative time of
60~per cent spent in the pre-stellar phase and 40~per cent in the protostellar, which is consistent with an analogous estimate of \citet{bat17}. For the population ratio of $\sim 1.2$ 
achieved at $10^4~M_\odot$ (over this value the curve starts to show significant scatter), the above time fractions change to 55~per cent and 45~per cent, respectively. Notice that, for
consistency with \citet{svo16} and \citet{bat17} papers, the above comparisons have been made by considering our entire sample, with no division in distance bins as done for Fig.~\ref{timescales}, bottom.

\section{Distance-independent parameters}\label{nodistparam} 
Given the large uncertainties existing on heliocentric distance estimates (M\`ege et al., 
accepted), the analysis of distance-independent source parameters is surely more 
robust, being formally unbiased. Actually, distance affects 
the meaning that we can assign to such observables, because they are single global/average
numbers describing entire complex but unresolved structures. 
For example, a fundamental difference exists between assigning an average 
temperature to a protostellar core and to a much larger clump, in which a 
wider variety of physical conditions can coexist, from active star-forming sites 
to quiescent regions. Notwithstanding, the analyses by \citet{bal17a,bal17b}
on how the distance bias affects temperature and the luminosity/mass
ratio, respectively, suggest that, in general, global estimates of these
parameters for distant clumps mirror the average behaviour for the 
same parameters in the underlying population of cores. 
This encourages us to discuss distance-independent parameters and to
propose evolutionary classification metrics based on them (Section~\ref{radarsec}).

\subsection{Luminosity-mass ratio}\label{lmr_par}
\begin{figure*}
\centering
\includegraphics[width=18cm]{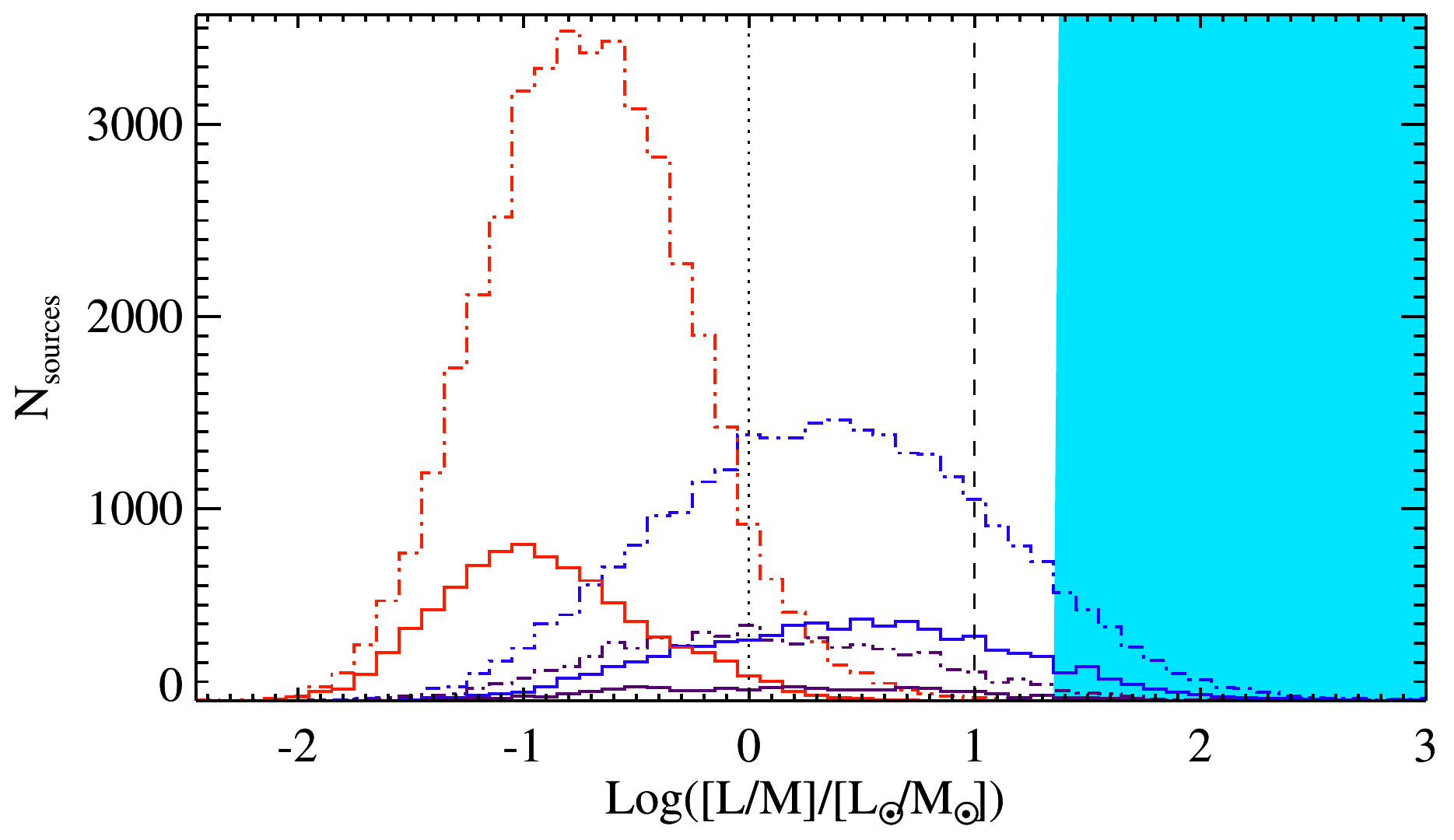}
\caption{Distributions of the ratio of bolometric luminosity to envelope mass for pre-stellar (red histograms), 
protostellar (blue histograms) and MIR-dark protostellar (dark purple histograms, arbitrarily enhanced by a factor of~2 to improve the readability) Hi-GAL sources 
considered for science analysis in this paper. Dot-dashed histograms are for the inner Galaxy 
and solid for the outer. The dotted and the dashed vertical lines represent the
thresholds identified by \citet{mol16b} for expecting star formation to be traced by line emission 
of methyl-acetylene CH$_3$C$_2$H(12-11) or by the presence of a zero age main sequence (ZAMS) star 
inside the clump, respectively. 
The light blue-shaded area contains clumps that are candidates to host H\,\textsc{ii} regions (see the text).}
\label{lmrfig}
\end{figure*}

We discuss first the ratio of bolometric luminosity to 
envelope mass, $L_\mathrm{bol}/M$. As we will see below, a threshold on this parameter
allows us to identify a sub-class of particularly evolved protostellar sources,
to be analysed subsequently in light of the additional distance-independent quantities.

Figure~\ref{lmrfig} shows the distribution of this ratio for both the pre-stellar and protostellar 
sources, in both the inner and outer Galaxy.
A good degree of segregation is seen between the two classes of objects, especially in the outer
Galaxy. We evaluate its extent by quantifying the fraction of the histogram
area of a source class overlapping the histogram of the other class, and vice versa, as follows:
given two generic histograms $H_1$ and $H_2$ defined over the same $N_\mathrm{bin}$ bins, the area 
of their overlap region is $\sum_{i=1}^{N_\mathrm{bin}} \min (H_1(i),H_2(i))$. 
The overlap fractions for the two 
histograms are this number divided
by the integral of $H_1$ and $H_2$, respectively. For the adopted histogram bin size of Fig.~\ref{lmrfig} 
(0.1 for $\log_{10}([L_\mathrm{bol}/M]/[L_{\odot}/M_{\odot}])$), for the inner Galaxy the 
overlap fractions are \percinterslminnpre~per cent for pre-stellar sources and 
\percinterslminnproto~per cent for protostellar sources. For the outer Galaxy 
they drop to \percinterslmoutpre~per cent and \percinterslmoutproto~per cent, respectively, i.e., more segregation. 

Correspondingly, a larger gap between median values is observed in the outer Galaxy: medians of 
$L_\mathrm{bol}/M$
for pre-stellar and protostellar sources are \medianlminnpre~$L_{\odot}/M_{\odot}$ and 
\medianlminnproto~$L_{\odot}/M_{\odot}$, respectively, in the inner Galaxy, and 
\medianlmoutpre~$L_{\odot}/M_{\odot}$ and \medianlmoutproto~$L_{\odot}/M_{\odot}$, respectively, 
in the outer Galaxy. 

All four values are lower than $10~L_{\odot}/M_{\odot}$,
around which the ATLASGAL sources of \citet{urq18} appear to have a concentration. But these are 
among the brightest far-infrared sources in the Galaxy 
and Hi-GAL, which is remarkably more sensitive, is able to detect
significantly fainter sources.
Table~\ref{evoltable} records these medians along with medians of
all distance-independent parameters, separately for different evolutionary classes 
and inner/outer Galaxy location.

\begin{table*}
\centering
\caption{Median values for distance-independent physical parameters, subdivided by evolutionary class and sub-class and inner/outer Galaxy location.}
\label{evoltable}
\begin{tabular}{lcccc c cccc}
\hline
 & \multicolumn{4}{c}{Inner Galaxy} & &\multicolumn{4}{c}{Outer Galaxy} \\
\cline{2-5} \cline{7-10}
 & Pre-stellar & \multicolumn{3}{c}{Protostellar} &&  Pre-stellar & \multicolumn{3}{c}{Protostellar} \\
\cline{3-5} \cline{8-10}
 &             & All & MIR-dark & H\textsc{ii} candidates &&             & All & MIR-dark & H\textsc{ii} candidates \\
\hline
$L_\mathrm{bol}/M\, [L_\odot/M_\odot]$ & 0.2 & 2.6 & 1.2 & 40.4 && 0.1 & 3.1 & 1.7 & 39.7\\
$T$\, [K]  & 11.4 & 15.2 & 14.6 & 24.6 && 10.5 & 15.3 & 15.3 & 23.9\\
$L_\mathrm{bol}/L_\mathrm{smm}$ & 5.7 & 30.4 & 15.6 & 191.9 && 4.4 & 36.9 & 30.9 & 199.6\\
$T_\mathrm{bol}$\, [K]  & 17.6 & 39.5 & 23.7 & 50.5 && 16.2 & 43.4 & 25.6 & 51.5\\
$\Sigma$\, [g~cm$^{-2}$]  & 0.14 & 0.21 & 0.24 & 0.07 && 0.10 & 0.12 & 0.15 & 0.05\\
\hline
\end{tabular}

\vspace{1ex}
The uncertainty of each median can be estimated as 
$(Q_3-Q_1)/(2\sqrt{N_\mathrm{tot}})$, i.e., as the half distance between the third and the first quartile of the 
distribution (also a surrogate for the standard deviation in the case of strongly asymmetrical distributions),
divided by the square root of the total number of counted objects, which can be quite different for different populations (Table~\ref{catstats}). 
Consequently, the uncertainty of the median of $L_\mathrm{bol}/M$ ranges from $0.001~L_{\odot}/M_{\odot}$ for pre-stellar sources in the inner Galaxy to $0.6~L_{\odot}/M_{\odot}$ for H\textsc{ii} region candidates in the outer Galaxy.
Similarly, uncertainties of the median range from 0.007 to 0.1~K for $T$, from 0.01 to 2 for $L_\mathrm{bol}/L_\mathrm{smm}$, from 0.01 to 0.02~K for $T_\mathrm{bol}$, and from 0.0005 to 0.001~g~cm$^{-2}$ for $\Sigma$.

\end{table*}

Detection of CH$_3$C$_2$H(12-11) line emission is considered a signature of ongoing star formation.
By cross-correlation, \citet{mol16b} proposed a threshold of 1~$L_{\odot}/M_{\odot}$ for detection.
Fig.~\ref{lmrfig} shows that in the inner Galaxy this threshold falls well inside the region of the 
pre-stellar/protostellar overlap, whereas in the outer Galaxy only a small fraction of 
pre-stellar clumps is found above this threshold. 
As seen by \citet{zha20}, quiescent massive clumps associated with H\,\textsc{ii} regions
can reach $L_\mathrm{bol}/M > 10~L_{\odot}/M_{\odot}$ because of significant external heating, 
suggesting a more advanced evolutionary state than is the case. 
The higher density of H\,\textsc{ii} regions in the inner
Galaxy compared to the outer Galaxy \citep{and14} and, in general, the stronger interstellar radiation 
field \citep[e.g.,][]{mat83} can create the higher degree of overlap in the inner Galaxy. 
Thus, this effect is probably the major cause
of the overlap of the pre-stellar distribution into the protostellar distribution, for this parameter and also the others
discussed in the following sections. 
A further 
check to address this is described in Section~\ref{radarsec}.

Additional contamination between the
two classes can arise from source misclassification due to the aforementioned distance bias,
as examined in \citetalias{eli17}. This is expected to be more of an issue in the inner Galaxy because of the 
larger heliocentric distances ($d > 12$~kpc) there.

\citet{mol16b} further proposed a threshold of 10~$L_{\odot}/M_{\odot}$ at which the temperature derived from CH$_3$C$_2$H(12-11) 
starts to increase monotonically for increasing $L_\mathrm{bol}/M$, and interpreted this as the 
appearance of one or more ZAMS stars in the clump. In our data for both the inner and 
the outer Galaxy a significant fraction of protostellar sources is found beyond this value
(\nlmrzamsinn~per cent and \nlmrzamsout~per cent, respectively), whereas the presence there of 
pre-stellar sources is negligible.

The light blue shaded area in Fig.~\ref{lmrfig} corresponds to the $22.4~L_{\odot}/M_{\odot}$
threshold for identifying candidate H\,\textsc{ii} regions, as introduced in textbf{Section~\ref{lifepar}} \citep[see also][]{eli20}. We find \nhiiinner~candidates in the inner Galaxy 
and \nhiiouter~in the outer Galaxy. These are expected to be the most evolved sources in our 
catalogue. Their corresponding median values are reported in Table~\ref{evoltable}.
Checking for further evidence of the H\,\textsc{ii} region nature of these objects
is not among the aims of this paper. Nevertheless, we cross-matched the positions of these sources 
with the WISE catalogue of H\,\textsc{ii} regions by \citet{and14} and found \andersonmatches~matches,
\andersonmatchesnoq~of which are associated with regions showing radio emission.

On the opposite side of the distribution for the protostellar class we expect to find the
MIR-dark sources \citepalias[cf.][]{eli17}, i.e., those having a detection at $70~\umu$m
but no detection at shorter wavelengths (MSX/WISE/MIPSGAL). However, as already shown
in \citetalias{eli17}, although evolutionary parameters of these sources do indicate, on 
average, an earlier stage with respect to the global population of protostellar sources,
they do not actually produce a clear ``left tail'' of the protostellar distribution. This is 
confirmed by the distributions plotted for this sub-class in Fig.~\ref{lmrfig}, which 
extend over a wide range of $L_\mathrm{bol}/M$ values both in the inner and in the 
outer Galaxy. Their corresponding median values are reported in Table~\ref{evoltable}.

\subsection{MBB temperature}\label{temppar}
\begin{figure}
\centering
\includegraphics[width=8.5cm]{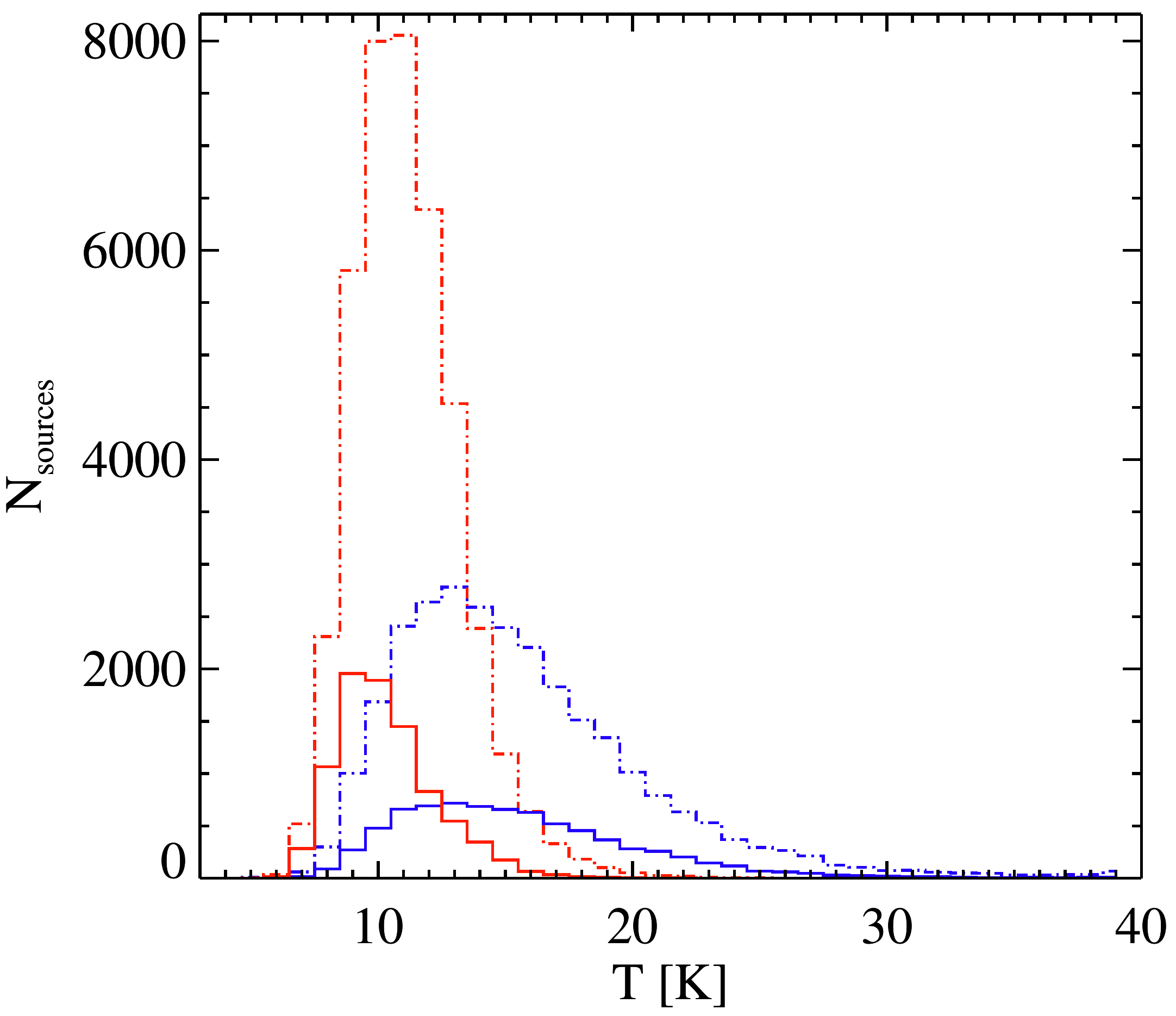}
\caption{Distributions of MBB temperature for pre-stellar (red histograms) and 
protostellar (blue histograms) sources. Dot-dashed histograms are
for the inner Galaxy and solid for the outer.}
\label{hist_temp}
\end{figure}

In \citetalias{eli17} it has been demonstrated that the dust temperature $T$ estimated by the MBB
fit of SEDs at $\lambda \geq 160~\umu$m shows quite different distributions for pre-stellar
and protostellar sources. Moreover, temperature itself is a reliable evolutionary parameter
for protostellar sources. It is reasonably well correlated with other evolutionary indicators, 
first of all $L_\mathrm{bol}/M$, although it produces a lower degree of segregation. 

These conclusions are corroborated by the extension of our analysis to the outer Galaxy.
In Fig.~\ref{hist_temp} the new temperature distributions for both pre-stellar and protostellar
sources in the outer Galaxy are shown, together with those in the inner Galaxy for comparison.
It is confirmed that also in the outer Galaxy the temperature of protostellar sources is higher, on average, than that of pre-stellar sources, as it was already found for the inner Galaxy in \citetalias{eli17} and also in \citet{svo16} and \citet{mer19}, based on independent ammonia observations.
Similarly to the behaviour seen for $L_\mathrm{bol}/M$ in Section~\ref{lmr_par}, a 
higher degree of segregation between the two distributions is found in the outer Galaxy, probably
due to a lower impact of the environment on the temperature of pre-stellar clumps (see 
Section~\ref{radarsec}). Using the same method to determine the overlap
of the two pre-stellar and protostellar histograms, we find that in the outer Galaxy the overlap fraction corresponds
to \percinterstoutpre~per cent and \percinterstoutproto~per cent, respectively, 
compared with \percinterstinnpre~per cent and \percinterstinnproto~per cent, respectively, in the inner Galaxy.
Correspondingly, the median values of $T$ are also found to be more distant from each other
in the outer Galaxy (\mediantempoutpre~K and \mediantempoutproto~K for pre-stellar and protostellar population, respectively), 
than in the inner Galaxy (\mediantempinnpre~K and \mediantempinnproto~K, respectively). See again Table~\ref{evoltable}. Notice that, despite the degree of overlap between pre-stellar and protostellar distributions, the gap between their medians is meaningful, being even broader than similar estimates: \citet{liu18} found 13.5~K and 15.5~K, respectively, in a sample of MALT90 \citep{jac13} clumps with Hi-GAL counterparts.

Here we briefly discuss whether and how the observed segregation among source classes (and sub-classes) in Fig.~\ref{hist_temp} can be affected by our choice of using a single and common opacity law for the MBB fit of all SEDs in the catalogue. Indeed, variations of the $\beta$ exponent are observed in the ISM \citep[e.g.,][]{sad16}, and generally interpreted as a consequence of dust grain evolution. Neglecting for a moment other dust parameters, it can be roughly said that grain growth produces a decrease of $\beta$ \citep[e.g.,][]{bec91,guz15,mer15,li17}. In our case, to possibly consider a smaller value of $\beta$ for protostellar sources with respect to pre-stellar sources would imply an increase of the temperature estimated through the MBB fit \citep{des08,mar12}, and consequently a higher level of separation in Figure~\ref{hist_temp} between the distributions of these two classes. However grain growth is observed to occur mostly in the vicinity of forming stars, then its most relevant effects can be observed mainly in resolved cores located in nearby regions as those studied by \citet{sad16}. This discourages us to consider a differentiation of $\beta$, rather than a common value, for Hi-GAL clump SEDs, which are dominated by the emission of the large-scale envelope.

\begin{figure}
\centering
\includegraphics[width=8.5cm]{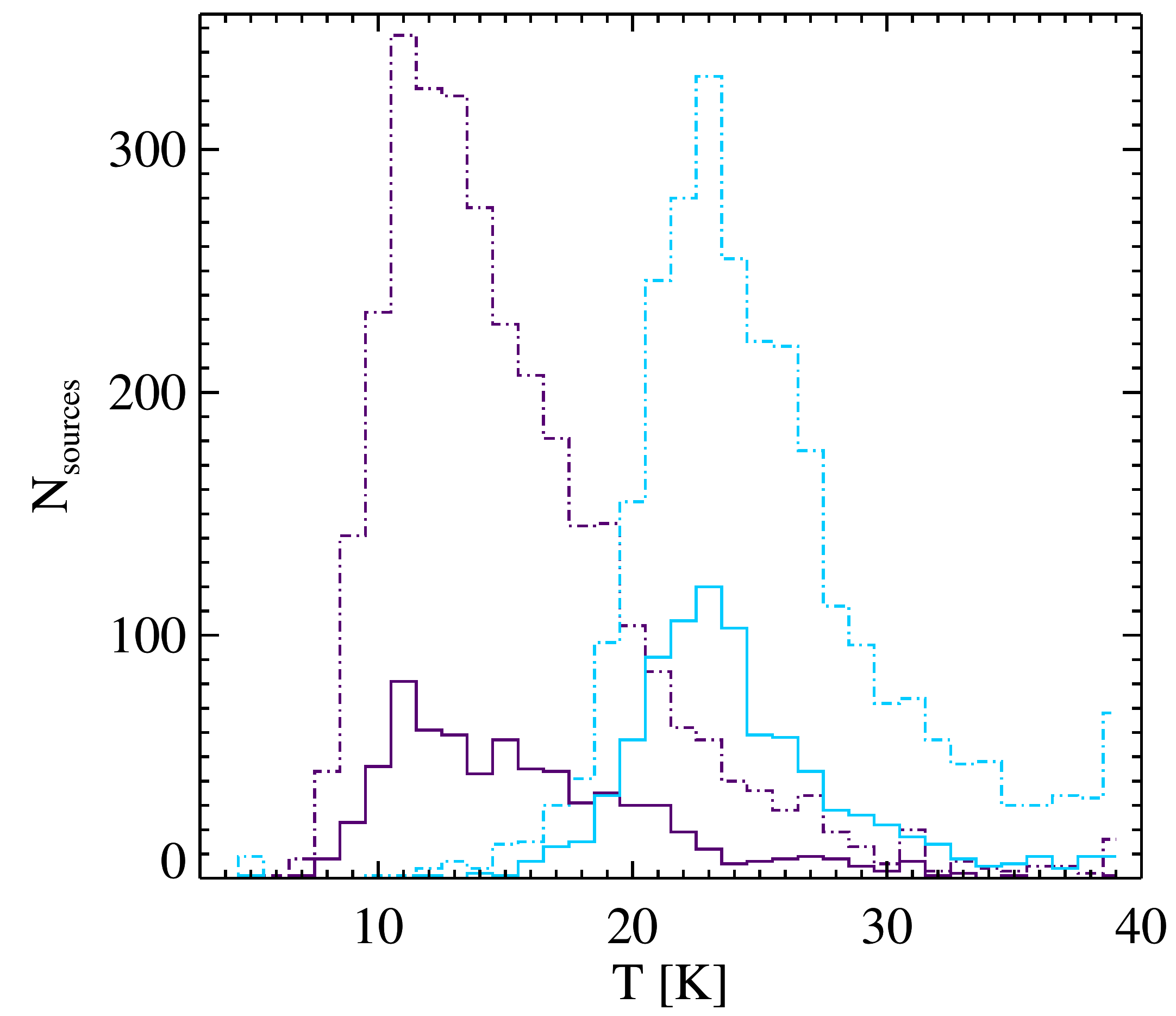}
\caption{Same as Fig.~\ref{hist_temp}, but for two particular sub-classes of the 
protostellar sample: MIR-dark sources (dark purple histograms) and H\,\textsc{ii} region 
candidates (light blue histograms).}
\label{hist_temp_hii}
\end{figure}

To explore in more detail the temperatures of sub-classes of the protostellar sample, namely MIR-dark
and candidate H\,\textsc{ii} regions, we plot separately their temperature
histograms in Fig.~\ref{hist_temp_hii}. 
As expected, candidate H\,\textsc{ii} regions are found at relatively high temperatures.

The median temperatures of candidate H\,\textsc{ii} regions in the inner and outer Galaxy are \mediantempinnhii~K and
\mediantempouthii~K, respectively, but both distributions are right-skewed and the 
10th percentiles are at \tenpercenttempinnhii~K and \tenpercenttempouthii~K, respectively. 
\citet{liu18}, \citet{guz15}, and \citet{urq13b} estimated a typical temperature of~22.5~K, 23.7~K and 25.0~K, respectively, for clumps hosting an ultra-compact H\,\textsc{ii} region, which is in good agreement with our 
statistics. Based on Hi-GAL data for twelve known H\,\textsc{ii} regions, \citet{par14} found temperatures in the range 22-45~K.
However, they included the $70~\umu$m data in their fits, both 
a simple MBB and a more refined model for dust grain emissivity, then accounted for 
measurements from a warmer dust component. Similarly, but considering only $\lambda > 70~\umu$m, \citet{pal12} found temperatures in the range 20-30~K for Hi-GAL counterparts of a sample of 16 evolved H\,\textsc{ii} regions. Finally, although again with a small sample of eight resolved H\,\textsc{ii}
regions observed in HOBYS survey \citep{mot10},
\citet{and12} highlighted that a fit to their entire SED yields, 
on average, a temperature of about $25$~K, but that considering their internal components individually 
average temperatures range from about $15$~K for infrared dark clouds to 
$26$~K for photodissociation regions. The compatibility of these previous values with median 
temperatures obtained for our H\,\textsc{ii} region candidates supports the reliability of the identification
criterion established in Section~\ref{lmr_par}.

The MIR-dark sub-class is expected to be less evolved among the protostellar sources and indeed the median values of $T$ are relatively low (\mediantempinnnomir~K
in the inner Galaxy and \mediantempoutnomir~K in the outer).
However, the distributions in 
Fig.~\ref{hist_temp_hii} show a significant range and are skewed towards higher temperatures, even overlapping the distributions for the H\,\textsc{ii} region candidates.
This behaviour, already highlighted in \citetalias{eli17}, is found also for the outer Galaxy.

\subsection{Ratio of bolometric to submillimetre luminosity}\label{lsubmm}

\begin{figure*}
\centering
\includegraphics[width=18cm]{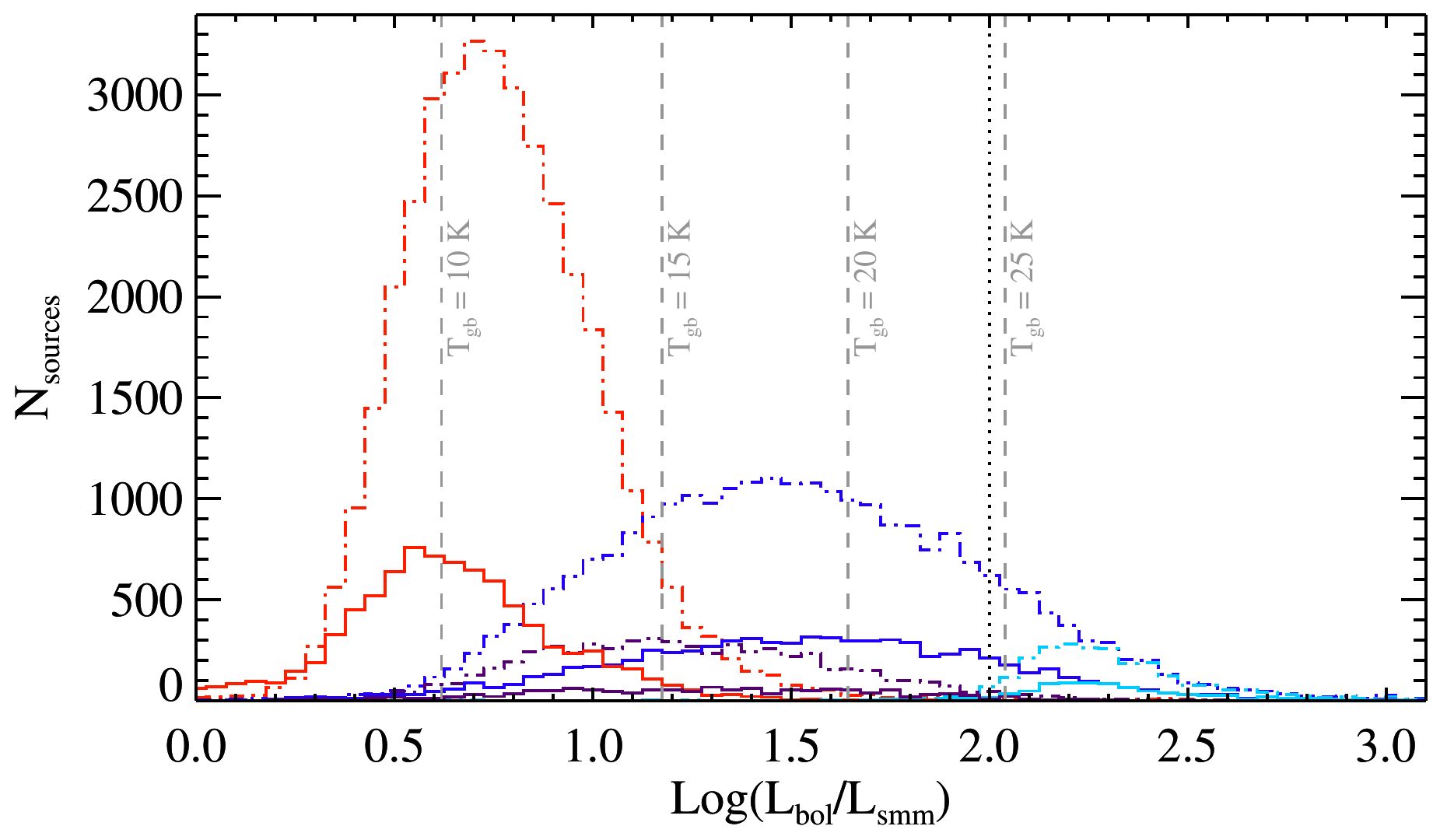}
\caption{Distributions of the ratio of the bolometric luminosity to the portion in the sub-millimetre ($\lambda \geq 350~\umu$m)
 for the pre-stellar (red histograms), protostellar (blue histograms), MIR-dark 
protostellar (dark purple histograms, arbitrarily enhanced by a factor of~2 to improve the readability), and H\,\textsc{ii}-region candidate (light blue histograms)
Hi-GAL sources, respectively. Dot-dashed histograms are for the inner Galaxy and solid for the 
outer. For a MBB with 
$\beta=2$, the dashed grey vertical lines mark ratios corresponding to the different MBB temperatures
indicated. The dotted black line at
$L_\mathrm{bol}/L_\mathrm{smm}=100$ marks the ratio separating Class~0 from Class~I low-mass YSOs 
as proposed by \citet{mau11}.}\label{lsub_hist}
\end{figure*}

We also use the ratio of the bolometric luminosity $L_\mathrm{bol}$ to the portion $L_\mathrm{smm}$ in the sub-millimetre ($\lambda \geq 350~\umu$m) as an 
evolutionary indicator. This was introduced by \citet{and00} for discriminating between Class~0 and Class~I young stellar
objects (YSOs) in the low-mass star formation regime \citep[with a separation threshold fixed at
$L_\mathrm{bol}/L_\mathrm{smm}=100$,][]{mau11}, but here it cannot be used for
such a classification because the sources being discussed are clumps containing 
entire star-forming regions, possibly including high-mass star formation. Nevertheless, 
as seen in \citetalias{eli17}, this parameter remains interesting because it also
ensures a good segregation among the evolutionary classes proposed for our sources.

The distributions of $L_\mathrm{bol}/L_\mathrm{smm}$ for the inner and the outer Galaxy can be compared in
Fig.~\ref{lsub_hist}. Similar to parameters analysed in previous sections, we 
note a stronger segregation between pre-stellar and protostellar populations in the outer Galaxy,
than in the inner Galaxy. This is quantified by the lower overlap fractions
(\percinterslsuboutpre~per cent for pre-stellar sources and \percinterslsuboutproto~per 
cent for protostellar sources in the outer Galaxy, against \percinterslsubinnpre~per cent and 
\percinterslsubinnproto~per cent, respectively, in the inner Galaxy), and by a larger gap between 
median values of the two populations (\medianlsuboutpre~and \medianlsuboutproto~in the outer
Galaxy, respectively, against \medianlsubinnpre~and \medianlsubinnproto~in the inner Galaxy). See again Table~\ref{evoltable}.

The values expected for a MBB \citep[cf.][]{eli16}
with $\beta=2$ are shown for four temperatures $T=10,~15,~20,~25$~K in Fig.~\ref{lsub_hist}, along with
the aforementioned $L_\mathrm{bol}/L_\mathrm{smm}=100$ (close to the one for $T=25$~K). We notice that almost all the H\,\textsc{ii} region candidates lie above 
100, suggesting this threshold as a necessary 
condition in searching for H\,\textsc{ii} region candidates among protostellar clumps.

\subsection{Bolometric temperature}\label{tbolpar}

The bolometric temperature $T_\mathrm{bol}$ has been found by \citetalias{eli17} 
to be the parameter for which the segregation between pre-stellar 
and protostellar sources is highest. It is defined as the average frequency of 
the SED, weighted with fluxes, and translated in terms of temperature of an 
equivalent blackbody \citep{mye93}. For an analytic MBB, the relation between $T_\mathrm{bol}$ and the MBB
temperature is linear \citep[e.g.][]{eli16}. However, here the MBB temperature is determined for data at $\lambda \geq 160~\umu$m and so for
sources with data at $\lambda < 100~\umu$m in excess of the fitted MBB,
$T_\mathrm{bol}$ is necessarily higher than in the linear relation, 
being particularly sensitive to MIR fluxes where detected (the impact of failure to detect a faint
MIR counterpart near the survey sensitivity limit on the estimate of $T_\mathrm{bol}$ is discussed in 
Appendix~\ref{mirappendix}).

\begin{figure}
\centering
\includegraphics[width=8.5cm]{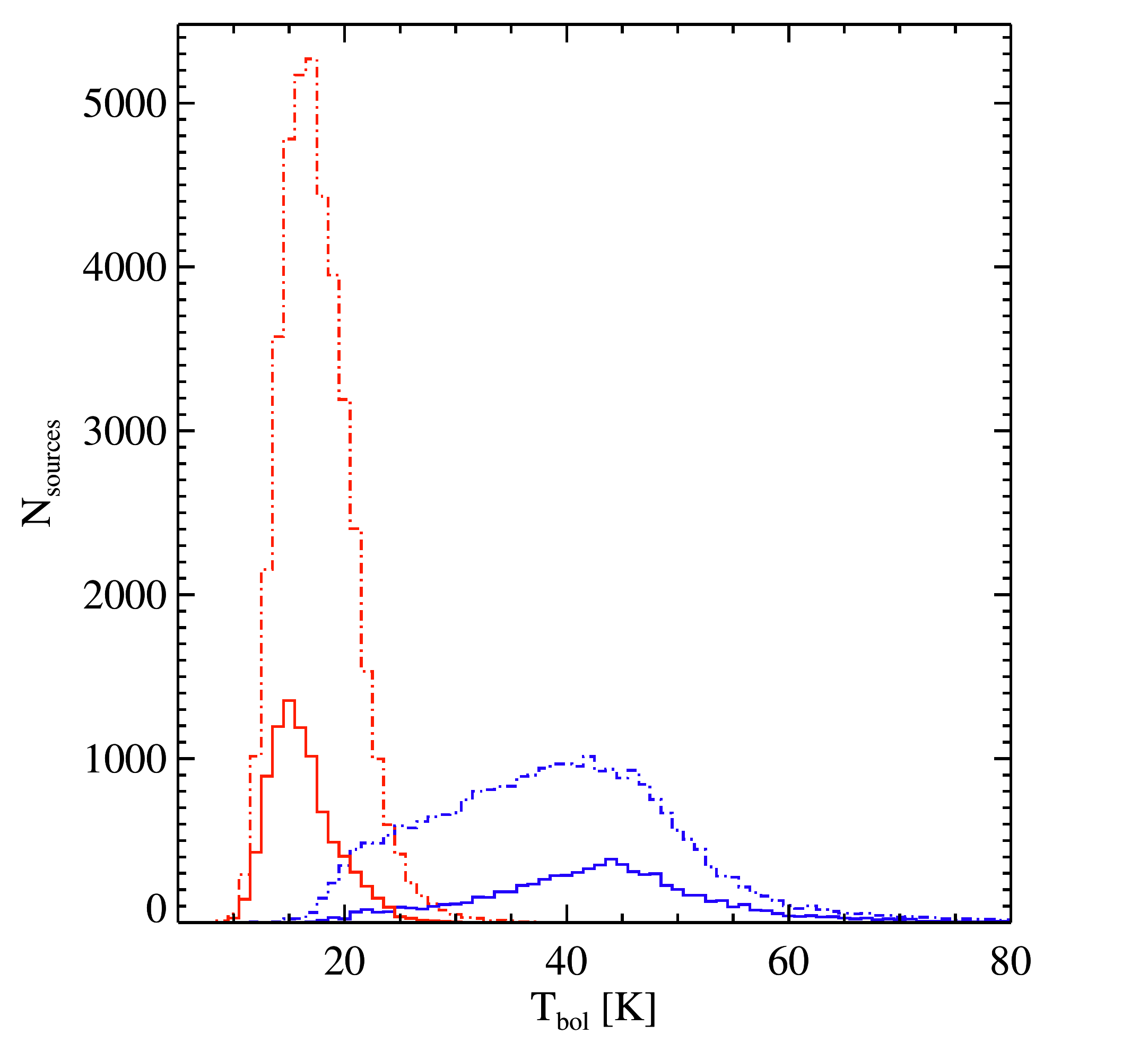}
\caption{Distributions of bolometric temperature for pre-stellar (red) and protostellar (blue) sources. Dot-dashed histograms are for the inner Galaxy and solid for the outer.}
\label{hist_tbol}
\end{figure}

As for other evolutionary indicators, the distributions of $T_\mathrm{bol}$ for pre-stellar and protostellar sources in the outer Galaxy appear 
better separated than in the inner Galaxy (Fig.~\ref{hist_tbol}). In the outer Galaxy, overlap fractions for the two
histograms are \percinterstboutpre~per cent and \percinterstboutproto~per cent for pre-stellar 
and protostellar clumps, respectively, and median values are \mediantboutpre~K and 
\mediantboutproto~K, respectively, whereas in the inner
Galaxy these quantities are \percinterstbinnpre~per cent 
and \percinterstbinnproto~per cent, and \mediantbinnpre~K and \mediantbinnproto~K. 

We can extend to the outer Galaxy two general considerations expressed in \citetalias{eli17} concerning the inner Galaxy. 
First, almost all sources have $T_\mathrm{bol}<70$~K, which 
was recognised as the threshold between Class~0 and Class~I objects by \citet{che95} in 
the low-mass star formation regime. Second, values we find are smaller than 
those of \citet{mul02} and \citet{ma13}, who considered SEDs more dominated by MIR fluxes.

\begin{figure}
\centering
\includegraphics[width=8.5cm]{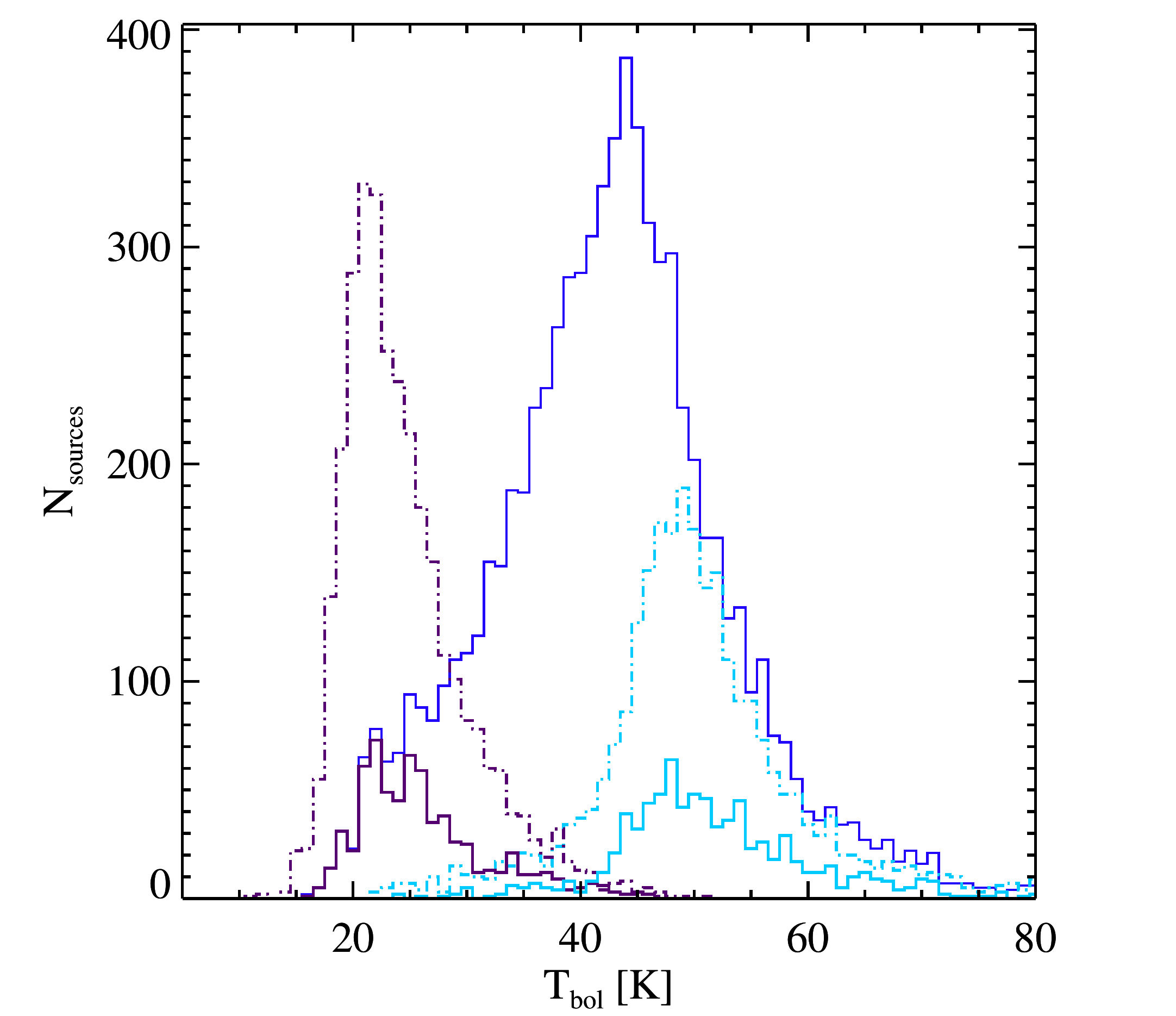}
\caption{Same as Fig.~\ref{hist_tbol}, but for MIR-dark sources (dark purple) and H\,\textsc{ii} region candidates (light blue). In addition,
the distribution for all protostellar sources in the outer Galaxy (solid blue, also shown in Fig.~\ref{hist_tbol}) is given for comparison.}
\label{hist_tbol_hii}
\end{figure}

The distributions of $T_\mathrm{bol}$ for both
MIR-dark and H\,\textsc{ii} region candidates are reported separately in Fig.~\ref{hist_tbol_hii}.
Unlike in the case of the MBB temperature (Fig.~\ref{hist_temp_hii}), here the two 
sub-classes appear well segregated. As already found in \citetalias{eli17} for
inner Galaxy sources, the MIR-dark sources in the outer Galaxy also produce the left tail of the 
entire protostellar distribution, as expected from the definition of $T_\mathrm{bol}$.
For H\,\textsc{ii} region candidates the distributions are shifted towards high $T_\mathrm{bol}$ 
(with median values \mediantbouthii~K and \mediantbinnhii~K in the outer 
and in the inner Galaxy, respectively); however, they make up only a subset of the right tail of the entire
protostellar distribution.

\subsection{Surface density}\label{surfdenspar}

\begin{figure}
\centering
\includegraphics[width=8.5cm]{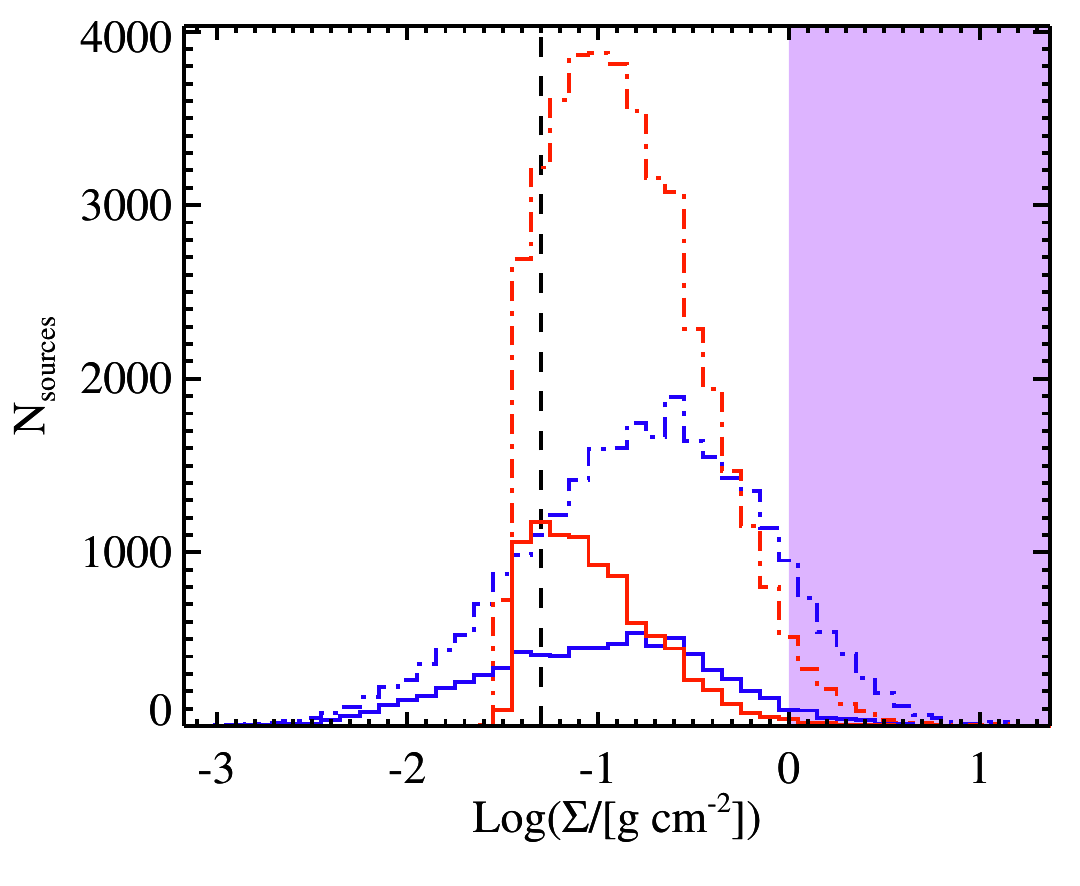}
\caption{Distributions of surface density for pre-stellar (red histograms) and
protostellar (blue histograms) in the inner Galaxy (dot-dashed histograms) and the outer
(solid histograms). The zone surpassing
the \citet{kru08} threshold of 1~g~cm$^{-2}$ is shaded purple, as in Fig.~\ref{masssize}. The lowest surface density at which massive star formation is found by \citet{urq14c}, namely $\Sigma=0.05$~g~cm$^{-2}$, is also reported as a dotted vertical line. Other thresholds for compatibility with massive star formation, discussed in Section~\ref{masses}, cannot be represented in this plot because they do not correspond to a constant value of $\Sigma$.}
\label{surfdens_hist}
\end{figure}

The surface density parameter $\Sigma$ summarises the mass-radius relation studied in Section~\ref{masses}.
Distributions of $\Sigma$ are shown in Fig.~\ref{surfdens_hist}. Unlike for the other parameters, the segregation of pre-stellar and protostellar sources is not obvious.

In \citetalias{eli17} we highlighted an increase of median surface density from pre-stellar
to protostellar sources \citep[see also][]{bat14}. Furthermore, the median surface density of MIR-dark 
sub-sample was even higher, suggesting that the highest density is achieved around this stage, before 
the source starts to emit in the MIR. At later times, stellar feedback can start to be relevant in producing
envelope dissipation and so a possible decrease of median surface density. The effect of this 
evolution should be most evident on the opposite side of protostellar class, i.e., for H\,\textsc{ii} 
region candidates \citep[cf.][]{guz15}, though this is complicated by the very large spread in that distribution 
around the median \citep[cf. also][]{feh17}. 

\begin{figure}
\centering
\includegraphics[width=8.5cm]{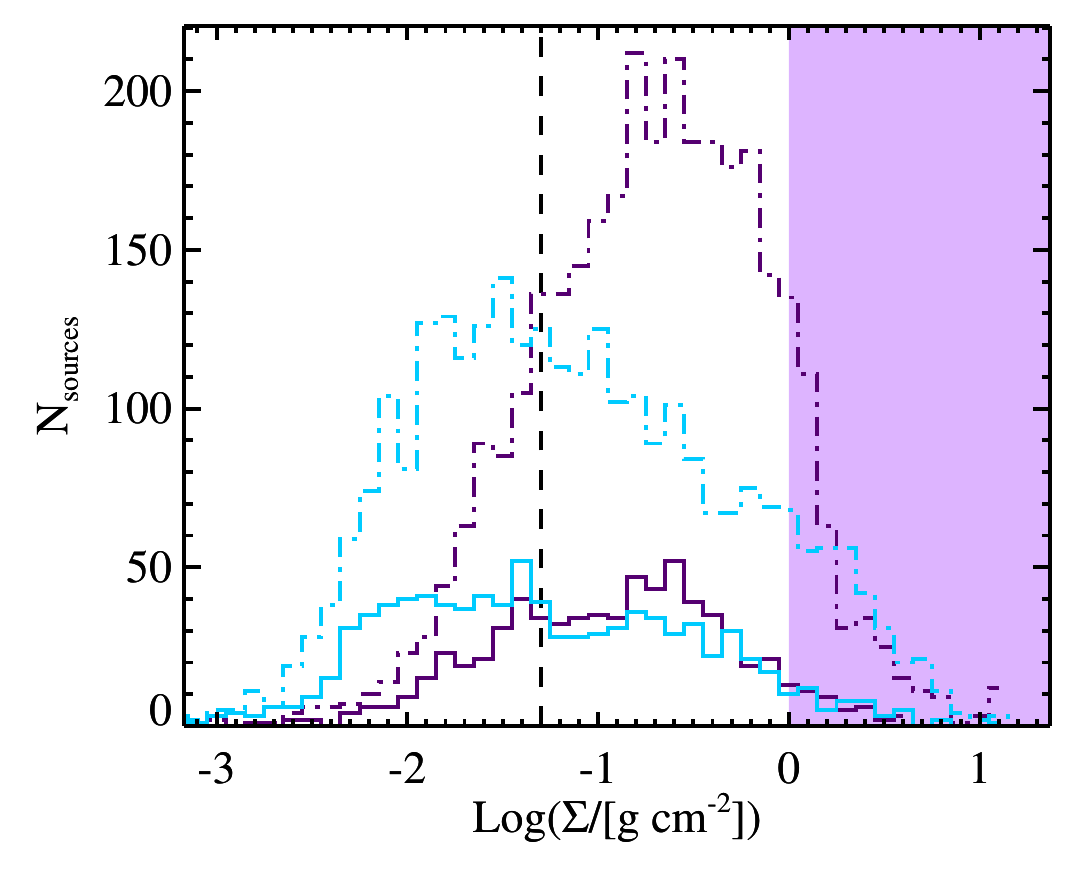}
\caption{Same as Fig.~\ref{surfdens_hist}, but for two sub-classes of the protostellar sample: MIR-dark sources 
(dark purple histograms) and H\,\textsc{ii} region candidates (light blue histograms).}
\label{surfdens_hist_hii}
\end{figure}

This trend of median surface densities in the inner Galaxy is confirmed in this work, with values
\mediansdinnpre, 
\mediansdinnnomir,
\mediansdinnproto, and
\mediansdinnhii~g~cm$^{-2}$ for pre-stellar, MIR-dark, protostellar, and H\,\textsc{ii} 
region candidates, respectively (Table~\ref{evoltable}).
In the outer Galaxy the sequence is 
\mediansdoutpre, 
\mediansdoutnomir,
\mediansdoutproto, and
\mediansdouthii~g~cm$^{-2}$.
Although the trend is the same,
the median surface densities are systematically lower for all classes, 
as can be appreciated also in Figs.~\ref{surfdens_hist} and \ref{surfdens_hist_hii}
\citep[cf. also][]{zah16}. This is connected directly to the different regimes of masses observed in the inner and outer
Galaxy (Fig.~\ref{massdist}).

To connect this information in a more systematic way, the surface density is discussed again 
in Section~\ref{galactotrends} as a function of the Galactocentric radius.

\subsection{Overall classification}\label{radarsec}
\begin{figure}
\centering
\includegraphics[width=8.5cm]{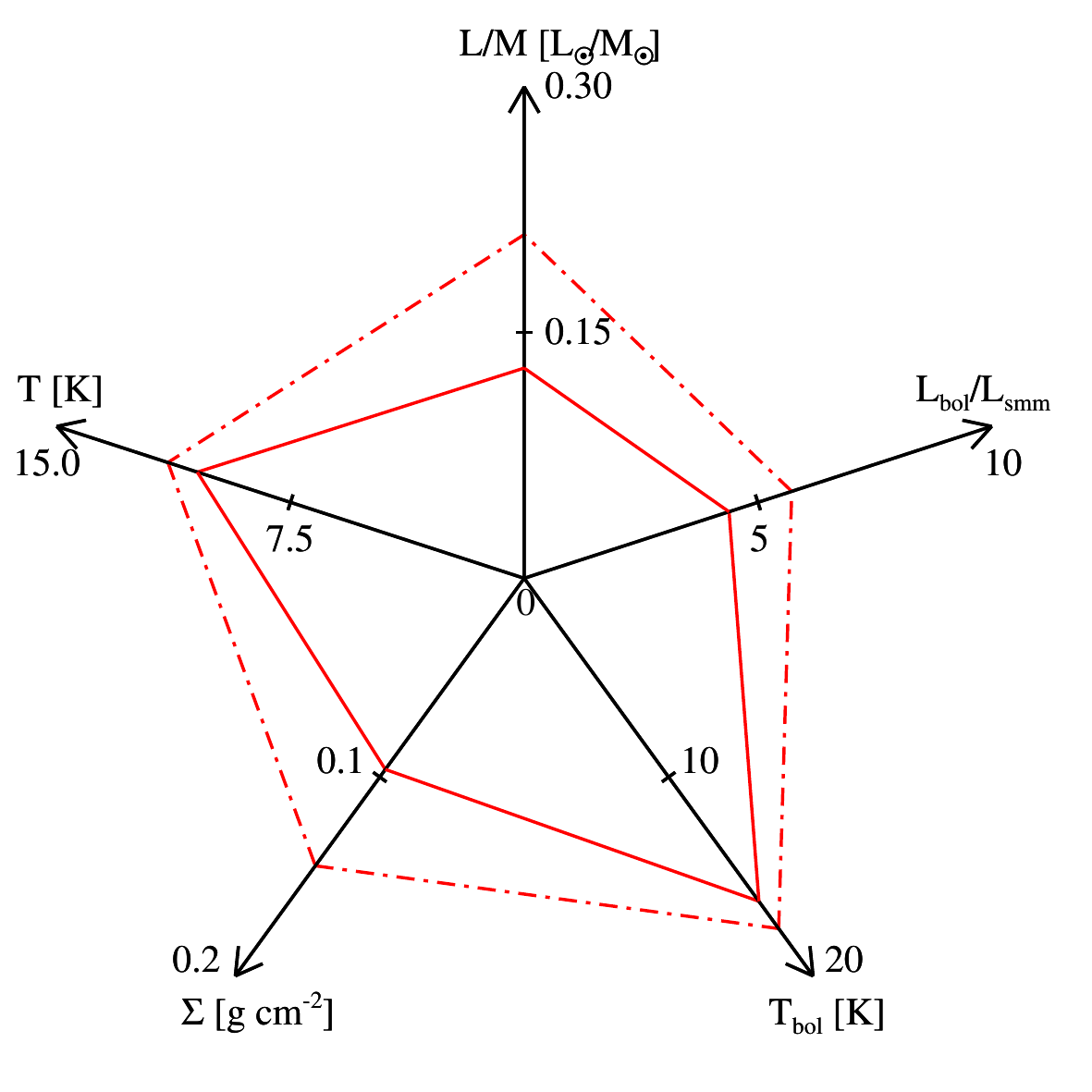}
\caption{Radar plot for the median values of five distance-independent physical parameters (see
text) for pre-stellar clumps, for the inner Galaxy (dot-dashed line) and the
outer (solid line). Scales on each axis are linear, ranging from 0 to the value 
specified at the end.}
\label{radarpre}
\end{figure}

To give a synoptic view of median values of the distance-independent observables discussed in previous
sections, we adopt the radar chart visualisation of $L_\mathrm{bol}/M$, $T$, $\Sigma$, $T_\mathrm{bol}$, and 
$L_\mathrm{bol}/L_\mathrm{smm}$ as introduced in \citetalias{eli17}. Five axes 
represent these parameters on the linear scales indicated. The median values are plotted on each axis and connected by lines between adjacent axes to form a polygon.
Medians taken for different source sub-samples, whether selected by classification and/or distances, are displayed in the same plot for the purpose of comparison.

\begin{figure}
\centering
\includegraphics[width=8.5cm]{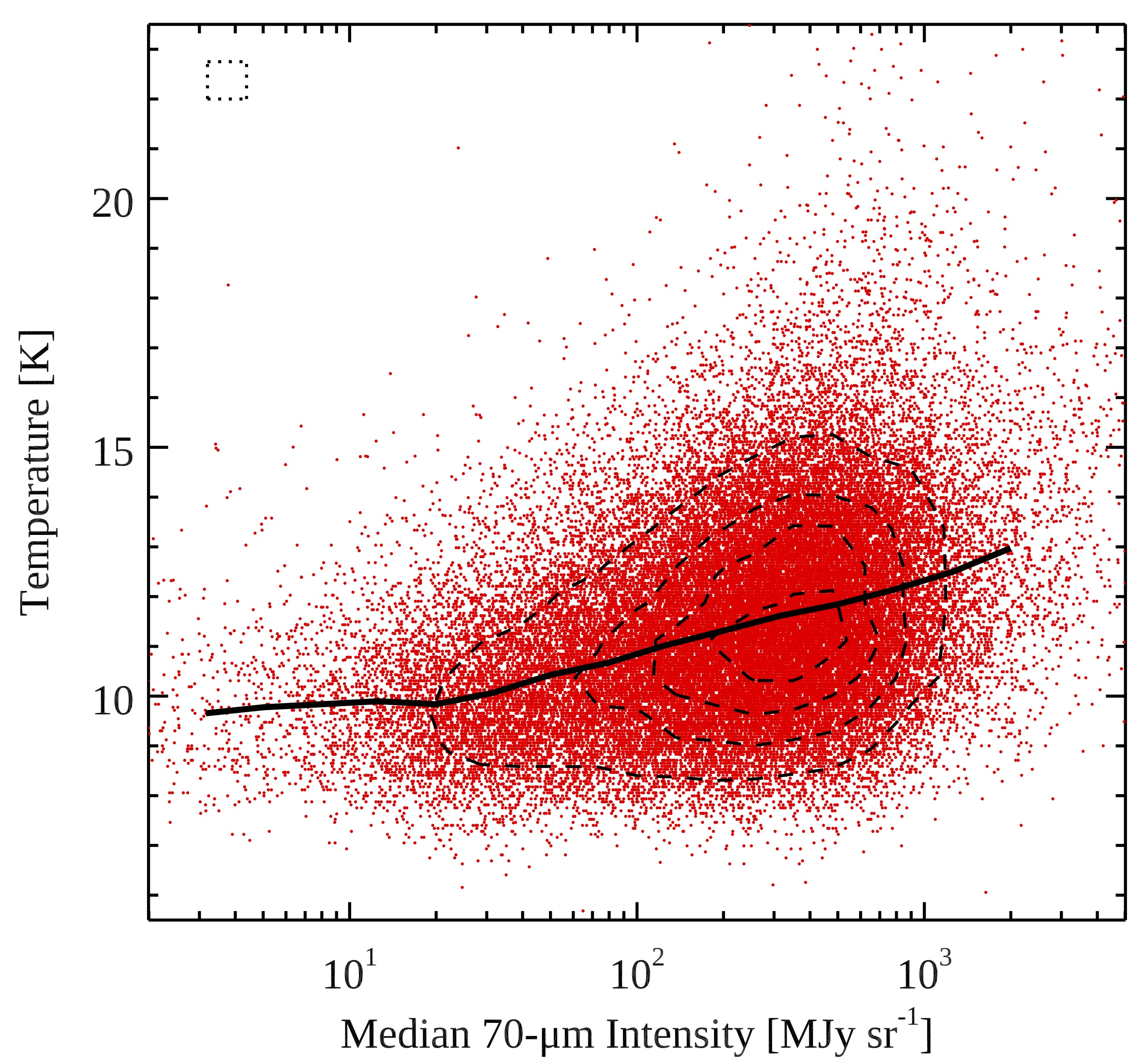}
\caption{Temperature $T$ of pre-stellar sources vs median intensity $\bar{I}_{70}$
evaluated in $61\times 61$ pixel boxes centered on clump 
centroids in PACS 70~$\umu$m maps. For clarity, the density of points in the plot, evaluated on a grid whose element has the size of the dotted box placed at the top left corner of the plot, is represented with dashed contours starting from 200 and in steps of 200.} Medians of $T$ in bins of $\log\bar{I}_{70}$ (bin width = 0.2) are connected by the solid line.
\label{radiation70}
\end{figure}

The first such plot is Fig.~\ref{radarpre} representing median values for pre-stellar sources,
for inner and outer Galaxy separately. All medians, 
and in particular the four evolutionary indicators $L_\mathrm{bol}/M$, $T$, $T_\mathrm{bol}$, and 
$L_\mathrm{bol}/L_\mathrm{smm}$, are larger in the inner Galaxy,
as already seen in Sections~\ref{lmr_par}-\ref{surfdenspar}.
A hasty interpretation might be that pre-stellar clumps in the inner Galaxy are ``more evolved'' on average. 
However, it is more plausible to ascribe this to 
different environmental conditions in the inner and outer Galaxy, because external irradiation 
\citep[][and references therein]{mez90} is the main source of heating for pre-stellar 
clumps \citep[e.g.,][]{eva01,lip16,yua17,mer19}. 

To assess the effects of external heating,
it is sufficient 
to analyse the pre-stellar clump temperature as a function of the Galactic position,
because for a MBB (as pre-stellar 
SEDs are expected to follow) the $L_\mathrm{bol}/M$, $L_\mathrm{bol}/L_\mathrm{smm}$, 
and $T_\mathrm{bol}$ quantities are expected to increase monotonically with increasing 
temperature, following precise analytic relations \citep{eli16}.
As a proxy for the interstellar 
radiation field \citep{com10,ber10}, we used the intensity of 
$70~\umu$m emission in the neighbourhoods of clumps: for each clump we found the average, $\bar{I}_{70}$, of the PACS $70~\umu$m 
intensity over a $61 \times 61$ pixel ($\sim 3.25\times 3.25$~arcmin$^2$)
sub-frame centered on the source centroid
(or smaller, if limited by proximity to tile border). 
We first explored the relationship of $\bar{I}_{70}$ to
source temperature (Fig.~\ref{radiation70}). Due to the relatively narrow temperature 
distribution of pre-stellar clumps seen in Fig.~\ref{hist_temp}, we do not expect 
large variations of the temperature as a function of 
$\bar{I}_{70}$. Furthermore, we also do not 
expect a straightforward relation between these two observables in all cases, because
environmental conditions can change locally. However, considering the entire 
sample in Fig.~\ref{radiation70} the median temperature in bins of 
$\bar{I}_{70}$ is seen to increase at increasing $\bar{I}_{70}$.

\begin{figure*}
\centering
\includegraphics[width=18cm]{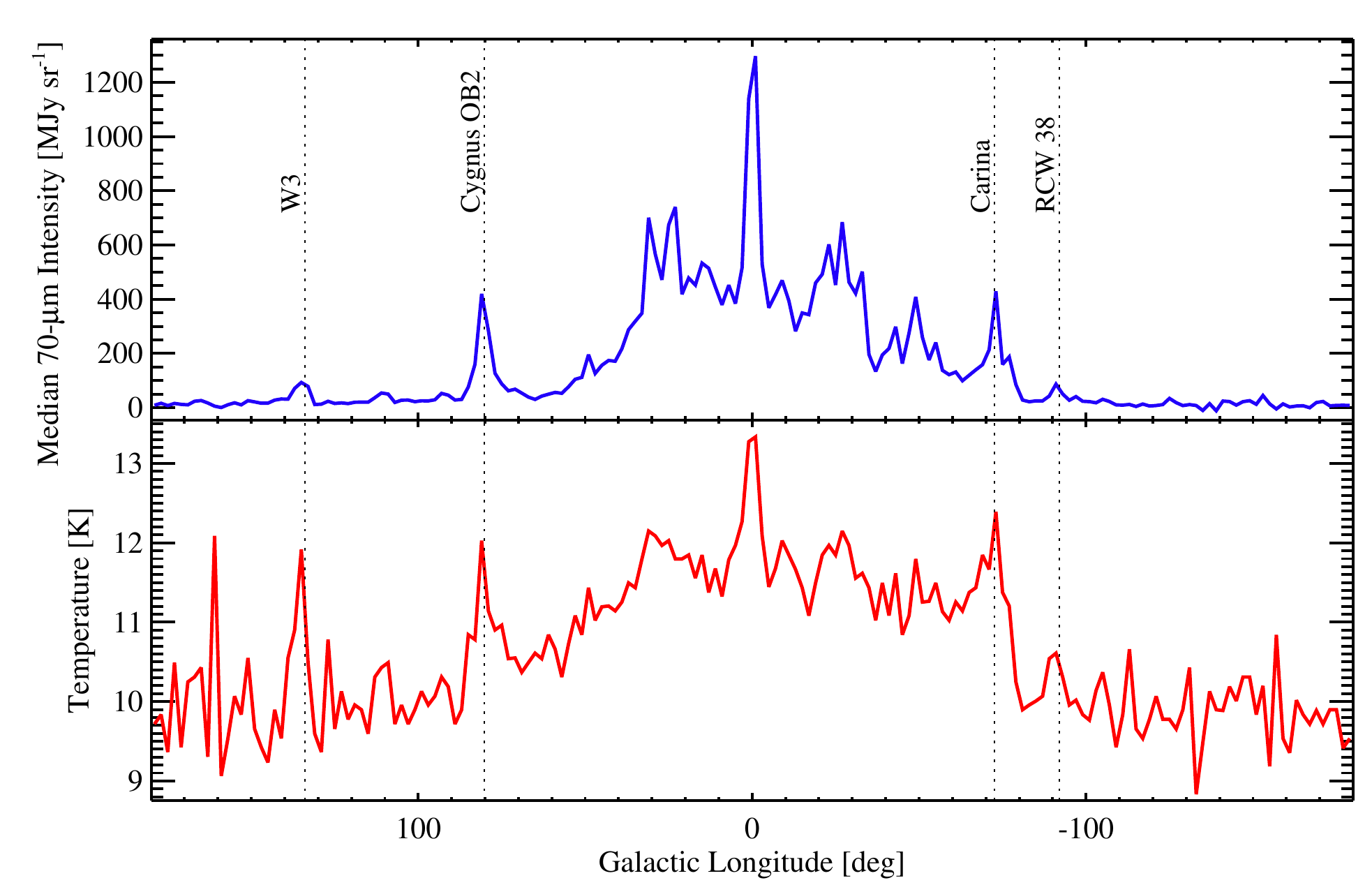}
\caption{Medians, in $2^\circ$-bins of Galactic longitude, of two quantities characterizing pre-stellar sources. Upper panel: median 
intensity $\bar{I}_{70}$ of the $70~\umu$m map intensity around pre-stellar sources 
(see Fig.~\ref{radiation70}). Lower panel: 
source temperature. Dashed vertical lines indicating the 
average longitudes of the labelled star-forming clouds in the outer Galaxy are coincident with local peaks in the curves in both panels.
}
\label{glonradiation70}
\end{figure*}

Fig.~\ref{glonradiation70} shows the behaviour of $T$ and $\bar{I}_{70}$ individually
as a function of Galactic longitude. In the central quadrants $\bar{I}_{70}$ is far more intense and the median $T$
of pre-stellar sources increases correspondingly. Furthermore, main
local peaks of the two quantities spatially coincide (cases of average longitudes 
of Cygnus OB2, W3, RCW~38, and Carina star-forming clouds). 
These correlations can not be simply casual, even if the variability range of median $T$ in the bottom panel of Fig.~\ref{glonradiation70} might appear relatively narrow. It should be considered, indeed, that the response of dust temperature to the ultraviolet field is expected by \citet{ber10} to follow a power law with an exponent as shallow as $1/(4+\beta)$.

Because a fraction
of outer Galaxy sources can be found at relatively low $|\ell|$, 
the observed decrease of pre-stellar source temperature has to be confirmed, more rigorously,
as a function of Galactocentric distance. This will be shown in Section~\ref{galactotrends}.
We can conclude that the 
general increase of temperature and other evolutionary parameters for pre-stellar 
clumps in the inner Galaxy is related mostly to the amount of irradiation to which they are exposed.

\begin{figure}
\centering
\includegraphics[width=8.5cm]{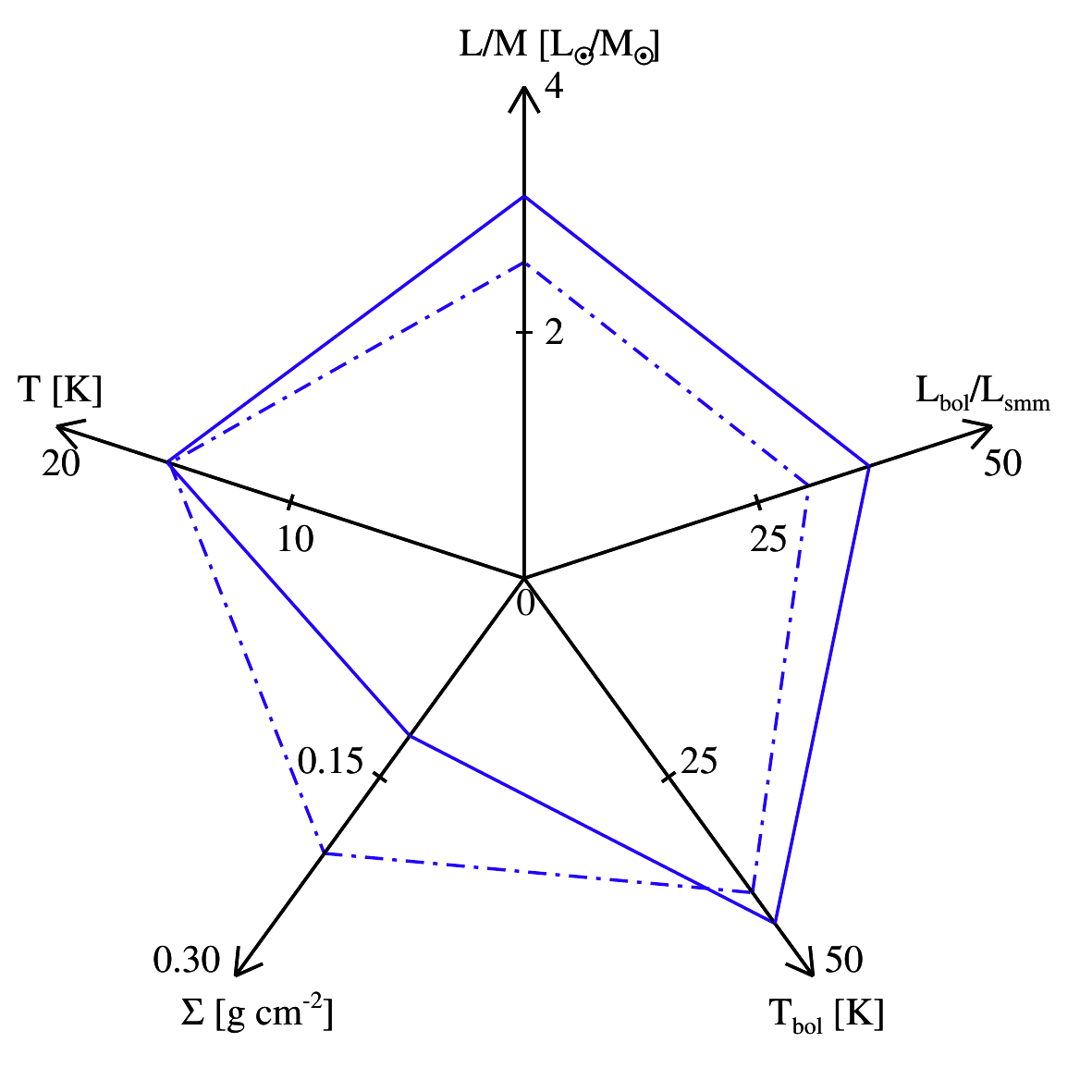}
\caption{Same as Fig.~\ref{radarpre}, but for protostellar
sources.}
\label{radarproto}
\end{figure}

\begin{figure}
\centering
\includegraphics[width=8.5cm]{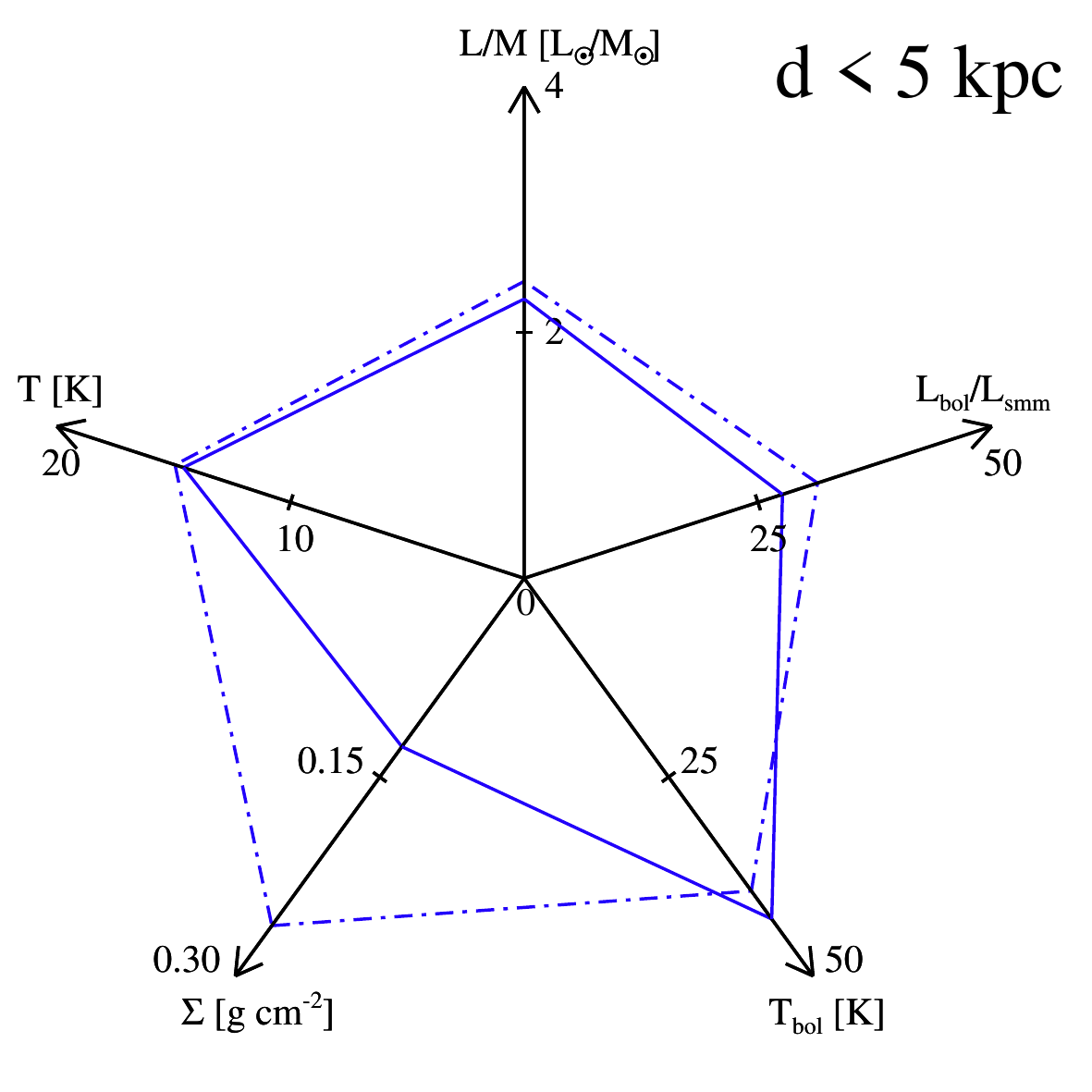}
\caption{Same as Fig.~\ref{radarproto}, but for the sub-set of protostellar
sources within heliocentric distance $d<5$~kpc.}
\label{radarprotolt5kpc}
\end{figure}

An opposite trend is shown by statistics of protostellar sources in Fig.~\ref{radarproto}: the medians of $T$ have become about equal and the medians of the other three evolutionary indicators are now higher in the outer Galaxy.
For this class, because the main source of clump heating is internal, the influence of the 
environment is less relevant, and one could attribute a genuine evolutionary 
meaning to this behaviour. However, it is necessary to take into account possible 
basic systematic differences between the samples of protostellar clumps from the inner 
and outer Galaxy. The biggest factor is probably represented 
by the different distribution of heliocentric distances, highlighted by Fig.~\ref{hist_dist},
because it produces a globally different distribution of physical sizes. The inner Galaxy
contains a large number of sources located at $d>5$~kpc, which are unresolved
structures of increasing complexity containing protostellar cores but also quiescent
cores and inter-core medium \citepalias[see][their Appendix~C]{eli17}. This can affect some evolutionary indicators. 
To check this, in Fig.~\ref{radarprotolt5kpc} we show a radar plot similar to that of Fig.~\ref{radarproto},
but limited to distances $d<5$~kpc. 
The largest effect is reductions of the medians of $L_\mathrm{bol}/M$ and 
$L_\mathrm{bol}/L_\mathrm{smm}$, making them along with the two median temperatures indistinguishable in the inner and outer Galaxy.
We therefore conclude that
the distance bias can significantly affect some source evolutionary 
indicators and consequently the classification too.

Like for the pre-stellar sources, the median $\Sigma$ for protostellar sources is higher in inner Galaxy.
Limiting 
to the $d<5$~kpc sub-samples, the medians increases in both the inner and 
outer Galaxy, but more so for the inner so that their discrepancy increases. This can
be explained with the intrinsically different regimes of density in the two zones, as already
highlighted in Section~\ref{surfdenspar} and further discussed, in terms of Galactocentric 
distance in Section~\ref{galactotrends}.

\section{Trends with the Galactocentric radius}\label{galactotrends}

\begin{figure*}
\centering
\includegraphics[width=18cm]{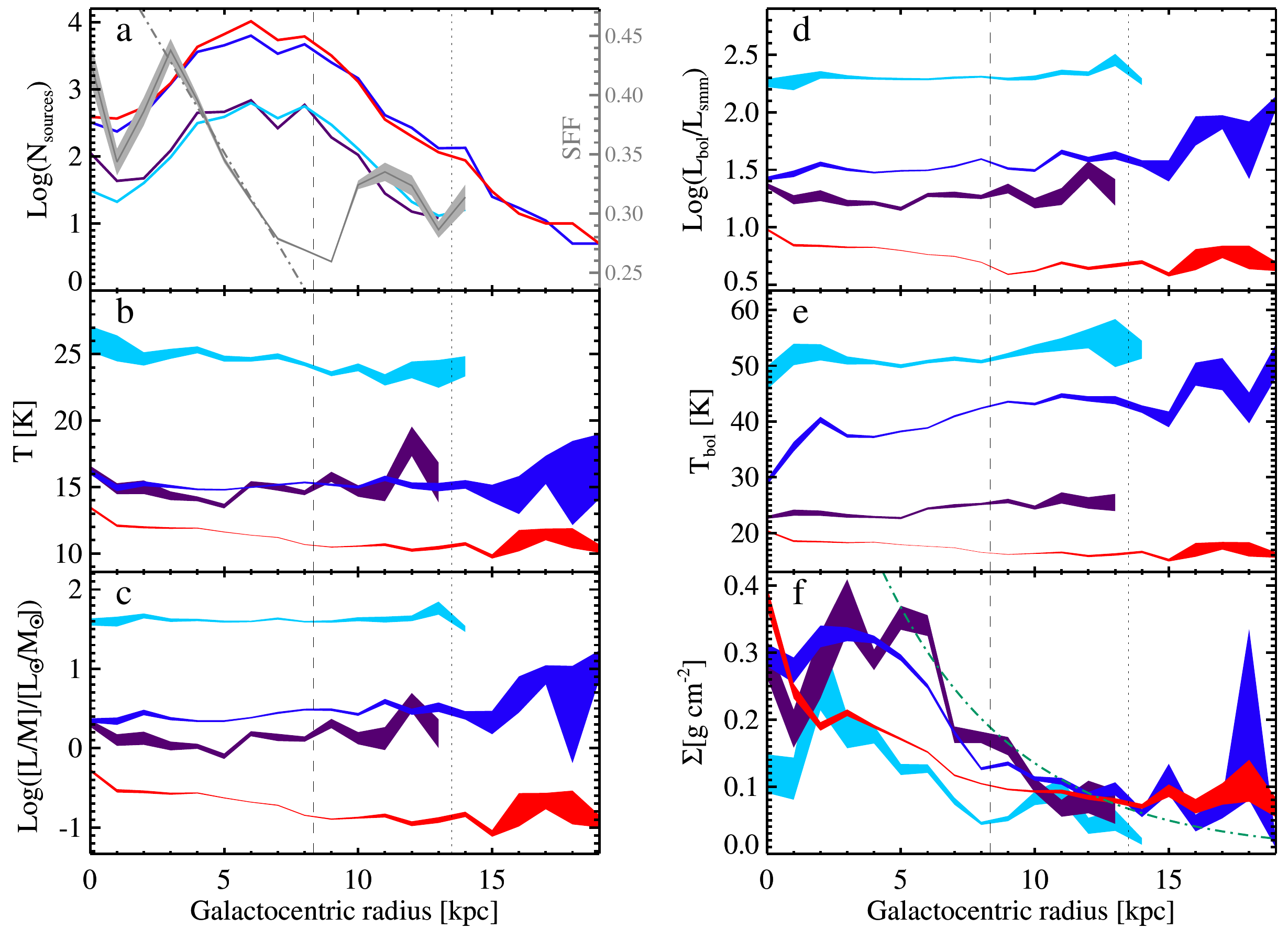}
\caption{Characterization of sources vs Galactocentric radius in bins of 1~kpc for various populations: pre-stellar sources (red), all protostellar sources (blue), H\,\textsc{ii}-region candidates (light 
blue), and MIR-dark protostellar clumps (dark purple). The vertical dashed 
and dotted lines mark the values adopted for the Solar circle and inner radius of the far outer Galaxy, respectively.
Panel~$a$: number of sources. The star-forming 
fraction (SFF, see the text) is also plotted in grey, referring to the grey $y$-axis on the
right. The grey dot-dashed line represents a fit to the SFF data
in the range $3~\mathrm{kpc} \leq R_\mathrm{GC} \leq 7~\mathrm{kpc}$. 
The remaining panels $b$ through $f$ are for $T$, $L_\mathrm{bol}/M$, $L_\mathrm{bol}/L_\mathrm{smm}$, 
$T_{\textrm{bol}}$, and $\Sigma$, as labelled.
In each bin the median and its uncertainty, calculated from interquartile ranges as explained in Table~\ref{evoltable}, are calculated and then the uncertainty intervals are displayed as shaded areas.
Values are shown only for bins with at least 10 sources. 
In panel $f$, the green dot-dashed curve represents the radial dependence of the inverse of the gas-to-dust ratio 
predicted by \citet{gia17}; the vertical scaling
of the curve is arbitrary.}
\label{plotgalactotrends}
\end{figure*}

In this section we move on from the inner-outer Galaxy dichotomy introduced in Section~\ref{innerouter}
to discuss all distance-independent parameters as a function
of $R_\mathrm{GC}$, similarly to \citetalias{eli17} 
(their Section~8.2), but over a wider range of $R_\mathrm{GC}$ and also including pre-stellar clumps.
Contrary to Section~\ref{nodistparam}, the analysis presented here is based only on the statistics of sources
with a known distance, as required to compute $R_\mathrm{GC}$.
Fig.~\ref{plotgalactotrends} shows the average behaviour of various parameters for different
evolutionary classes. 

\subsection{Number counts and star formation fraction}\label{numsff}
The source number in bins of Galactocentric radius is 
shown in panel~$a$ to give an idea of the statistical relevance of curves shown in the 
subsequent panels. Unlike the bottom panel of Fig.~\ref{hist_dist},
this does not separate between inner and outer Galaxy populations.

It also reports the curves corresponding to the two sub-classes of 
MIR-dark and candidate H\,\textsc{ii} regions, for which here we make some further considerations 
in addition to those in Section~\ref{heliocentric}. The 
fraction of these two sub-classes with respect 
to the whole protostellar population is quite constant. The only exception is 
an excess of MIR-dark sources in the $R_\mathrm{GC}=0$~kpc bin, a bias produced by 
severe saturation affecting both the MIPSGAL and WISE-W4 band observations towards the 
Galactic centre. No MIR-dark sources are found at $R_\mathrm{GC}>14$~kpc and there are few of either major class in the far outer Galaxy (FOG).
A local peak for H\,\textsc{ii} region candidates is found around 6~kpc, corresponding approximately to
the enhancement of H\,\textsc{ii} regions found by \citet{and09}; however, this peak also coincides
with local peaks of our pre-stellar and overall protostellar populations, with no particular
excess there compared with other $R_\mathrm{GC}$. Therefore the local peak of H\,\textsc{ii} 
region candidates seems to be related to a global increased 
availability of molecular material near 6~kpc \citep{dob12} rather than to other specific conditions
of the ISM. 

Given the separate distributions of pre-stellar and protostellar sources with $R_\mathrm{GC}$,
it is straightforward to investigate their fraction of the total. In particular, \citet{rag16}
discussed the ``star-forming fraction'' (hereafter SFF), namely the fraction of sources in
the Hi-GAL catalogue of \citetalias{eli17} with a detection at 70~$\umu$m (i.e., 
the definition of protostellar source adopted here\footnote{Possible sources of misclassification between pre-stellar and protostellar clumps, particularly related to heliocentric distance and possibly affecting the derived SFF, are summarised in Section~\ref{building}.}).
\citet{rag16} found a slightly decreasing behaviour of SFF over
$3.1~\mathrm{kpc}< R_\mathrm{GC}< 8.6~\mathrm{kpc}$, with a linear fit slope of 
$(-0.026 \pm 0.002)$~kpc$^{-1}$. That analysis was carried out including all Hi-GAL 
sources with a known distance (including starless unbound).
To enable comparison with \citet{rag16}, we have to consider sources in
both of our catalogues, including unbound clumps. 

The SFF curve obtained from our data, considering only 1-kpc bins containing at least 100 sources in 
total, is shown in panel $a$ of Fig.~\ref{plotgalactotrends}, 
referring to the $y$-axis on the right side. The behaviour of SFF is quite
scattered, though confined to a relatively narrow range from \sffmin~to \sffmax. A
decreasing trend is confirmed in the range $3~\mathrm{kpc} \leq R_\mathrm{GC} \leq 7~\mathrm{kpc}$,
over which we derive a linear
fit slope of $(\sffslope \pm \sffsloperr)~\mathrm{kpc}^{-1}$, similar to that of
\citet{rag16}.\footnote{The binning in \citet{rag16} is finer than we use and here the behaviour of SFF at 
$R_\mathrm{GC} \gtrsim 8$~kpc is not decreasing.}
However, embedded in a larger range of $R_\mathrm{GC}$, here such behaviour
does not seem global anymore. Indeed, isolated increases of the fraction of protostellar 
sources are seen, due to local conditions, e.g., the Galactic centre position, around
$R_\mathrm{GC}=3$~kpc \citep[cf.][]{lun06} and, in the outer Galaxy,
12 and 14~kpc, neither related to spiral arms (see 
Section~\ref{heliocentric}) and possibly affected by relatively poor statistics. These fluctuations constitute a
departure from a simple scenario in which the SFF, in turn related to star formation efficiency, 
decreases systematically from the centre to the periphery of Milky Way, which was 
considered already by \citet{rag16} to be not easily explicable. 

\subsection{Evolutionary indicators}\label{evolindic}
Looking at the behaviour of the evolutionary indicators $T$, $L_\mathrm{bol}/M$, $L_\mathrm{bol}/L_\mathrm{smm}$, and
$T_\mathrm{bol}$ in panels $b$ to $e$, respectively, first of all
we notice that for each parameter the degree of segregation among classes/sub-classes and their ranking is preserved across the range of $R_\mathrm{GC}$ explored.
For example, as seen in Section~\ref{nodistparam}, the average temperatures of MIR-dark 
sources are not distinguishable from those of the global protostellar population, while 
other indicators, especially $L_\mathrm{bol}/L_\mathrm{smm}$ and $T_\mathrm{bol}$, show 
a better separation.
Moreover, for all these parameters, we notice common global trends for the different source
classes. 

The parameters for pre-stellar sources show a shallow decrease with $R_\mathrm{GC}$,
as expected from the discussion in Section~\ref{radarsec} about the relationship
with the interstellar radiation field, which drops at increasing $R_\mathrm{GC}$ \citep{mat83}.
A less clear trend is seen beyond the boundary we assume for the FOG.
Around $R_\mathrm{GC}=14$~kpc, a local growth 
is seen in all parameters, corresponding to a local 
increase of the source number (panel $a$). At these Galactocentric radii, the source statistics 
become relatively poor, and contributions from single regions can dominate the estimate of 
median indicators. In this case, the main contributions to higher median temperature come 
from two groups of a few tens of sources: one located in the innermost part of the plane
($\ell\sim 357^\circ$) and with large heliocentric distances ($d>15-20$~kpc) probably
affected by large uncertainties, and another one corresponding to the neighbourhood of the 
Sh2-284 H\,\textsc{ii} nebula, namely $\ell\sim 212^\circ$, $V_\mathrm{LSR}\sim 45$~km~s$^{-1}$
\citep[cf.][]{bli82}, $d\sim 6.6$~kpc \citep[cf.][]{mof79}. Finally, another smoother peak 
is found at $R_\mathrm{GC}>15$~kpc, essentially due to sources located around the Galactic 
anticentre, again a region characterized by high uncertainties on distance estimates.

For protostellar sources, an almost constant $T$ is seen, 
compatible with the trend found by \citet{rig19} for the excitation temperature of 
clumps from the CHIMPS survey \citep{rig16} in the Galactic longitude range $27\pdeg8 \lesssim \ell \lesssim 46\pdeg2$. The sub-sample of MIR-dark sources follows approximately the same 
behaviour, but with more scatter due to the lower number of sources. Actually, one
can glimpse a decreasing trend for MIR-dark sources in the range $R_\mathrm{GC} \leq 5$~kpc, which is supported
in the two subsequent panels. The sub-class of
H\,\textsc{ii} region candidates seems to show a slight decreasing trend 
overall.

\citet{urq18} found a slightly increasing trend of the temperature of their $\sim8000$ ATLASGAL 
clumps at increasing $R_\mathrm{GC}$, up to 10~kpc 
(bins at larger radii are not statistically meaningful). 
The majority of their 
sources are located between 3 and 7~kpc, where the average temperature is quite constant 
($\sim 20$~K). 
About 88~per cent are associated with star formation activity, and so a comparison 
with our protostellar class should be made.
However, a direct comparison with our data in absolute terms is difficult 
to perform: following \citet{kon17}, in \citet{urq18} temperatures are estimated
including the Hi-GAL flux at $70~\umu$m, which leads to 
higher temperatures, on average, compared to ours.

The $L_\mathrm{bol}/M$, $L_\mathrm{bol}/L_\mathrm{smm}$, and 
$T_\mathrm{bol}$ parameters for pre-stellar sources show a globally decreasing trend with 
$R_\mathrm{GC}$, as discussed for $T$. This is to be expected because for these sources the SEDs are modeled as simple 
MBBs, for which all of these quantities are correlated analytically \citep{eli16}
and show the same kind of monotonic behaviour.

The median behaviour of $L_\mathrm{bol}/M$ for protostellar sources seems, instead, globally
constant, at least with respect to the expanded $y$-axis range chosen to accommodate the
curves of all evolutionary classes. In \citetalias{eli17} the curve appeared to be increasing 
in $6~\mathrm{kpc} \leq R_\mathrm{GC} \leq 9~\mathrm{kpc}$, going from 
$\log[(L_\mathrm{bol}/M)/(L_\odot/M_\odot)]=0.2$ to 0.6, which is confirmed here. However, now such an increase can be considered a weak 
fluctuation around the practically constant value exhibited over a significantly larger range of 
$R_\mathrm{GC}$. Between
2 and 9~kpc \citet{urq18} found a fairly constant value $10~L_\odot/M_\odot$, which is 
higher than our average value (see Section~\ref{lmr_par}) and understood for the same systematic reasons discussed for $T$. Because this observable is considered a reliable proxy
for star formation efficiency \citep[e.g.,][]{ede15}, this suggests that the SFE, at least as 
traced through Hi-GAL and averaged in rings of $R_\mathrm{GC}$, appears not to show radial substantial 
variations across the Milky Way up to at least $R_\mathrm{GC}=15$~kpc.

For H\,\textsc{ii} region candidates, $L_\mathrm{bol}/M$ is also almost constant. Interestingly, \citet{djo19}, considering 445 ATLASGAL clumps hosting bona-fide 
compact and ultra-compact H\,\textsc{ii} regions, highlight a drop of $L_\mathrm{bol}/M$ from 
$R_\mathrm{GC}\simeq 2$~kpc to $\simeq 14$~kpc. However, these authors warn that the masses 
could turn out to be systematically overestimated at large $R_\mathrm{GC}$ as a result of 
their choice of adopting a constant temperature of 27~K instead of the MBB fit-derived
temperatures of \citet{urq18}. 

Medians of the remaining evolutionary indicators, namely
$L_\mathrm{bol}/L_\mathrm{smm}$ and $T_\mathrm{bol}$ (panels $d$ and $e$, respectively),
show a slight trend with $R_\mathrm{GC}$ that is decreasing for pre-stellar sources and 
increasing for protostellar sources.
The trend for protostellar sources confirms that highlighted in \citetalias{eli17} 
in the range 4~kpc~$<R_\mathrm{GC}<10$~kpc. This is not found for 
the sub-class of H\,\textsc{ii} region candidates, for which these indicators are
substantially flat. Finally, for the sub-class of MIR-dark sources, $L_\mathrm{bol}/L_\mathrm{smm}$
behaves qualitatively like $L_\mathrm{bol}/M$, while for
$T_\mathrm{bol}$ the trend is roughly constant up to $R_\mathrm{GC}=13$~kpc.

To summarise the Galactocentric behaviour of median evolutionary indicators plotted in 
panels $b$ to $e$ of Fig.~\ref{plotgalactotrends}, we note by class: ($i$) for pre-stellar 
sources, a systematic decrease, as a possible consequence of lower intensity in the interstellar
radiation field; ($ii$) for all protostellar sources, a flat behaviour for temperature and 
a slightly increasing one for remaining indicators; ($iii$) for MIR-dark protostellar
sources, a mildly decreasing behaviour up to 5~kpc (with the exception of $T_\mathrm{bol}$, which is substantially flat), followed by an increasing one at larger
radii, but more scattered; and $(iv)$ for H\,\textsc{ii} region candidates, a flat behaviour
for all indicators but $T$, which is slightly decreasing. Based on these
considerations, we cannot affirm that there are clear evolutionary trends from the centre
to the periphery of Milky Way, or evident signatures relating to spiral arms.
This confirms the result of \citet{urq18}, which was based on a smaller coverage of the Galaxy, and of \citetalias{eli17}, based on a specific source-to-arm association.

We conclude this analysis of evolutionary indicators with a note about the region of the 
panels from $b$ to $e$ of Fig.~\ref{plotgalactotrends} corresponding to the FOG. Although
statistics become less certain at these values of $R_\mathrm{GC}$, we notice some increase:
indicators of pre-stellar sources stop decreasing in the FOG,
and the flat or slightly increasing behaviour seen for protostellar sources gets
steeper. It would be suggestive to link these increases to the higher star formation 
efficiency suggested by \citet{bra01b} as a result of the predominance of gravity over turbulence
in this area of the Galaxy.
Verifying this hypothesis would require detailed study of single sources or regions, which lies 
outside the aims of this work and will be addressed in a future paper. 

\subsection{Surface density}\label{surdensity}

Median surface density in panel~$f$ of Fig.~\ref{plotgalactotrends} has a
a completely different behaviour with respect to the clump properties in previous panels, for all source classes and sub-classes. This confirms that surface density cannot be used straightforwardly as
an evolutionary indicator for clumps. As already suggested in Sections~\ref{surfdenspar} 
and~\ref{radarsec} in
the inner vs outer Galaxy comparison (following the longitude-based definition for the two 
zones), a strong difference between median densities within and outside the Solar circle is found,
for all classes. The curve corresponding to pre-stellar sources appears to be the most regular,
with a peak at the centre of the Galaxy and a monotonic decrease with
three different overall slopes, changing at $R_\mathrm{GC}\simeq 2$~kpc and $\simeq 7$~kpc.
The protostellar source class and its two sub-classes share a qualitatively similar
behaviour among them (although over different ranges of density): a bump between 
$R_\mathrm{GC}\simeq 1$~kpc and about $5-7$~kpc (roughly corresponding to
the ``molecular ring'') and a shallower decrease at larger radii,
with a degree of scatter higher than for the pre-stellar class.

A decrease of clump surface density with Galactocentric radius was highlighted already 
by \citet{zet18}, but based on a smaller source numbers on a restricted range of longitudes ($10^\circ < \ell < 56^\circ$)
and radii ($3.5~\mathrm{kpc} < R_\mathrm{GC} < 7$~kpc), 
and with no distinction between the starless or protostellar
nature of the clumps. The decreasing trend can be understood in terms of the local availability of matter in 
hosting molecular clouds, which generally decreases towards the periphery of the Galaxy.
\citet{rom10} show a peak of molecular cloud surface density at $R_\mathrm{GC}\simeq 3$~kpc, 
followed by a drop that they trace out to $R_\mathrm{GC}\simeq 8$~kpc. The curves of H$_2$ volume 
density by \citet{bro88} and \citet{nak06}, collected and shown together by \citet{hey15}, 
have a similar behaviour, with a peak at $R_\mathrm{GC}\simeq 4.5$~kpc and 5~kpc, respectively.
Interestingly, some features present in the curves of panel $f$ can be recognised qualitatively
in the theoretical curve of H$_2$ volume density produced by \citet{sof16}, their Fig.~6, namely
a peak close to the Galactic centre, a bump with secondary peaks for
$3~\mathrm{kpc} \lesssim R_\mathrm{GC} \lesssim 8$~kpc, and another smaller bump
around $R_\mathrm{GC}\simeq 11$~kpc. 

Another effect, which can overlap with the previous ones, may be related to the 
variation of the local value of the gas-to-dust ratio. This ratio is observed 
to increase at increasing Galactocentric radius \citep{gia17}, while in this analysis it has been
kept at a constant value of 100. Ignoring such a change might lead to a systematic overestimation of densities 
and masses at small $R_\mathrm{GC}$, and underestimation in the outer Galaxy. 
The radial dependence of such a change is shown in Fig.~\ref{plotgalactotrends}, panel~$f$, proportional to the inverse of the gas-to-dust 
ratio as a function of $R_\mathrm{GC}$ as predicted from \citet[their Eq.~2]{gia17}.
Compared with the behaviour of the median $\Sigma$ of both pre-stellar and protostellar clumps, the prediction appears 
generally steeper. In more quantitative terms, for pre-stellar sources the median of $\Sigma$ 
between 3 and 10~kpc is observed to decrease by a 
factor~\sjumpe, against a predicted factor of 
$\gjumpe \pm \egjumpe$. For protostellar sources, there is a decrease by a 
factor~\sjumpo~between 4 and 12~kpc, to be compared to a predicted factor of 
$\gjumpo \pm \egjumpo$. 
Nevertheless, the decrease
in both trends at increasing $R_\mathrm{GC}$ is interestingly of the same order of magnitude, even if not identical. This
suggests that variation of the gas-to-dust ratio is probably playing a role in what is being observed and interpreted, although weaker than expected. A full accounting 
would also require understanding the variation of dust opacity with position in the Galaxy, which is beyond the scope of this work. 

\section{Summary}\label{summary}

In this paper we presented the 360$^\circ$-catalogue of physical properties of more than 
$1.5 \times 10^5$ Hi-GAL compact sources. We divided the sources into ``inner'' and ``outer'' Galaxy sets, according to source position
inside or outside the Solar circle, respectively. 
As in \citetalias{eli17}, we emphasized which information can
be obtained by using photometric data alone (from both Hi-GAL and ancillary surveys), reserving spectroscopic
data exclusively to derive heliocentric distances. In this respect, we achieved three main goals:
\begin{itemize}
\item To deliver, for the first time, an unbiased catalogue of clumps in the outer Galaxy,
at the quality level enabled by \textit{Herschel}, in terms of both resolution and sensitivity, completing a homogeneous catalogue of the entire Galactic plane.
\item To refine the description and analysis of the inner Galaxy clump population provided in \citetalias{eli17}, 
by applying a new set of heliocentric distances.
\item To combine all of this information to discuss similarities and differences between 
inner and outer Galaxy populations, and also to show possible trends of median clump properties as 
a function of distance from the Galactic centre.
\end{itemize}

The availability of this large and rich database on
clumps in the Galactic disc, 
hosting star formation or possible progenitors of it, allowed us to carry out a systematic
statistical analysis from which we can draw several main conclusions:
\begin{enumerate}
\item The majority (\perclumps~per 
cent) of sources with a known distance estimate correspond to the definition of clump,
and a significant fraction of them fulfils different thresholds for compatibility with massive
star formation based on the mass-radius relation, including in the outer Galaxy.
\item The mass versus bolometric luminosity diagram confirms segregation between pre-stellar and protostellar
clumps. For both these quantities, intrinsically lower values are encountered, on average, in the 
outer Galaxy compared to the inner. Correspondingly, the clump surface density drops
considerably from the inner to outer Galaxy, which can be understood in terms of a greater availability of matter in molecular clouds in the inner Galaxy, and/or a systematic 
bias introduced by adopting a gas-to-dust ratio that is constant instead of increasing with
Galactocentric radius.
\item For distance-independent quantities -- such as the MBB temperature, the ratio of luminosity 
to mass, the ratio of luminosity to sub-millimetre luminosity, and the bolometric temperature -- 
the distributions for pre-stellar and protostellar populations appear different, confirming the utility of
these quantities in delineating an evolutionary picture for the sources.
\item In particular, the aforementioned distributions appear better separated in the outer Galaxy
than in the inner Galaxy, with lower average values for pre-stellar clumps and higher for protostellar 
clumps. The effect seen for pre-stellar sources can be explained by the stronger interstellar radiation
field in the inner Galaxy, whereas the effect seen for protostellar sources can be reconciled
in terms of resolution/distance bias effects (blending of distant protostellar sources with 
starless sources and quiescent inter-clump emission).
\item As in \citetalias{eli17} we identified and discussed two sub-classes of 
protostellar sources. The sources lacking a detection at MIR wavelengths do not appear
as a well-confined group with respect to evolutionary parameters, 
except for bolometric temperature. On the other hand, sources with a high ratio of luminosity to mass
that are identified as H\,\textsc{ii}-region candidates generally show high values in all evolutionary
indicators, both in the inner and outer Galaxy.
\item Our statistics of source properties as a function of the Galactocentric radius are meaningful 
within at least the first 15~kpc. We notice a clearly 
decreasing trend in source number outside the Solar circle. On the other hand, the star formation fraction, defined as the ratio
of the number of protostellar clumps to the total number of clumps, does not decrease monotonically with increasing 
$R_\mathrm{GC}$ as suggested by previous literature, but instead shows a dip around $R_\mathrm{GC}=7$~kpc
and a subsequent increase beyond the Solar circle.
\item Looking at the Galactocentric distributions of the four evolutionary indicators 
$L_\mathrm{bol}/M$, $T$, $\Sigma$, and $T_\mathrm{bol}$,
we find that the degree of segregation among different classes remains roughly constant across 
the Galaxy, regardless of possible spiral arm positions. In all cases indicators for pre-stellar sources show 
a decrease at increasing $R_\mathrm{GC}$, explained again with a lower interstellar radiation field, 
while the indicators for protostellar sources remain flat or slightly increasing.
Indicators for MIR-dark sources show a behaviour similar to that of the overall protostellar class, 
while those for H\,\textsc{ii} region candidates remain substantially flat, except for the temperature,
which decreases slightly at increasing $R_\mathrm{GC}$. In summary, we do not find striking 
differences in median evolutionary stage across different Galactocentric radii, and/or in 
correspondence with spiral arms, whose role seems not to be crucial for triggering star formation, 
but rather for gathering matter.
\item A few hundred sources have been identified in the far outer Galaxy ($R_\mathrm{GC}>13$~kpc), 
providing a solid 
base for future studies of star formation in the outskirts of the Galactic disc. From these 
sources we notice a slight rise of median values for evolutionary indicators at
$R_\mathrm{GC}=12-13$~kpc, which might point to local conditions that enhance the star 
formation efficiency in the far outer Galaxy.
\end{enumerate}

\section*{Acknowledgements}
PACS has been developed by a consortium of institutes led by MPE (Germany) 
and including UVIE (Austria); KU Leuven, CSL, IMEC (Belgium); CEA, LAM (France); 
MPIA (Germany); INAF-IFSI/OAA/OAP/OAT, LENS, SISSA (Italy); IAC (Spain). This 
development has been supported by the funding agencies BMVIT (Austria), 
ESA-PRODEX (Belgium), CEA/CNES (France), DLR (Germany), ASI/INAF (Italy), 
and CICYT/MCYT (Spain).
SPIRE has been developed by a consortium of institutes led by Cardiff University (UK) and 
including Univ. Lethbridge (Canada); NAOC (China); CEA, LAM (France); IFSI, Univ. Padua (Italy); 
IAC (Spain); Stockholm Observatory (Sweden); Imperial College London, RAL, UCL-MSSL, 
UKATC, Univ. Sussex (UK); and Caltech, JPL, NHSC, Univ. Colorado (USA). This development 
has been supported by national funding agencies: CSA (Canada); NAOC (China); CEA, CNES, 
CNRS (France); ASI (Italy); MCINN (Spain); SNSB (Sweden); STFC, UKSA (UK); and NASA (USA).
The ATLASGAL project is a collaboration between the Max-Planck-Gesellschaft, the European 
Southern Observatory (ESO) and the Universidad de Chile. It includes projects E-181.C-0885, 
E-078.F-9040(A), M-079.C-9501(A), M-081.C-9501(A) plus Chilean data.
This research is supported by INAF, through the
Mainstream Grant 1.05.01.86.09 ``The ultimate exploitation
of the Hi-GAL archive and ancillary infrared/mm data''
(P.I. D. Elia), by the H2020-EU.1.1. - EXCELLENT SCIENCE - European Research Council program through the ECOGAL Synergy Grant, and by the Agenzia
Spaziale Italiana (ASI) through the research contract 2018-31-HH.0. AZ thanks the support of the Institut Universitaire de France.
PP acknowledges support from FCT through the research grants UIDB/04434/2020 and 
UIDP/04434/2020. PP receives support from the fellowship SFRH/BPD/110176/2015 
funded by FCT (Portugal) and POPH/FSE (EC).
\section*{Data Availability}
The data underlying this article are available in VIALACTEA project, at
\url{http://vialactea.iaps.inaf.it/vialactea/public/HiGAL_360_clump_catalogue.tar.gz}. The catalogue of Hi-GAL clump physical properties is also hosted in the VIALACTEA knowledge base
\citep[VLKB,][]{mmol16}.


\bibliographystyle{mnras}
\bibliography{2pipaperarxiv}



\appendix

\appendix
\section{Description of physical catalogue}\label{catdescription}

The Hi-GAL physical catalogue for the inner Galaxy is arranged in two tables (high- and low-reliability SEDs), each
containing the same columns. Most columns in the new full catalogue coincide with those of the
\citetalias{eli17} catalogue and so for these we give a concise description, recommending that the
reader consult Appendix~A of \citetalias{eli17} for further details. However, for the columns specifically introduced in this paper we give
a detailed description here. 
\begin{itemize}
\item Column [1], \textit{ID}: running number of the entry.
\item Column [2], \textit{DESIGNATION}: string containing the Galactic coordinates of the source.
\item Columns [3], \textit{GLON}, and [4], \textit{GLAT}: source Galactic longitude and latitude, respectively.
\item Columns [5], \textit{RA}, and [6], {DEC}: same as columns [3] and [4], respectively, 
but for Equatorial coordinates.
\item Column [7], \textit{DESIGNATION\_70}: designation of the PACS $70~\umu$m counterpart (if available), 
as introduced in the catalogue of \citet{mol16a}.
\item Column [8], \textit{F70}: flux density (hereafter flux) of the PACS $70~\umu$m counterpart (if available), 
in Jy, as quoted by \citet{mol16a}. The null value is 0.
\item Column [9], \textit{DF70}: uncertainty of the flux in column [8].
\item Column [10], \textit{F70\_TOT}: sum of fluxes of all PACS $70~\umu$m counterparts (if available) lying
inside the half-maximum ellipse of the source detected by CuTEx \citep{mol11a} in the SPIRE $250~\umu$m maps. 
This is the flux at $70~\umu$m actually used to estimate the source bolometric luminosity and temperature.
\item Column [11], \textit{DF70\_TOT}: uncertainty of the total flux in column [10].
\item Column [12], \textit{F70\_ADD\_TOT}: sum of fluxes of all PACS $70~\umu$m counterparts (if available) 
found through targeted source extraction using a detection threshold lower than 
in \citet{mol16a}, and lying inside the ellipse at $250~\umu$m. 
This is the flux at $70~\umu$m (if any) actually used to estimate
the source bolometric luminosity and temperature where $F_\mathrm{70,tot}=0$.
\item Column [13], \textit{DF70\_ADD\_TOT}: uncertainty of the flux in column [12].
\item Column [14], \textit{ULIM\_70}: 5-$\sigma$ upper limit in the PACS $70~\umu$m band, estimated if
both $F_\mathrm{70,tot}=0$ and $F_\mathrm{70,add,tot}=0$.
\item Columns [15], \textit{DESIGNATION\_160}, [16], \textit{F160}, and [17], \textit{DF160}: 
the same as columns [7], [8], and [9], respectively, but for the PACS $160~\umu$m band.
\item Column [18] \textit{F160\_ADD}: flux of the closest PACS $160~\umu$m counterpart (if available) found,
in the $F_\mathrm{160}=0$ case, through targeted source extraction using a detection threshold lower than 
in \citet{mol16a} .
\item Column [19], \textit{DF160\_ADD}: uncertainty of the flux in column [18].
\item Column [20], \textit{ULIM\_160}: 5-$\sigma$ upper limit in the PACS $160~\umu$m band, estimated 
where both $F_\mathrm{160}=0$ and $F_\mathrm{160,add}=0$.
\item Columns [21], \textit{DESIGNATION\_250}, [22], \textit{F250}, and [23], \textit{DF250}: the 
same as columns [7], [8], and [9], respectively, but for the SPIRE $250~\umu$m band.
\item Columns [24], \textit{DESIGNATION\_350}, [25], \textit{F350}, and [26], \textit{DF350} : the 
same as columns [7], [8], and [9], respectively, but for the SPIRE $350~\umu$m band.
\item Column [27], \textit{FSC350}: SPIRE $350~\umu$m flux ``scaled'' as described in 
\citetalias{eli17}. Further details of the method are provided, e.g., in \citet{gia12}.
\item Column [28], \textit{DFSC350}: uncertainty of the flux in column [27].
\item Columns [29], \textit{DESIGNATION\_500}, [30], \textit{F500}, [31], \textit{DF500},
[32], \textit{FSC500}, and [33], \textit{DFSC500}: the same as columns~[24], [25], [26], [27],
and [28], respectively, but for the SPIRE $500~\umu$m band.
\item Column [34], \textit{DESIGNATION\_21}: designation of the MSX $21~\umu$m counterpart 
(if available), as defined in the MSX point source catalogue.
\item Column [35], \textit{F21}: flux of the closest MXS $21~\umu$m counterpart.
\item Column [36], \textit{DF21}: uncertainty of the flux in column [35]. 
\item Column [37], \textit{F21\_TOT}: sum of fluxes of all MXS $21~\umu$m counterparts (if available) 
lying inside the ellipse at $250~\umu$m.
\item Column [38], \textit{DF21\_TOT}: uncertainty of the flux in column [37], computed
as for column~[11].
\item Columns [39], \textit{DESIGNATION\_22}, [40], \textit{F22}, [41], \textit{DF22}, 
[42], \textit{F22\_TOT}, and [43], \textit{DF22\_TOT}: the same as columns [34], [35], [36], [37], 
and [38], but for the WISE $22~\umu$m band.
\item Column [44], \textit{DESIGNATION\_24}: designation of the MSX $24~\umu$m counterpart. 
A string beginning with ``MG'' indicates a source taken from the catalog
of \citep{gut15}, while a string beginning with ``D'' identifies a source specifically 
detected to complement this catalogue work \citepalias[see][]{eli17}. In addition, 
lack of a counterpart due to saturation is identified with the ``satutared'' string. 
\item Columns [45], \textit{F24}, [46], \textit{DF24}, [47], \textit{F24\_TOT}, and [48], 
\textit{DF24\_TOT}: the same as columns [35], [36], [37], and [38], respectively, but for 
the MIPSGAL $24~\umu$m band. 
\item Column [49], \textit{DESIGNATION\_870}: designation of the ATLASGAL $870~\umu$m counterpart. 
A string beginning with ``G'' indicates a source taken from the catalog
of \citet{cse14} while the string "CuTEx" identifies a source specifically 
detected for this work \citepalias[see][]{eli17}.
\item Columns [50], \textit{F870}, and [51], \textit{DF870}: the same as columns [35] and [36],
respectively, but for the ATLASGAL $870~\umu$m band. 
\item Column [52], \textit{DESIGNATION\_1100}: designation of the BGPS $1100~\umu$m counterpart 
(if available), as defined in the BGPS catalogue \citep{gin13}.
\item Columns [53], \textit{F1100}, and [54], \textit{DF1100}: the same as columns [35] and [36],
respectively, but for the BOLOCAM $1100~\umu$m band. 
\item Column [55], \textit{DFWHM250}: circularized and beam-deconvolved size (if the circularised size exceeds the 
instrumental beam size\footnote{Although at the detection step CuTEx rejects sources smaller than $1 \times$ instrumental point spread function, the subsequent photometry step grants little additional tolerance to the 2-D Gaussian fit, to better suit the source profile. This can result in an angular size estimate slightly smaller than the intrumental beam: in the present catalogue, at $250~\mu$m this is found for 214~sources only. For these sources, the originally observed size is quoted.}), of the source as estimated 
by CuTEx in the $250~\umu$m band, in arcseconds. 
\item Columns [56], \textit{DIST}: heliocentric distance of the source, in kpc, from \citet{meg21}.
 The null value, absent a distance estimate, is -999.
\item Column [57], VLSR: $V_{\mathrm{LSR}}$, in km~s$^{-1}$, assigned to the source by 
\citet{meg21}. It is important to note that the distance in column [56] is not necessarily consistent 
with this velocity, depending on the distance assignment decided by the algorithm of \citet{meg21}. See flag in column [58].
\item Column [58], DFLAG: flag indicating whether the distance in column [56] is derived directly from
the $V_{\mathrm{LSR}}$ through the Galactic rotation curve (value ``1''), or either assigned based on other 
criteria by the decision tree of \citet{meg21} or not available at all (value ``0'').
\item Column [59], \textit{R\_GAL}: Galactocentric radius of the source, in kpc. The null value, absent a 
distance estimate, is -999.
\item Column [60], \textit{DIAM}: source linear diameter, in pc, obtained combining columns~[55]
and [56].
\item Column [61], \textit{M\_LARS}: Larson's mass, in Solar masses, evaluated as described in 
Section~\ref{sedsel}. The null value, absent a distance, is -999.
\item Column [62], \textit{FIT\_TYPE}: flag indicating whether the expression of the MBB fitted 
to the source SED is given by Equation~3 (optically ``thick'' case, ``tk'' flag) or 
Equation~8 (``thin'' case, ``tn'' flag) of \citet{eli16}, respectively.
\item Column [63], \textit{EVOL\_FLAG}: flag indicating the evolutionary classification of the
source (0: starless unbound; 1: pre-stellar; 2: protostellar).
\item Column [64], \textit{MASS}: clump total mass, in units of Solar masses. Absent a distance, 
the fit is evaluated for a hypothetical distance of 1~kpc and the corresponding mass is quoted as a negative value.
\item Column [65], \textit{DMASS}: uncertainty of the mass in column [64].
\item Column [66], \textit{TEMP}: dust temperature of the clump, in~K, derived from the MBB 
fit.
\item Column [67], \textit{DTEMP}: uncertainty of the temperature in column [66].
\item Column [68], \textit{LAM\_0\_TK}: value of $\lambda_0$ \citep[see Equation~3 of][]{eli16}, 
in $\umu$m, derived from the MBB fit. The null value, in correspondence with the value ``tn'' 
for the flag FIT\_TYPE in column [62], is 0.
\item Column [69], \textit{L\_BOL}: bolometric luminosity, in units of Solar luminosity, estimated as 
described in \citetalias{eli17}. Absent a distance estimate,
it is calculated for a hypothetical distance of 1~kpc and quoted as a negative 
value. 
\item Column [70], \textit{LRATIO}: ratio of the bolometric luminosity in column~[69] to
the luminosity computed in the sub-millimetre ($\lambda \geq 350 \umu$m). Being 
distance-independent, it is evaluated also for sources without a distance estimate.
\item Column [71], \textit{T\_BOL}: bolometric temperature, in~K.
\item Column [72], \textit{SURF\_DENS}: surface density, in g~cm$^{-2}$, calculated by dividing 
the mass in column~[64] by the area of a circle having the diameter in column~[60]. Being 
distance-independent, it is evaluated also for sources without a distance estimate.
\end{itemize}

\section{Statistics of MIR ancillary photometry}\label{mirappendix}

In Section~\ref{building} the Galactic plane surveys used for ancillary photometry near
20~$\umu$m were presented. MIPSGAL certainly offers a better sensitivity compared to WISE (W4 band) 
and MSX, but suffers from a lower saturation limit and, moreover, covers only the central third of 
the Galactic plane. For \citetalias{eli17} the saturation issue made it necessary to 
complement MIPSGAL data with WISE and MSX data, while additionally for this paper the issue of coverage 
has to be taken into account. For example, this implies that our ability to complement the SEDs of Hi-GAL 
sources in the outer Galaxy is limited by the WISE sensitivity. 

\begin{figure}
\centering
\includegraphics[width=8.5cm]{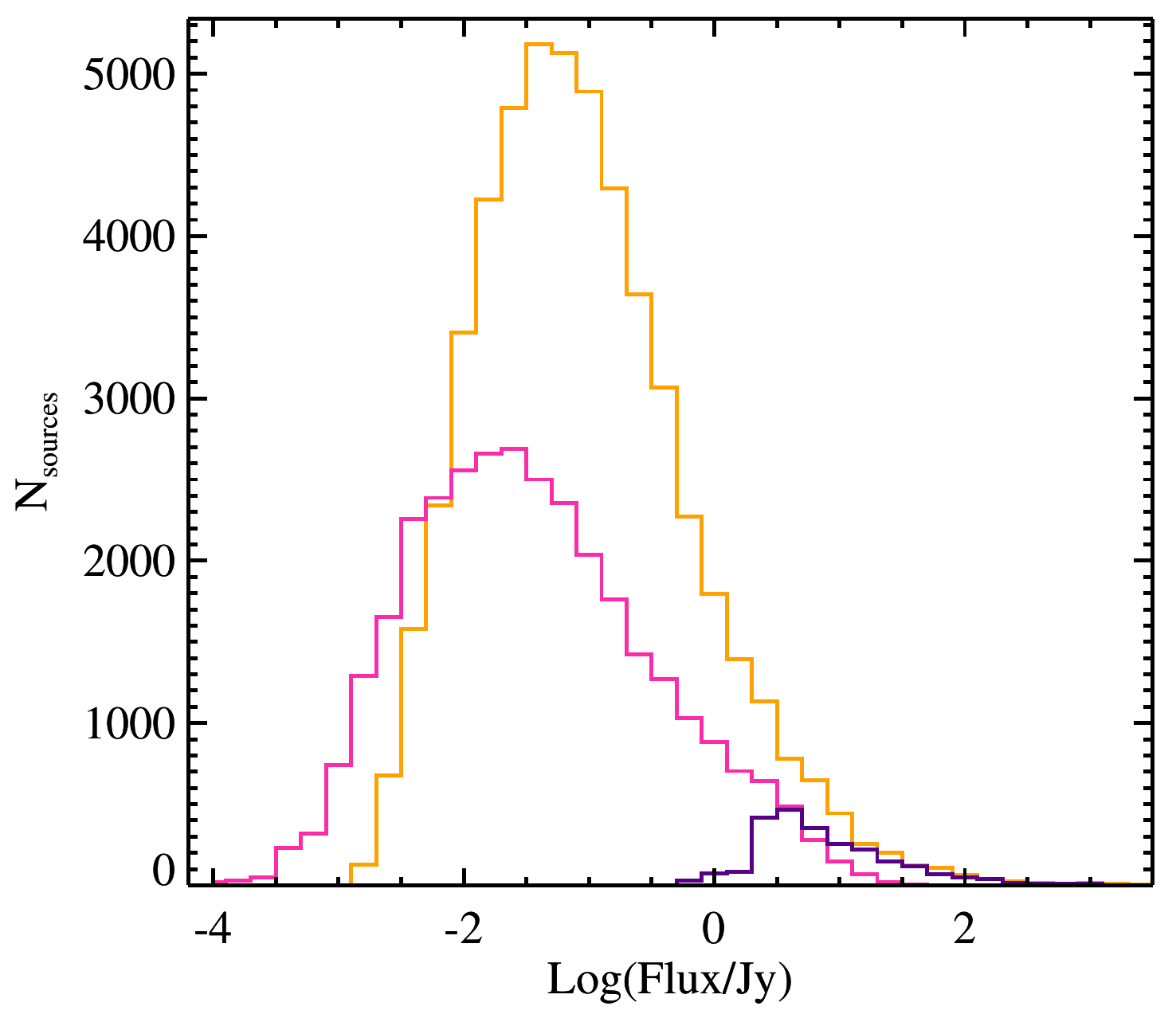}
\caption{Distributions of fluxes of counterparts of the Hi-GAL sources 
in the catalogue, at 24~$\umu$m (MIPSGAL, magenta), 22~$\umu$m (WISE, orange), 
and 21~$\umu$m (MSX, purple).}
\label{mirfluxes}
\end{figure}

To quantify this bias between 
inner and outer Galaxy SEDs, Fig.~\ref{mirfluxes} shows the distributions of MIPSGAL, WISE,
and MSX fluxes associated with the Hi-GAL sources in the catalogue. In our sample, 
MIPSGAL seems to be around 3 times (0.5 in logarithm) more sensitive than WISE. From the standpoint of completeness, in 
the inner Galaxy sources with $F_{24} > 0.03$~Jy have a good chance of being detected in MIPSGAL,
whereas in the outer Galaxy sources need to have $F_{22} > 0.1$~Jy. Thus in the outer Galaxy 
many more potential source counterparts will remain undetected in the
MIR. Finally, it can be seen that MSX fluxes are all above 1~Jy.

\begin{figure}
\centering
\includegraphics[width=8.5cm]{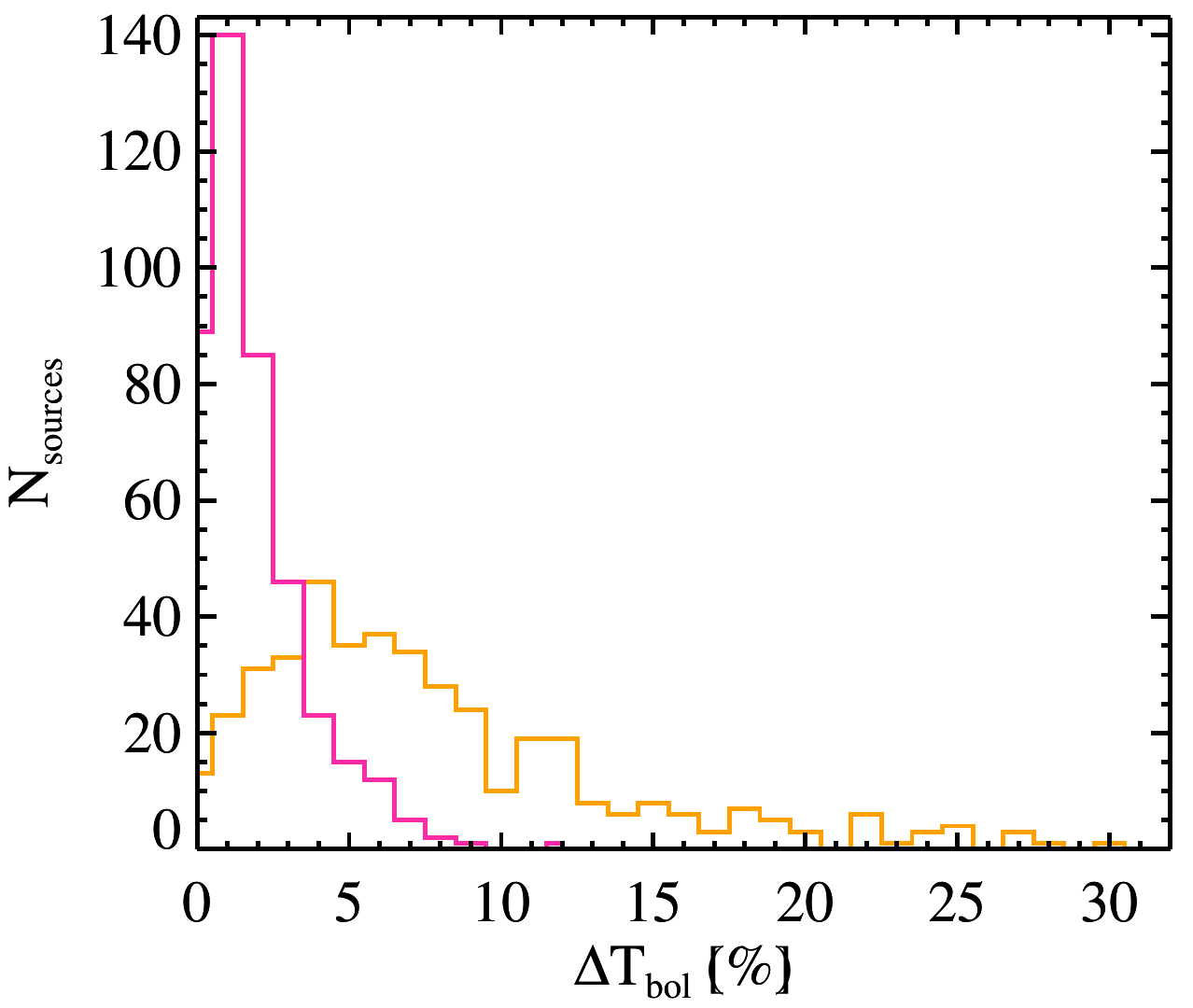}
\caption{Distributions of the percentage increase of $T_{\textrm{bol}}$ for relatively faint
($F_{70} < 2$~Jy) and MIR-dark Hi-GAL sources produced by extending their SED with a MIR flux at the corresponding survey completeness limit (cf. Fig.~\ref{mirfluxes}):
MIPSGAL, $F_{24} = 0.03$~Jy at 24~$\umu$m (magenta) and WISE, $F_{22} = 0.1$~Jy at 22~$\umu$m (orange).
}
\label{tboltest}
\end{figure}

Consequently, it is important to quantify how the bolometric 
temperature (Section~\ref{tbolpar}) is underestimated by the failure to detect a faint MIR counterpart. A simple way to evaluate this for a Hi-GAL SED without a MIR flux is to artificially add a flux close to the apparent completeness
limit of WISE or MIPSGAL at 22 or 24~$\umu$m, respectively. 
To consider only realistic cases, 
we chose relatively faint MIR-dark protostellar sources by imposing $F_{70} < 2$~Jy (\ntesttbol~sources).
We recalculated the bolometric temperature after adding a hypothetical detection of $F_{22} = 0.1$~Jy at 22~$\umu$m, 
or $F_{24} = 0.03$~Jy at 24~$\umu$m, respectively, to their SED. The percentage increases in $T_{\textrm{bol}}$ are given in Fig.~\ref{tboltest}.

The case of WISE is relevant to the entire sky area covered by Hi-GAL and as expected shows a larger effect than for MIPSGAL.
Although the increase can reach 30~per cent, the 90th percentile is \testtbolninetyttwo~per cent, and the median \testtbolfiftyttwo~per cent. The latter is equivalent to the statement that failure to detect the MIR source in WISE near the sensitivity limit leads typically to an underestimate of $T_{\textrm{bol}}$ by about 7~per cent.
For sources with 
Herschel fluxes larger than those used in this test, this effect would be even lower because an even greater portion of the integral of the SED is produced at lower frequencies.

Sources at longitudes $-68^\circ < \ell < 69^\circ$ (cf. Fig.~\ref{surveycoverage})
can benefit from the better sensitivity
achieved by MIPSGAL at 24~$\umu$m. In this case, our test predicts a median underestimate of $T_{\textrm{bol}}$ 
by just \testtbolfiftytfour~per cent. 
Starting from the definition of $T_{\textrm{bol}}$
\citep{mye93}, it is possible to show that the variations produced by introducing a flux at relatively similar frequencies (as is the case for WISE and 
MIPSGAL) are in a nearly constant ratio,
dependent on the fluxes set for the test. The ratio is about 3.67 in this specific case, in agreement with the median values found for the two cases.

\bsp	

\clearpage

\noindent
Author afilliations\\
\iaps INAF-IAPS, via del Fosso del Cavaliere 100, I-00133 Roma, Italy\\
\chil Departamento de Astronom\'ia, Universidad de Chile, Casilla 36-D, Santiago, Chile\\
\lamm Aix Marseille Univ., CNRS, LAM, Laboratoire d'Astrophysique de Marseille, Marseille, France\\
\iuf Institut Universitaire de France, Paris, France\\
\toron Canadian Institute for Theoretical Astrophysics, University of Toronto, McLennan Physical Laboratories, 60 St. George Street, Toronto, Ontario, Canada\\
\oafi INAF, Osservatorio Astrofisico di Arcetri, Largo E. Fermi 5, I-50125, Firenze, Italy\\
\calg Department of Physics \& Astronomy, University of Calgary, AB, T2N 1N4, Canada\\
\cardiff School of Physics and Astronomy, Cardiff University, Cardiff CF24 3AA, Wales, UK\\
\liver Astrophysics Research Institute, Liverpool John Moores University, Liverpool Science Park Ic2, 146 Brownlow Hill, Liverpool, L3 5RF, UK\\
\stsi Space Telescope Science Institute, 3700 San Martin Dr., Baltimore, MD, 21218, USA\\
\caltech Infrared Processing Analysis Center, California Institute of Technology, Pasadena, CA 91125, USA\\
\porto Instituto de Astrof\'isica e Ci{\^e}ncias do Espa\c{c}o, Universidade do Porto, CAUP, Rua das Estrelas, PT4150-762 Porto, Portugal\\
\esa European Space Agency (ESA), European Space Research and Technology Centre (ESTEC), Keplerlaan 1, 2201 AZ Noordwijk, The Netherlands\\
\irabo Italian ALMA Regional Centre, INAF-IRA, Via P. Gobetti 101, I-40129, Bologna, Italy\\
\koln I. Physikalisches Institut, Universit\"at zu K\"oln, Z\"ulpicher Str. 77, D-50937 K\"oln, Germany\\
\unile Dipartimento di Matematica e Fisica, Universit\`a del Salento, I-73100, Lecce, Italy\\
\inafle INAF - Sezione di Lecce, via Arnesano km 5, 73100, Lecce, Italy\\
\chalm Dept. of Space, Earth \& Environment, Chalmers University of Technology, Gothenburg, Sweden\\
\charlot Dept. of Astronomy, University of Virginia, Charlottesville, Virginia 22904, USA\\
\nwest Center for Interdisciplinary Exploration and Research in Astrophysics and Department of Physics and Astronomy, Northwestern University,\\2145 Sheridan Road, Evanston, IL 60208-3112, USA\\
\colo Center for Astrophysics and Space Astronomy, University of Colorado, Boulder, CO, 80309, USA\\
\buda Konkoly Observatory, Research Centre for Astronomy and Earth Sciences, Konkoly-Thege Mikl\'os \'ut 15-17, 1121 Budapest, Hungary\\
\oar INAF - Osservatorio Astronomico di Roma, via di Frascati 33, I-00078, Monte Porzio Catone, Italy\\
\torv Dipartimento di Fisica, Universit\`a di Roma ``Tor Vergata'', Via della Ricerca Scientifica 1, I-00133, Roma, Italy\\
\diet DIET, Universit\`a di Roma ``La Sapienza'', I-00185 Roma, Italy\\


\label{lastpage}
\end{document}